\documentclass[notitlepage, superscriptaddress, 11pt]{revtex4-1}
\usepackage{enumitem}
\usepackage{mathtools}
\usepackage{graphicx,graphics,epsfig,subfigure,times,bm,bbm,amssymb,amsmath,amsfonts,amsthm,mathrsfs,MnSymbol}
\usepackage[matrix,frame,arrow]{xypic}
\usepackage[normalem]{ulem}
\usepackage{slashed}
\usepackage{color}
\usepackage[usenames,dvipsnames,svgnames,table]{xcolor}
\definecolor{darkblue}{rgb}{0.0,0.0,0.3}
\usepackage[colorlinks=true,
            linkcolor=red,
            urlcolor= darkblue,
            citecolor=blue]{hyperref}

\usepackage{enumerate}
\usepackage{epstopdf}
\usepackage{relsize}

\newtheorem{definitionenv}{Definition}
\newtheorem{remarkenv}[definitionenv]{Remark}

\newtheorem{exampleenv}{Example}

\newcommand{\bes} {\begin{subequations}}
\newcommand{\ees} {\end{subequations}}
\newcommand{\bea} {\begin{eqnarray}}
\newcommand{\eea} {\end{eqnarray}}

\newcommand{\beq}{\begin{equation}}

\newcommand{\eeq}{\end{equation}}

\newcommand{\ignore}[1]{}

\def\t{\theta}

\def\>{\rangle}
\def\<{\langle}

\newcommand{\bk}[2]{\<{#1}|{#2}\>}

\newcommand{\bra}[1]{\langle#1|}
\newcommand{\ket}[1]{|#1\rangle}

\begin{document}
\title{Unitary-Coupled Restricted Boltzmann Machine Ansatz for Quantum Simulations}
\author{Chang Yu Hsieh}
\email{kimhsieh@tencent.com}
\affiliation{Tencent Quantum Lab, Shenzhen, Guangdong 518057, China}
\author{Qiming Sun} 
\affiliation{Tencent America, Palo Alto, CA 94306 , United States}
\author{Shengyu Zhang} 
\affiliation{Tencent Quantum Lab, Shenzhen, Guangdong 518057, China}
\author{Chee Kong Lee}
\email{cheekonglee@tencent.com}
\affiliation{Tencent America, Palo Alto, CA 94306 , United States}
\begin{abstract}
Neural-Network Quantum State (NQS) has attracted significant interests as a powerful wave-function ansatz to model quantum phenomena. In particular, a variant of NQS based on the restricted Boltzmann machine (RBM) has been adapted to model the ground state of spin lattices and the electronic structures of small molecules in quantum devices.  Despite these progresses, significant challenges remain with the RBM-NQS based quantum simulations. In this work, we present a state-preparation protocol to generate a specific set of complex-valued RBM-NQS, that we name the unitary-coupled RBM-NQS, in quantum circuits. This is a crucial advancement as prior works deal exclusively with real-valued RBM-NQS for quantum algorithms. 
With this novel scheme, we achieve (1) modeling complex-valued wave functions, (2) using as few as one ancilla qubit to simulate $M$ hidden spins in an RBM architecture, and (3) avoiding post-selections to improve scalability. 
%
\end{abstract}

\maketitle


Hybrid quantum-classical (HQC) algorithms\cite{mcclean2016theory} offer an exciting avenue to explore the potential of a noisy intermediate-scale quantum\cite{preskill2018quantum} (NISQ) device without quantum error corrections. The HQC algorithms run on parametrized quantum circuits aiming to minimize an objective function, such as the average energy. The Variational Quantum Eigensolver\cite{peruzzo_natcomm_14} (VQE) and Quantum Approximate Optimization Algorithm\cite{Farhi2014} (QAOA) are two prominent examples leading the current wave of HQC algorithm developments. In particular, VQE has been experimentally demonstrated on several leading platforms of quantum computations\cite{peruzzo_natcomm_14,kandala2017hardware,hempel2018quantum,colless2018computation,hempel2018quantum,sagastizabal2019experimental}. These encouraging experimental outcomes strengthen our anticipation that quantum simulations\cite{aspuru2005simulated,li2019variational,cao2018quantum,mcardle2018quantum,childs2018toward} should be among the first set of applications to benefit from quantum computations. Nevertheless, it is also becoming increasingly clear that further developments\cite{wecker_pra2015,kuhn2018accuracy,li2019electronic,reiher2017elucidating,kivlichan2018quantum,babbush2018low} are required to improve VQE and similar HQC algorithms if the goal is to establish an unambiguous quantum advantage for problems of realistic interests. For instance, many recent developments address the following aspects: (1) design novel wave function ansatz\cite{wecker_pra2015,ryabinkin2018qubit,ryabinkin2019iterative,dallaire2019low,kandala2017hardware,romero2018strategies,o2019calculating,benfenati2019extended} with efficient usage of variational parameters and circuit depth, (2) reduce the number of required measurements\cite{zhao2019measurement,rubin2018application,crawford2019efficient,izmaylov2019revising,izmaylov2019unitary,verteletskyi2019measurement,
huggins2019efficient,mitarai2019methodology}, and (3) overcome the challenge of high-dimensional optimization\cite{mcardle2018img,yang2017optimizing,zhu2018training,shaydulin2019multistart,guerreschi2017practical,nakanishi2019sequential,parrish2019jacobi,
parrish2019hybrid,schuld2019evaluating,moseley3bayesian} needed for training parametrized circuits. Any of these technical challenges could potentially become a computational bottleneck for a HQC algorithm beyond the small-scale testings reported in the recent literature.

In this work, our primary focus is to investigate whether neural-network quantum states (NQS) \cite{sarma2019machine,melko2019restricted,jia2019quantum} can be tailored to better fit the paradigm of hybrid quantum-classical algorithms. We focus on a particular form of NQS based on the restricted Boltzmann Machine (RBM) architecture. Within the communities of computational many-body physics, quantum information and condensed matter physics, there is a growing trend of adopting neural networks techniques, such as the RBM architecture, for various applications. Notable examples include identification of different phases of matter\cite{torlai2016learning,carrasquilla2017machine, kaubruegger2018chiral,koch2018mutual,czischek2018quenches,lu2019efficient}, scalable quantum state tomography\cite{xu2018neural,torlai2019integrating}, efficient sampler to accelerate Monte Carlo simulations\cite{huang2017recommender,wang2017exploring,inack2018projective}, quantum error correction codes\cite{torlai2017neural,bausch2018quantum,zhang2018efficient,jia2019efficient}, and variational ansatz for many-body simulations\cite{carleo2017solving,deng2017machine,saito2018method,hartmann2019neural,nagy2019variational,yoshioka2019constructing,luo2019backflow, torlai2018latent,nomura2017restricted,jonsson2018neural,glasser2018neural}. Especially, the work \cite{carleo2017solving} by Carleo and Troyer demonstrated that a complex-valued RBM model can efficiently model manybody wavefunctions with fewer variational parameters than tensor-network methods for few spin-lattice models. Subsequent investigations\cite{deng2017quantum,huang2017neural,Gao2017,clark2018unifying,chen2018equivalence} have further clarified and affirmed the usefulness of variants of Boltzmann Machines to model many-body quantum states of complex systems. These encouraging results and rapidly accumulated knowledge about RBM-NQS motivated us to investigate whether it is suitable to apply this family of ansatz in the context of quantum simulation algorithms.  In the most part of this work, we should use RBM-NQS and NQS interchangeably when there is no risk of ambiguity,.

While there are already quantum simulation algorithms\cite{xia2018rbm,gardas2018dwave} using NQS as variational ansatz with encouraging results, some fundamental obstacles limit their scope of applications. For instance, the existing approaches require the preparation of an extended wave function composed of all visible and hidden spins. As each hidden spin is explicitly modeled with an ancilla qubit, these prior methods consume too many qubits, which are expensive resources for near-term quantum devices.  Furthermore, there is no scalable strategy for preparing a general NQS in quantum circuits. This is because many NQS can only be obtained via non-unitary transformations on an input state. Existing NQS based simulation algorithm\cite{xia2018rbm} relies on a probabilistic post-selection to achieve the non-unitary operations. 
Finally, the lack of complex parameters severely limits the usefulness of the NQS for quantum simulations. For instance, (1) Complex-valued wave functions allow us to simulate fermions in time reversal symmetry breaking systems such as electrons in the presence of a magnetic field. (2) To simulate quantum dynamics, it is a necessity to account for the accumulation of dynamical phase factors.  

The aforementioned limitations certainly have cast doubts on whether the NQS should be used in the cquantum circuits; despite many of their theoretical merits as wave-function ansatz and convincing demonstrations in numerical studies (i.e. classical simulations) in a broad range of physical systems cited above. To address these deficiencies, we propose a state-preparation protocol for creating complex-valued NQS in a quantum circuit. In particular, the state-preparation protocol does not use $N+M$ qubits to model $N$ visible spins and $M$ hidden spins explicitly. This is because every term in a RBM Hamiltonian commutes with each other, we can explicitly arrange the order in which the unitary gates acting on the hidden spins. Hence, a single ancilla qubit (representing one hidden spin) can be recycled upon measurement and be reused to represent another hidden spin in a subsequent stage. This qubit-recyle scheme \cite{liu2019recycle,Huggins_2019} tremendously reduces the number of physical qubits (down to just one extra ancilla at the bare minimum) needed to execute the proposed state-preparation protocol. This advantage cannot be underestimated as typical RBM-NQS might use as many as $M=$Poly($N$) hidden spins.
To avoid the probabilistic scheme to mediate the non-unitary couplings between the visible and hidden layers, we further propose to only consider unitary couplings between visible and hidden spins under the RBM architecture in the main text.

As extensively discussed in Supplementary II and illustrated with numerics reported in the Result section, we show that arbitrary NQS can be systematically mapped onto these unitary-coupled RBM-NQS without sacrificing expressive power to model a variety of quantum systems in our empirical studies.  In Supplementary V, we also provide an extension of the state-preparation protocol to handle arbitrary complex-valued coupling parameters across layers. Furthermore, we propose a novel approach to avoid post-selections on hidden-spin states.
In this scheme, we decompose a single NQS into an ensemble of modified NQS, which are essentially outputted by the quantum circuit when the post-selection fails (i.e. the outcome of the projective measurement is not the desired one). Under the existing scheme, one would either discard the present result and restart after a failed post-selection or attempt to recover the input quantum state from a failed post-selection. In this ensemble scheme, one does not have to go through either of these repetitive procedures. In our case, estimates of observables on the original NQS can be extracted from the ensemble of ``post-selection failures" within a Monte Carlo framework.

Finally, due to the ``diagonal structure" of the NQS ansatz, the simulation algorithm based on the imaginary-time dependent variational principle \cite{mcardle2018img} can be dramatically simplified as one just needs to perform measurement on one quantum circuit instead of working with $\mathcal{O}(N_{\text{var}}^2)$ different quantum circuits where $N_{\text{var}}$ is the number of total variational paramters.  For large-scale simulations, $N_{\text{var}}\propto \mathcal{O}(MN)$ could be a huge number. In summary, the present work has both expanded the scope of application for NQS based hybrid quantum-classical simulations and has lowered the barrier for experimental implementations. In order to frame the significance of this work in a better perspective, we summarize and clarify the challenges we set out to address in prior works in Supplementary I.

\section*{Results}
\textbf{Preparing a complex-valued NQS in a quantum circuit.} A brief introduction of NQS may be found in the Method section. All NQS may be obtained from at least one bipartite Ising Hamiltonian $
\hat{H}_{RBM}(\theta) = \sum_{i} b_i \hat{v}^z_i +  \sum_{j} m_j \hat{h}^z_j + \sum_{ij}W_{ij} \hat{v}^z_i \hat{h}^z_j$, 
where $\hat v^z_i$ or $\hat h^z_j$ is the Pauli Z operator for the visible or hidden qubit, respectively. We also denote complex-valued RBM parameters as $\theta = [b_1,\cdots, b_N, m_1 \cdots, m_M, W_{11} \cdots, W_{NM}]$, and use subscripts $R$ and $I$ to denote the real and imaginary parts, respectively. The complex-valued NQS can be created with a two-step approach. First, entangle $N+M$ qubits (including all visible and hidden spins of an RBM architecture) according to
\begin{eqnarray}\label{eq:vhwf}
    |\Psi_{vh} (\theta)  \rangle &=& \left[\frac{e^{\hat{H}_{RBM}(\theta)} }{\sqrt{\prescript{}{vh}{\bra{++ \dots +}}
    e^{2\hat{H}^R_{RBM}(\theta)}\ket{++\dots+}_{vh}}}\right]\ket{++ \dots +}_{vh}, 
\end{eqnarray}
where $\ket{+}=\frac{1}{\sqrt{2}}(\ket{0}+\ket{1})$, $\hat H^R_{RBM}(\theta)$ is the Hermitian part of the RBM Hamiltonian and the subscript $vh$ denotes all visible and hidden (ancilla) qubits. Equation \ref{eq:vhwf} gives a conceptually simple wave function that could be generated by first applying single-qubit transformations $\exp(b_i\hat v^z_i )$ and $\exp(m_j\hat h^z_j )$ on individual qubits followed by
$\exp(W_{ij}\hat v^z_i \hat h^z_j)$ to couple qubits. The quantum operations are non-unitary when RBM parameters take on real parts, i.e. $b_i^R \neq 0$, $m_j^R \neq 0$ or $W^{R}_{ij} \neq 0$.  In general, the non-unitary two-qubit operation mediating entanglement across the visible-hidden layer are difficult to implement. One could adopt the probabilistic scheme introduced in Ref.~\cite{xia2018rbm} to generate the inter-layer couplings with an extra ancilla qubit. However, for complex-valued wave function, this probabilistic approach in Ref.~\cite{xia2018rbm} is difficult to scale with the number of qubits involved.

Once the extended wave function $\ket{\Psi_{vh}(\theta)}$ is generated, all ancilla qubits (i.e. hidden spins) are post-selected for $\ket{+}_h$ and the desired NQS in Eq.~\ref{eq:nnqs} is reconstructed in the quantum circuit,   
\begin{eqnarray}\label{eq:bmcircuit}
     |\Psi_v (\theta)  \rangle &=& \frac{\prescript{}{h}{\left\langle ++ \dots + \vert \Psi_{vh}(\theta) \right\rangle}}{\sqrt{ \langle \Psi_{vh}(\theta) | \hat{P}^{(h)}_+  | \Psi_{vh}(\theta)\rangle }} 
      \nonumber \\
     & = & \frac{1}{N_v} \sum_{\mathbf{h}} e^{\hat H_{RBM}(\theta,\mathbf{h})} \ket{++\dots+}_v,
\end{eqnarray}
where $\hat{P}_+^{(h)} = \ket{++\dots+}_h\bra{++\dots+}$,  $N_v = \sqrt{\prescript{}{vh}{\langle ++\cdots+ |}e^{\hat{H}_{RBM}(\theta)} \hat{P}^{(h)}_+ e^{\hat{H}_{RBM}(\theta)} | ++\cdots+ \rangle_{vh}}$ and $e^{\hat H_{RBM}(\theta,\mathbf h)}$ is
an operator acting on the visible spins only as we replace the Pauli operator $\hat h^z_j$ with a binary value ($\pm 1$) of $h_j \in \mathbf{h}=[h_1,\cdots , h_M]$. From the second line of Eq.~\ref{eq:bmcircuit}, it is clear that the hidden spins jointly mediate a specific quantum transformation, $\sum_{\mathbf{h}} e^{\hat H_{RBM}(\theta,\mathbf{h})}$, on the visible-spin wave function.  Since this transformation is non-unitary in general, the amplitude-amplification type of techniques\cite{brassard2002quantum,Berry_oaa_2014} is not immediately applicable to enforce the post-selection. This inconvenient fact creates another obstacle to prepare complex-valued NQS in quantum circuits. 

\begin{figure}[h!]
\centering
\includegraphics[width=.8\linewidth]{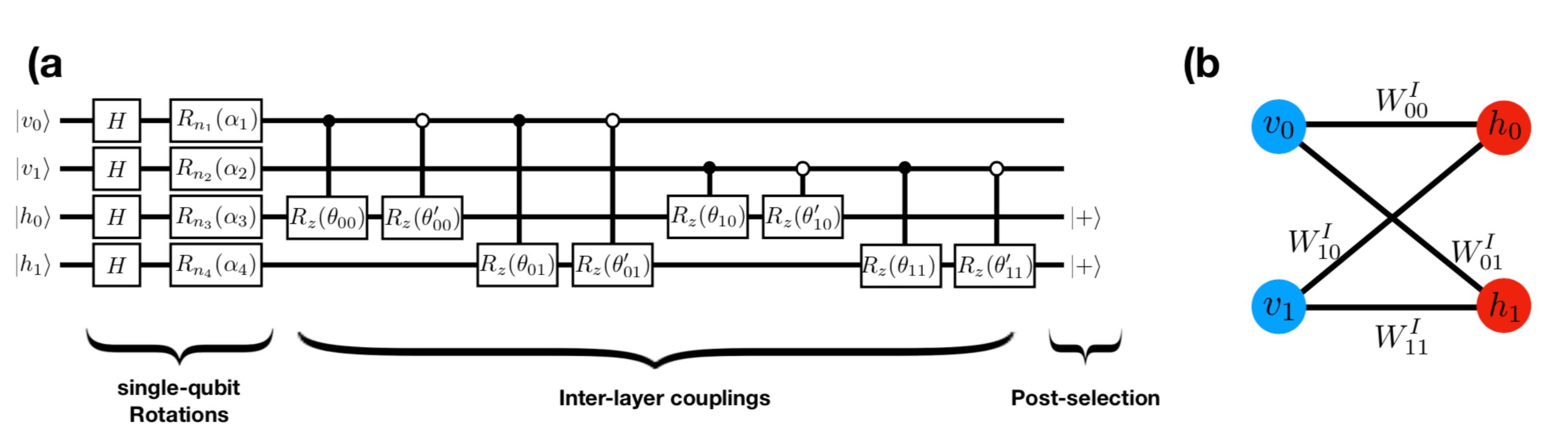}
\caption{Unitary-Coupled RBM-NQS. (a): quantum circuit to prepare a two-qubit unitary-coupled RBM state having two hidden spins.
The inter-layer coupling is mediated by unitary gates. Note $\theta^\prime_{ij}=-\theta_{ij}$. (b): schematic of the RBM state generated
by the quantum circuit in panel (a).}
\label{fig:circuit}
\end{figure}

In the rest of this section, we describe a scalable state-preparation protocol that overcomes the two obstacles. In particular, we implement complex-valued NQS using as few as one ancilla qubit and entirely avoids the post-selection. To simplify presentations, we illustrate how to prepare a subset of NQS that we dub the unitary-coupled RBM-NQS, which only allow purely imaginary-valued inter-layer couplings $W_{ij}=iW^I_{ij}$. Generation of unitary-coupled RBM-NQS bypass the inherent challenge to mediate entanglement via non-unitary transformations. Figure \ref{fig:circuit}a gives a circuit diagram of preparing a unitary-coupled RBM-NQS composed of two visible spins with inter-qubit couplings mediated by two hidden spins. The Hadarmard gates prepare the $\ket{+}$ state, and the parametrized single-qubit rotations are not fixed along the $z$ axis because of the non-unitary operations, $\exp\left(b_i^R \hat v^z_i\right)$ and $\exp\left(m_i^R \hat h^z_i\right)$. In the circuit diagram of panel a, the single-qubit rotations $R_n(\theta)$ are determined via relations of the form $\exp\left(b_i^R \hat v^z_i\right)\ket{+}/c = R_n(\theta)\ket{+}$ with the normalization factor $c=\sqrt{\bra{+}\exp\left(2 b_i^R \hat v^z_i\right)\ket{+}}$.
Figure \ref{fig:circuit}b gives a schematic depicting the RBM state generated by the circuit in panel a.
More importantly, as explained in the supplementary II, the unitary-coupled RBM-NQS does not necessarily suffer loss of expressive power. While we claim it is better to model quantum systems with unitary-coupled RBM-NQS for near-term applications; there is a straightforward extension of the current protocol to generate arbitrary complex-valued NQS in case it is desired. We defer the discussion of this extension to Supplementary V.

\textbf{Scalable preparation of a unitary-coupled RBM-NQS in a quantum circuit.}
To begin, we note Eq.~\ref{eq:bmcircuit} can be cast in an alternative form
\begin{eqnarray}\label{eq:qrecycle}
\ket{\Psi_v(\theta)} & = & \frac{1}{N_v}
       \left[\prescript{}{}{\bra{+}}\left[e^{\hat h^z_M \left(m_M + \sum_i iW^I_{iM}\hat v^z_i\right)}\right]\ket{+}\right]_{M}
       \left[\prescript{}{}{\bra{+}}\left[e^{\hat h^z_{M-1}  \left( m_{M-1} + \sum_i iW^I_{iM-1} \hat v^z_i\right)}\right]\ket{+}\right]_{M-1} \cdots 
       \nonumber \\
       & & \times \left[\prescript{}{}{\bra{+}}\left[e^{\hat h^z_1  \left(m_1 + \sum_i iW^I_{i1}\hat v^z_i\right)}\right]\ket{+}\right]_{1}\,\,
       e^{\sum_i b_i \hat v^z_i} \ket{++\cdots+}_v.
\end{eqnarray}
Each block, $\left[\prescript{}{}{\bra{+}}[ \cdots ]\ket{+}\right]_{j}$, encodes j-th hidden spin's effects on all visible ones. Clearly, one can use a single ancilla qubit
to implement these transformations sequentially. As shown in Eq.~\ref{eq:qrecycle}, we specifically consider
$W_{ij}=iW^I_{ij}$ for the unitary-coupled RBM-NQS.
Our proposed approach to bypass the post-selection of $\ket{+}$ on all hidden spins is inspired by the
following observation,
\begin{eqnarray}\label{eq:block-decomp}
 \prescript{}{}{\bra{+}}\left[e^{\hat h^z_j \left(m_j + \sum_i iW^I_{ij}\hat v^z_i\right)}\right]\ket{+}
& = & \sum_{s=\pm} \bra{+} e^{m^R_j \hat h^z_j} \ket{s} \bra{s} e^{\left(im^I_j+\sum_i iW^I_{ij}\hat v^z_i\right) \hat h^z_j} \ket{+} \nonumber \\
& = & \sum_{s=\pm} R_{s}(m_j^R) \bra{s} e^{\left(im^I_j+\sum_i iW^I_{ij}\hat v^z_i\right) \hat h^z_j} \ket{+},
\end{eqnarray}
where 
a resolution of identity $\sum_{s_j=\pm}\ket{s_j}\bra{s_j}$ for the ancilla qubit is inserted in the middle and 
$R_{s}(m^R_j)=\bra{+} e^{m^R_j \hat h^z_j} \ket{s} $ can be computed classically as it is the transformation matrix element associated with a single qubit. Not getting $R_{s}(m_j)$ experimentally is the key to avoid post-selection. Note that the decomposition of $\exp\left(\hat h^z_j \left(m_j + \sum_i iW^I_{ij}\hat v^z_i\right)\right)$ introduced in Eq.~\ref{eq:block-decomp} is exact as these operators commute.
Using Eq.~\ref{eq:block-decomp}, we re-write Eq.~\ref{eq:qrecycle},
\begin{eqnarray}\label{eq:qrecycle2}
\ket{\Psi_v(\theta)} & = &  \sum_{s_M=\pm} \cdots \sum_{s_1=\pm}  \frac{1}{N_v}\left(\prod_{j=1}^MR_{s_j}(m^R_j)\right)
       \prescript{}{}{\bra{s_M}}\left[e^{\hat h^z_M \left(i m^I_M + \sum_i iW^I_{iM}\hat v^z_i\right)}\right]\ket{+} \cdots \nonumber \\
       \nonumber \\
       & & \times \prescript{}{}{\bra{s_1}}\left[e^{\hat h^z_1  \left(i m^I_1 + \sum_i iW^I_{i1}\hat v^z_i\right)}\right]\ket{+} \,\,
       e^{\sum_i b_i \hat  v^z_i} \ket{++\cdots+}_v \nonumber \\
& = & \sum_{s_M=\pm} \cdots \sum_{s_1=\pm} \frac{N_{\vec s}}{N_v}\left(\prod_{j=1}^MR_{s_j}(m^R_j)\right )\ket{\Psi_v^{\vec{s}}(\theta)}, 
\end{eqnarray}
where $\vec{s}=[s_1,\cdots,s_M]$. $\ket{\Psi_v^{\vec{s}}(\theta)}$ is a visible-spin wave function created by projecting hidden spins onto basis states $\ket{s_1\cdots s_M}_h$ instead of enforcing the post-selection $\ket{++\cdots+}_h$. Due to the decomposition introduced in Eq.~\ref{eq:block-decomp}, only a portion of $\exp(\hat H_{RBM})$ contributes to the generation of $\ket{\Psi_v^{\vec{s}}(\theta)}$ and $N_{\vec s}$ is the normalization to keep
$\bk{\Psi_{v}^{\vec{s}}}{\Psi_{v}^{\vec{s}}}=1$. While there is no post-selection in Eq.~\ref{eq:qrecycle2}, there is now a summation over all possible $\vec{s}$.  

Instead of working directly with Eq.~\ref{eq:qrecycle2}, we look for an alternative approach based on the fact that we are primarily interested in the expectation value of an observable $\hat O$, which can be formulated as
\begin{eqnarray}\label{eq:expo}
\bra{\Psi_v(\theta)} \hat O \ket{\Psi_v(\theta)} & = & \int dz \vert\bk{\mathbf z}{\Psi_v(\theta)}\vert^2 \left[\int dz^\prime  O(z,z^\prime) \frac{\bk{\mathbf z^\prime}{\Psi_v(\theta)}}{\bk{\mathbf z}{\Psi_v(\theta)}}\right]. 
\end{eqnarray}
The equation above can be interpreted as follows. The expectation value of an observable $\hat O$ can be turned into the average of the expression inside the square bracket if we can efficiently sample $z$
according to the probability density $|\bk{\mathbf z}{\Psi_v(\theta)}|^2$, i.e. projecting the NQS in some basis. 
We further analyze this probability density by exposing the details of the RBM architecture,
\begin{eqnarray}\label{eq:mc_obs}
\vert \bk{\mathbf z}{\Psi_v(\theta)} \vert^2 & = & \frac{\vert \bk{\mathbf z}{\Psi_v(\theta)} \vert^2 }{ \bk{\Psi_v(\theta)}{\Psi_v(\theta)}\hfill} \nonumber \\
& = & 
\frac{\sum_{\vec{s}} \frac{N_{\vec s}^2}{N_v^2}\left(\prod_{j=1}^MR^2_{s_j}(m^R_j)\right) \vert \bk{\mathbf z}{\Psi_v^{\vec{s}}(\theta)} \vert^2}{\sum_{\vec{s}} \frac{N_{\vec s}^2}{N_v^2}\left(\prod_{j=1}^M R^2_{s_j}(m^R_j)\right)},
\end{eqnarray}
where $\sum_{\vec{s}} = \sum_{s_1} \cdots \sum_{s_M}$ and it is implicitly assumed that $\bk{\Psi_v(\theta)}{\Psi_v(\theta)}=1$. Instead of performing the exact summation over all $\vec s$ in Eq.~\ref{eq:mc_obs} (this is essentially the same summation in Eq.~\ref{eq:qrecycle2} mentioned at the end of the last paragraph), we should simply sample $\vec s$ in a Monte Carlo fashion. We note that $N_{\vec{s}}^2$ is the probability of observing $\{s_1,\cdots,s_M\}$ upon measuring those $M$ hidden spins during the construction of the state $\ket{\Psi_{v}^{\vec{s}}(\theta)}$. Hence, samples of $\vec{s}$ are effectively drawn from the probability density $N_{\vec s}^2$ during the construction of $\ket{\Psi_v^{\vec{s}}(\theta)}$ . In short, we replace the exact summation according to
\begin{eqnarray}
\sum_{\vec {s}} N^2_{\vec s} f(\vec s) & \xrightarrow[\text{according to } N^2_{\vec s}]{\text{Monte Carlo sampling of } \vec s} & \frac{1}{N_{\text{exp}}}  \sum_{k=1}^{N_{\text{exp}}} f(\vec{s}_k),
\end{eqnarray}
where $f(\vec{s})$ is some arbitrary function of $\vec{s}$ and $N_{\text{exp}}$ is the number of sampling experiments performed.

The factor $\left(\prod_j R^2_{s_j}(m^R_j)\right)$ in Eq.~\ref{eq:mc_obs} should be calculated classically. This probability $\vert \bk{\vec z}{\Psi_v^{\vec{s}}(\theta)} \vert^2$ is again sampled from projective measurements on visible spins. The only thing that is prohibitively expensive to estimate either classically or experimentally is the normalization constant $N_v$. This is the reason we introduce the denominator (which is really just $1$) in Eq.~\ref{eq:mc_obs} that carries another $N_v$ to cancel the one in the numerator.  By using Eqs.~\ref{eq:expo}-\ref{eq:mc_obs} together,  the challenging post-selection is replaced with the Monte Carlo framework that needs to sample multiple $\ket{\Psi_v^{\vec{s}}(\theta)}$ according to Eq.~\ref{eq:mc_obs}. Additional details on the ensemble state preparation method may be found in the supplementary III.

\textbf{Quantum simulations with NQS-Imaginary Time Evolution (NQS-ITE).}  Next, we discuss the imaginary-time dependent variational principle (ITDVP) to find a ground state of a Hamiltonian $\hat H$. The idea is to propagate a trial wave function $\ket{\Psi_v(\theta_\tau)}$ in the imaginary time domain. If the trial wave function $\ket{\Psi_v(\theta_0)}$ at time $\tau=0$ has a non-zero overlap with the ground state $\ket{\Psi_{gs}}$, then it should converge to an ansatz closest to $\ket{\Psi_{gs}}$ when $\tau \gg 1$. With a variational ansatz, the time-evolved $\ket{\Psi_v(\theta_\tau)}$ can be prepared in a quantum circuit if $\theta_\tau$ is given. In the method section, we summarize the standard ITDVP. The equations of motion for $\theta_\tau$ are given in Eq.~\ref{eq:IT-EOM}. In this section, we adapt the standard NQS-ITE algorithm to make it compatible with the state preparation protocol introduced earlier. Details of the derivation may be found in Supplementary IV. As a result, the approach presented here implements the imaginary-time propagation (for NQS ansatz) more efficiently than the original algorithm reported in the Method section.


In short, the modified equations of motion for $\theta$ assume the following form 
\begin{eqnarray}\label{eq:modified_ite}
\dot \theta_n =\sum_{m} A_{nm}^{-1} C_m.
\end{eqnarray}
The matrix A and vector C read,
\begin{eqnarray}\label{eq:mtxAvecC}
A_{nm} = \text{Re}\left(
\langle \hat O^\dag_n \hat O_m \rangle  - \langle \hat O^\dag_n \rangle_v  \langle \hat O_m \rangle_v\right),
\text{  and  }
C_m = \text{Re}\left(
\langle \hat O^\dag_m \hat H \rangle_v - \langle \hat O^\dag_m \rangle_v\langle \hat H \rangle_v  \right), 
\end{eqnarray}
%
where $\langle \cdots \rangle_v = \bra{\Psi_v(\theta)} \, \cdots \,  \ket{\Psi_v(\theta)}$.  The $O_n$ operators are defined as follows,
\begin{eqnarray}\label{eq:stochreconfigO}
O_n = \left\{ \begin{array}{ll}
i^{1-\delta} \hat v^z_i, & \text{ if } \theta_n = b_i, \\
i^{1-\delta} \tanh\left(m_j + \sum_{i} iW^{I}_{ij} \hat v^z_i\right), & \text{ if } \theta_n = m_j \\
i \hat v^z_i \tanh\left(m_j + \sum_{i} i W^{I}_{ij} \hat v^z_i\right), & \text{ if } \theta_n = W^{I}_{ij}, \\
\end{array}\right.
\end{eqnarray}
where $\delta = 0$ if $\theta_n = b^{I}_i$ or $\theta_n = m^{I}_j$ and $\delta=1$ if $\theta_n = b^{R}_i$ or $\theta_n = m^{R}_j$.
We note that Eqs.~\ref{eq:modified_ite}-\ref{eq:stochreconfigO} essentially give the stochastic reconfiguration method for the variational Monte Carlo framework. One should compare Eq.~\ref{eq:mtxAvecC} and Eq.~\ref{eq:IT-EOM} to see that the standard ITDVP approach (in the Method section) requires preparing $\mathcal{O}(N_{\text{var}}^2)$ quantum states, underlying the matrix element $A_{mn}=\bk{\partial_{\theta_m}\Psi_v(\theta)}{\partial_{\theta_n}\Psi_v(\theta)}$. On other hand, $A$ matrix elements correspond to $\mathcal{O}(N_{\text{var}}^2)$ measurements with respect to the state $\ket{\Psi_v(\theta)}$. In fact, one can simultaneously estimate all $\mathcal{O}(N_{\text{var}}^2)$ matrix elements with every given sample of $\mathbf z$ according to the definition of A matrix element in Eq.~\ref{eq:mtxAvecC}. This could be a tremendous advantage for large-scale simulations when $N_{\text{var}}\propto \mathcal{O}(MN)$ could be a huge number. 

Next, we point out an interesting observation.  The standard gradient-based energy minimization, as done within the VQE approach, updates $\theta_\t$ with the equation of motion 
$\dot \theta_n = C_n$
where $C_n$ being the gradient vector for the energy function $E_\theta = \langle \Psi_v(\theta) \vert \hat H \vert \Psi_v(\theta) \rangle$. The comparison of Eq.~\ref{eq:modified_ite} to the equation of motion above reveals that the NQS-ITE introduces a preconditioner $A^{-1}$ to the gradient vector in order adjust the step size to account for the intrinsically curved metric for the NQS manifold. However, the evaluations of the matrix $A$ requires no more experimental efforts for NQS-ITE than for a standard gradient-descent based approach as explained. As the imaginary-time method tends to give better result than the gradient descent; one should always adopt the imaginary-time propagation whenever NQS is used as the trial wave function.

\textbf{Numerical Results.} To demonstrate the effectiveness of RBM-NQS ansatz and NQS-ITE algorithm, we report numerical simulations on three different types of systems: molecules, spin chains and nanostructures (a triple quantum dots). Throughout the paper, we use a hidden and visible spin ratio of $\alpha=M/N =1$, unless otherwise specified. We first present results with the standard complex-valued NQS ansatz (which requires the extended state preparation protocol described in Supplementary V), then we repeat the simulations in the last subsection to demonstrate that the unitary-coupled RBM-NQS could achieve the same level of accuracy with the same number of hidden spins, i.e. $\alpha=1$. 

3a. Molecular Systems.  We first test the NQS-ITE algorithm on common molecular benchmarks: the dissociation curves of
$\text{H}_2$ and LiH molecules.
The molecular Hamiltonians are first projected onto a discrete set of molecular orbitals. Here, we use the conventional STO-3G basis set, which constitutes a minimal set in that it represents the minimum number of orbitals required to represent a given atomic shell. 
The resulting fermionic Hamiltonians are subsequently mapped onto qubit Hamiltonians using the Bravyi-Kitaev transformation~\cite{seeley2012bravyi} . 
The computation of the integrals in second quantization and transformation of the Hamiltonians are done using PySCF~\cite{sun2018pyscf} and OpenFermion~\cite{mcclean2017openfermion}. 

\begin{figure}[h!]
\centering
\includegraphics[width=1.0\linewidth]{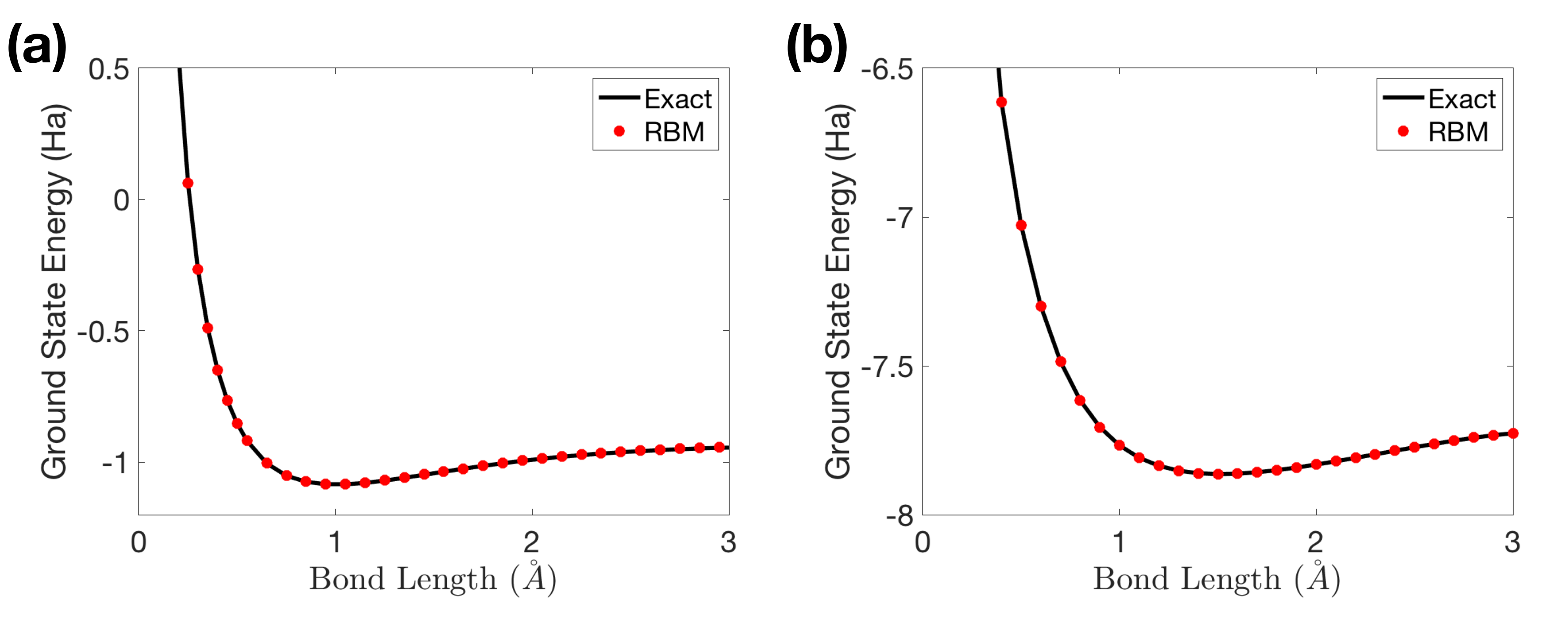}
\caption{ The ground state energy of H$_2$ and LiH molecules computed with the NQS-ITE algorithm (solid red dots) and the exact diagonalization (solid black lines). (a) Results for $H_2$ ($N=2$, $M=2$) as a function of inter-nuclear distance. (b) Results for LiH as a function of inter-nuclear distance($N=4$, $M=4$).}
\label{fig:molecule}
\end{figure}

In the current simulations, the hydrogen molecule requires 2 visible spins
whereas lithium hydride requires 4 visible spins.
The numerical results from NQS ansatz in computing the ground state energy as a function of inter-nuclear distance are plotted in Fig.~\ref{fig:molecule}. It can be seen that NQS is capable of reproducing nearly exact results despite using a modest number of hidden spins. 

\begin{figure}[h!]
\centering
\includegraphics[width=1.0\linewidth]{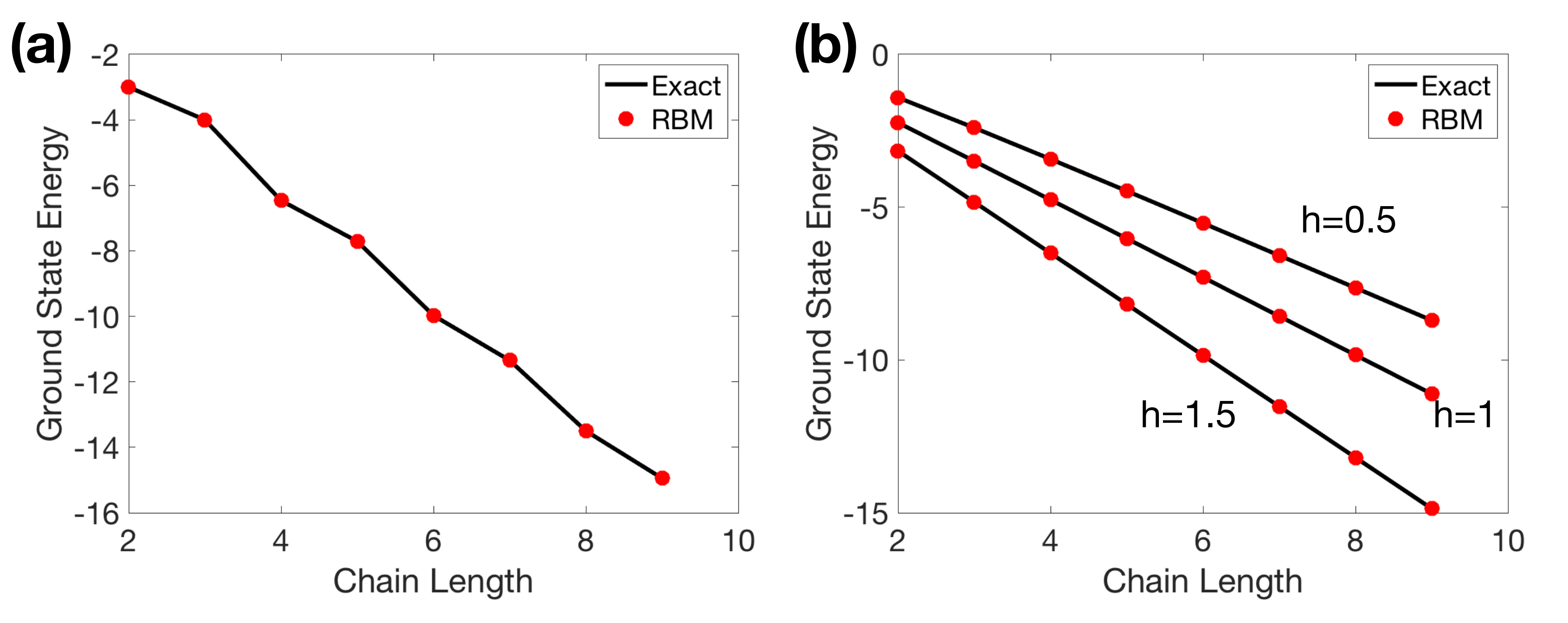}
\caption{ The ground state energy of spin chains computed with the NQS-ITE algorithm (solid red dots) and the exact diagonalization (solid black lines). 
(a) Results for Heisenberg chain as a function of chain length. 
(b) Results for Ising chain as a function of chain length at different values of transverse field.}
\label{fig:spins}
\end{figure}

3b. Spin systems.  Next, we consider the problem of finding the ground state of two prototypical spin models, i.e. the transverse-field Ising (TFI) model and the antiferromagnetic Heisenberg (AFH) model. The spin Hamiltonians can be written as, 
\begin{eqnarray}
    H_{TFI} &=& - h \sum_{i}\hat \sigma_i^{x} - \sum_{ij}\hat \sigma_i^{z}\hat \sigma_j^{z}\\
    H_{AFH} &=& \sum_{ij}\hat \sigma_i^{x}\hat \sigma_j^{x} + \hat \sigma_i^{y}\hat \sigma_j^{y} + \hat \sigma_i^{z}\hat \sigma_j^{z}
\end{eqnarray}
where $h$ is the transverse field strength and we use open boundary condition. 
In Fig.~\ref{fig:spins} we again compute the energies using the NQS-ITE algorithm and compare the results with those from the exact diagonalization, and we observe excellent agreements as expected.

3c. Triple Quantum Dots (TQD). A lateral TQD is an artificial, fully tunable molecule constructed using metallic gates localizing electrons in a semiconductor field-effect transistor (FET). A TQD allows one to study new phenomena not present in a single or double quantum dot, e.g. topological effects \cite{Hsieh2012,Delgado2008}. 
The Hamiltonian of the TQD subject to a uniform perpendicular magnetic field, ${\bf B}=B {\bf \hat{z}}$, is given by

\begin{eqnarray}
      H&=&
   \sum\limits_{i,\sigma}  E_{i\sigma} \hat d_{i\sigma}^{\dag} \hat d_{i\sigma}
+ \sum\limits_{\sigma,i,j;\;i\neq j}
             \tilde{t}_{ij} \hat d_{i\sigma}^{\dag} \hat d_{j\sigma}+ \sum\limits_{i} U_{i}\hat n_{i\downarrow} \hat n_{i\uparrow}
\end{eqnarray}
where $\hat d_{i\sigma}$ ($\hat d_{i\sigma}^\dag$) is fermionic annihilation (creation)
operator with spin $\sigma=\pm 1/2$ on orbital $i$.
$\hat n_{i\sigma} = \hat d_{i\sigma}^\dag \hat d_{i\sigma}$ and $\hat \varrho_{i} = \hat n_{i\downarrow} + \hat n_{i\uparrow}$
are the spin and charge density on orbital level $i$.
Each dot is represented by a single orbital with energy 
$E_{i\sigma}= E_i+ g^* \mu_B B \sigma$, where
$g^*$ is the effective Land\'e $g$-factor, $\mu_B$ is the Bohr magneton.
The dots are connected by magnetic field dependent hopping matrix elements
$\tilde{t}_{ij}=t_{ij}e^{2\pi i \phi_{ij}}$.
The details of the model can be found in the Method section.

\begin{figure}[h!]
\centering
\includegraphics[width=0.5\linewidth]{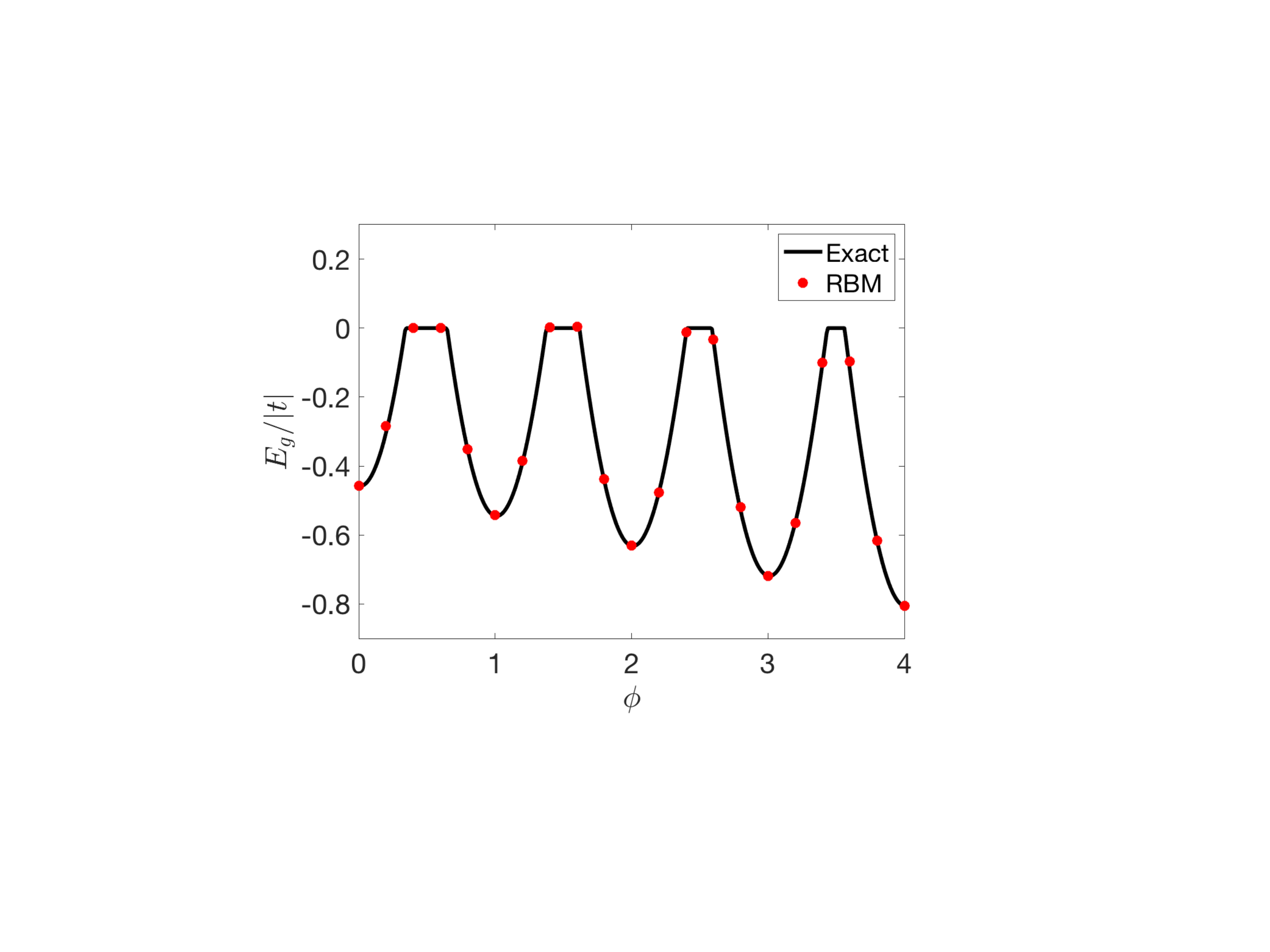}
\caption{The ground state energy of a lateral TQD ($N=6$) as a function of magnetic field obtained from the NQS-ITE algorithm (solid red dots) and the exact diagonalization (solid black lines).}
\label{fig:QDs}
\end{figure}

The ground state energy of the TQD as a function of magnetic field is plotted in Fig.~\ref{fig:QDs}, we observe excellent agreement between the results from the exact diagonalization and NQS-ITE.    
It is worth noting that, at non-zero magnetic field, the ground state wave function of a TQD could be complex, thus a complex RBM ansatz is necessary for accurate representation of the ground state wavefunction. 

\begin{figure}[h!]
\centering
\includegraphics[width=1.0\linewidth]{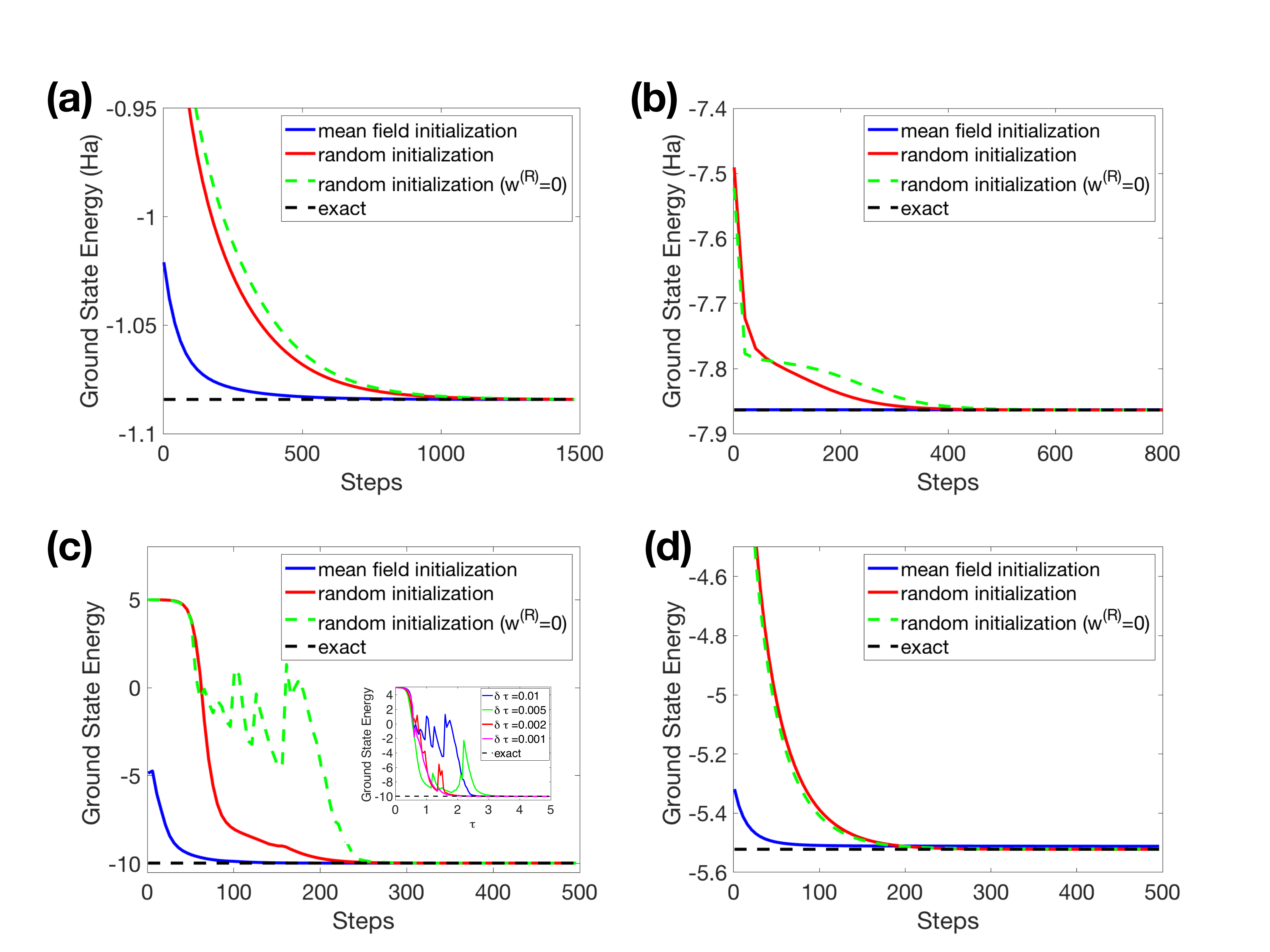}
\caption{Simulations of the ground state energy using NQS-ITE with random initialization (solid red lines), mean field initialization (solid blue lines), and random initialization for the unitary-coupled RBM ansatz (dashed green lines). 
The dashed black lines show the ground state energy from exact diagonalization.  
Results for (a)  H$_2$ at $1.05\AA$, 
(b) LiH at $1.5 \AA$, 
(c) Heisenberg chain ($N=6$), and 
(d) Ising chain ($N=6$, $h=0.5$). 
The inset in Figure (c) displays the results for the unitary-coupled RBM ansatz using different time-step size, $\delta\tau$, where $\tau = \text{Steps}*\delta \tau$.}
\label{fig:mean_field}
\end{figure}

3d. Results with unitary-coupled RBM-NQS. Finally in Fig.~\ref{fig:mean_field}, we repeat the same set of model studies as above but we now apply the unitary-coupled RBM-NQS ansatz (green dahsedlines) to the NQS-ITE. The key insight revealed by Fig.~\ref{fig:mean_field} is that, in most cases, the unitary-coupled RBM-NQS provide comparable performance to that of the standard complex NQS. In the case of Heisenberg model, the approximated RBM ansatz sometimes fails to encode the the imaginary time operator and this leads to non-monotonic behaviour during the imaginary time evolution, though our algorithm eventually finds the true ground state wave function. This non-monotonicity can mostly be mitigated by using a smaller update time steps ( Fig.~\ref{fig:mean_field}c inset). 

Furthermore,  we also present another set of results (solid red lines) in Fig.~\ref{fig:mean_field}. 
These results are obtained with RBM ansatz whose parameters are initialized with a mean-field solution to the original problems.
As shown, a better initial guess improves the convergence rate in comparison to randomly initialized cases. Particularly, mean-field initial wave function already provides nearly exact ground state wave function for lithium hydride molecule, as it can be seen from Fig.~\ref{fig:mean_field}b that the ground state energy from the mean-field approximation is nearly identical to the exact ground state energy.
The accuracy of the mean-field approximation for lithium hydride molecule is possibly due to the low level of electron correlations among the orbitals in minimal basis set.
To test this hypothesis, we solve for the ground state of lithium hydride using Hartree-Fock method and found that it already provides a very accurate ground state solution. In the method section, we briefly comment how one can systematically make an intelligent guess on a good initial state under various situations.

\section*{Discussion}
In conclusion, we present a practical approach to exploit a popular machine-learning model for quantum simulations on a digital quantum computer. Before fault-tolerant quantum computation becomes readily available, the hybrid quantum-classical algorithms will prevail as a popular approach for investigating novel applications of a quantum computer. Successful experimental demonstrations of hybrid quantum-classical algorithms have certainly attracted attentions and boosted confidence in quantum computations. Nevertheless, many obstacles still prevent a clear demonstration of unambiguous quantum advantages for these hybrid quantum-classical algorithms. One possible path towards this goal is to investigate more powerful wave-function ansatz that can achieve a good tradeoff between expressive power and number of variational parameters. With fewer parameters, potentially, one may deal with a shorter-depth state preparation and deals with a simpler optimization problem.

From this perspective, the NQS certainly seems a promising option to investigate. For instance, it is known that a fully-connected RBM ansatz satisfy an entanglement volume scaling, it is very intriguing to further investigate whether one can exploit this property to minimize number of variational parameters under a realistic setting. While the long-range connectivity between qubits is not necessarily easy to realize in every kind of quantum hardwares at the moment; it is at least experimentally feasible with one of the leading hardware architecture, the ion-trap based quantum computers\cite{brown2016iontrap,bruzewicz2019iontrapreview} having all-to-all connectivity among qubits.  As the coherence time of quantum hardware continues to improve, issues of qubit connectivity could potentially be mitigated with advanced techniques such as the fermionic swap network\cite{kivlichan2018quantum}. Theoretically, one may also design quantum algorithms based on the deep Boltzmann machines\cite{gao2017efficient} that further elevates the expressive power of Boltzmann-machine architectures with only short-range couplings, suitable for quantum hardware featuring local connectivity among qubits.  

In this current work, we set out to improve the existing NQS state-preparation protocol in the quantum circuits. As mentioned in the introduction, prior quantum algorithms using RBM ansatz suffer from several obstacles that prevents simulations for complex systems with many degrees of freedom.
Our proposed state-preparation protocols have significantly expanded the scope of applications for NQS as an ansatz for quantum simulations and lowered the experimental barriers. Without the complex-valued parameters, one cannot simulate some important quantum materials and quantum dynamics.  Our numerical testings manifest encouraging signs that the NQS ansatz performs remarkably well across a variety of systems of practical and theoretical interests.  Due to the qubit-recycling scheme, we reduce the number of required qubits from $O(N+M)$ in previous works down to $O(N)$ with sequential implementations of visible-hidden layer interactions. Avoiding the probabilistic preparation of inter-layer couplings (by imposing $W_{ij}=iW^I_{ij}$) also further improves the practicality of NQS-based simulations. The ensemble state preparation bypass the post-selection on hidden spins. Finally, it has been previously shown that imaginary time algorithm offers superior performance compared to VQE\cite{mcardle2018img}, but at the expense of more state preparations and measurements. In this work, we exploit the properties of RBM architecture and show that the number of different quantum states required in the imaginary time algorithm could be reduced from $\mathcal{O}(N_{var}^2)$ down to $1$. Hence, the experimental costs for NQS-ITE is comparable to what VQE demands. Since Boltzmann machine is a widely used machine learning model with many applications, we expect this HQC paradiagm building on the Monte Carlo framework and the NQS ansatz can be adopted for solving other important problems, such as discrete optimizations and machine learning.

\section*{Method}

\textbf{Restricted Boltzmann Machine As Trial Wave Function.}  Recently, Troyer and Carleo \cite{carleo2017solving} used the Restricted Boltzmann Machine (RBM) neural-network architecture as wave-function ansatz to manybody physics and attained impressive results.  Since then, several other wave function ansatz inspired by neural-network architecture have been explored. Collectively, we now refer to this set of wave function ansatzs as the neural-network quantum states (NQS). Although, in this work, we only consider RBM-NQS, which is particularly convenient to model quantum systems composed of two-level systems (TLSs) such as spin lattice commonly studied in conednsed matter physics and electronic structures problems formulated in terms of qubits. For these systems, each TLS is directly identified with a visible spin in the corresponding RBM model. The entanglement between these TLSs (or visible spins) is mediated by the pairwise interactions between visible and hidden spins. In short, a manybody wave function in the NQS form reads
\begin{eqnarray}\label{eq:nnqs}
\ket{\Psi_v(\theta)} 
& = & \frac{1}{N_v} \sum_{\bf{v}}\left(\sum_{\bf{h}} e^{E_\theta(\bf{v},\bf{h}) }\right) |\bf{v} \rangle,
\end{eqnarray}
with energy function $E_\theta(\mathbf{v}, \mathbf{h}) = \sum_{i} b_i v_i + \sum_{j} m_j h_j + \sum_{ij}W_{ij} v_i h_j$, its complex conjugate $\bar{E}_\theta(\mathbf{v},\mathbf{h})$, and
the normalization constant $N_v= \sqrt{\sum_{\mathbf{v}}\left(\sum_{\mathbf{h}}e^{\bar E_\theta(\mathbf{v},\mathbf{h})}\right)
\left(\sum_{\mathbf{h}}e^{E_\theta(\mathbf{v},\mathbf{h})}\right)}$. 
The RBM parameters are collectively denoted by $\theta = \{\mathbf b, \mathbf m, \mathbf W\}$.
As shown in Eq.~\ref{eq:nnqs}, the hidden spins are summed over in the bracket on the right-hand side of Eq.~\ref{eq:nnqs} to give a wave function $\ket{\Psi_v(\theta)}$ for the visible spins, which represent the physical system of interest. In principle, $\theta$ should possess non-vanishing imaginary components to describe complex-valued wave function. 

To prepare an NQS in a quantum circuit, we should take the energy function $E_\theta(\bf{v},\bf h)$ and promote it to an Hermitian operator by replacing the binary values of $\bf v$ and $\bf h$ with the corresponding Pauli operators. The quantum-circuit analog of Eq.~\ref{eq:nnqs} is decomposed into Eqs.~\ref{eq:vhwf}-\ref{eq:bmcircuit}: first entangle the visible and hidden spins then post-selects the hidden spins to mediate the desire non-unitary transformation on the visible-spin wave functions.

\textbf{Imaginary-time dependent variational principle for an NQS} 
In this section, we summarize a recently proposed imaginary-time dependent variational principle (ITDVP) \cite{mcardle2018img} for NQS ansatz.
Under the Wick rotation $t \rightarrow i \tau$, the time-independent Schrodinger equation reads,
\begin{eqnarray}\label{eq:ischeq}
\frac{\partial |\Phi(\tau) \rangle}{\partial \tau} = -(\hat H -E_\tau) |\Phi(\tau) \rangle
\end{eqnarray}
where the additional energy term $E_\tau = \langle \Phi(\tau) | \hat H|\Phi(\tau) \rangle$ arises from the normalization condition for $\ket{\Phi(\tau)}$. As $\tau \rightarrow \infty$, the stationary solution of Eq.~\ref{eq:ischeq} is the lowest-energy eigenstate that has a non-zero overlap with the initial state $\ket{\Phi(0)}$. The imaginary-time dynamical simulation is a well-established approach to obtain the ground state of complex Hamiltonians.

Through time dependent variational principle , an NQS ansatz $\ket{\Psi_v(\theta_\tau)}$ approximates a time-evolved quantum state 
$\ket{\Phi_\tau}$ via the following equation of motion,
\begin{eqnarray}
\delta|| (\partial /\partial \tau  + \hat H  - E_\tau ) | \Psi_v(\mathbf{\theta}_\tau)\rangle || = 0,
\end{eqnarray}
where $\theta(\tau)$ are determined via,
\begin{eqnarray}\label{eq:tdvp}
\sum_j A_{ij} \dot{\theta}_j = C_i.
\end{eqnarray}
The A and C matrices are defined as follows,
\begin{eqnarray} \label{eq:IT-EOM}
A_{mn} = \text{Re} \left\langle \frac{\partial \Psi_v(\theta)}{\partial \theta_m} \right\vert \left. \frac{\partial \Psi_v(\theta)}{\partial \theta_n}
\right\rangle, \text{    and      } 
C_{n} = -\text{Re}\left\langle \partial_{\theta_n} \Psi_v(\theta) \right\vert \hat H \left\vert  \Psi_v(\theta) 
\right\rangle.  
\end{eqnarray}
Therefore, one updates the variational parameters according to ${\theta}(\tau + \delta \tau) = 
{\theta}(\tau) + \dot{{\theta}}(\tau) \delta \tau =
{\theta}(\tau) + A^{-1}(\tau)C(\tau) \delta \tau$, where $\delta \tau$ is the update timestep.

For the variational parameter update in Eq.~\ref{eq:IT-EOM}, we need to compute the gradients of $|\Psi_v \rangle$  with respect to $\theta$, i.e. $|\frac{\partial \Psi_v}{\partial \theta_n} \rangle$.  As the properly normalized $\ket{\Psi_v(\theta)}$ carries a $\theta$-dependent normalization factor $N_v$ in Eq.~\ref{eq:nnqs}, the chain-rule derived formula for the gradient reads
\begin{eqnarray}\label{eq:chainrule1}
    \left \vert \frac{\partial \Psi_v}{\partial \theta_n} \right \rangle = \frac{_h\langle ++\dots + |\partial_{\theta_n} \Psi_{vh}\rangle}{\mathcal{N}_v} 
    - \text{Re} \left( \langle \Psi_{vh} | \hat{P}_+^{(h)}|\partial_{\theta_n} \Psi_{vh}  \rangle \right)   \frac{_h\langle ++\dots + | {\Psi_{vh}}  \rangle}{\mathcal{N}_v^3}
\end{eqnarray}
where $\mathcal{N}_v = \sqrt{ \langle \Psi_{vh}(\theta) | \hat{P}^{(h)}_+  | \Psi_{vh}(\theta)\rangle}$. The above expression depends on the gradients of the extended RBM wave function, $|\partial_\theta \Psi_{vh}  \rangle$, which can be derived analytically
\begin{eqnarray}\label{eq:psivh_deriv}
\frac{ \partial |\Psi_{vh} \rangle }{\partial b^{R}_i} &=&  (\hat{v}_i -  \langle \hat{v}_i \rangle_{vh} )  |\Psi_{vh} \rangle ,\,\,\,\,
\frac{ \partial |\Psi_{vh} \rangle }{\partial b^{I}_i} =  i\hat{v}_i |\Psi_{vh} \rangle, \nonumber \\
\frac{ \partial |\Psi_{vh} \rangle }{\partial m^{R}_j} &=&  (\hat{h}_j -  \langle \hat{h}_j \rangle_{vh} )  |\Psi_{vh} \rangle, \,\,\,\,
\frac{ \partial |\Psi_{vh} \rangle }{\partial m^{I}_j} =  i\hat{h}_j  |\Psi_{vh} \rangle, \nonumber \\
\frac{ \partial |\Psi_{vh} \rangle }{\partial W^{R}_{ij}} &=&  (\hat{v}_i \hat{h}_j -  \langle \hat{v}_i \hat{h}_j  \rangle_{vh} )  |\Psi_{vh} \rangle, \,\,\,\,
\frac{ \partial |\Psi_{vh} \rangle }{\partial W^{I}_{ij}} =  i \hat{v}_i \hat{h}_j |\Psi_{vh} \rangle,
\end{eqnarray}
where the superscript $(R, I)$ is used to denote the real and imaginary parts of the coefficients and $\langle \hat{O}\rangle_{vh} = \langle \Psi_{vh} | 
\hat{O}| \Psi_{vh}\rangle $.  
Substituting the expressions given by Eq.~\ref{eq:psivh_deriv} into Eq.~\ref{eq:chainrule1}, one obtains expressions like the following,
\begin{eqnarray}\label{eq:f1}
\left\vert \frac{\partial\Psi_v(\theta)}{\partial m^{I}_j} \right\rangle = \frac{\mathcal{N}_v^\prime}{\mathcal{N}_v} \left( \ket{\Psi_v(\theta^\prime)}  + 
\text{Re}\left[  \langle \Psi_{v}(\theta) |\Psi_{v}(\theta^\prime)  \rangle  \right] \ket{\Psi_v(\theta)} \right),
\end{eqnarray}
where $\theta^\prime=[\cdots, m^{I}_j+\frac{\pi}{2}, \cdots]$ and $\theta=[\cdots, m^{I}_j, \cdots]$ differ only by the value of $m^{I}_j$ and the normalization factor
$\mathcal{N}^\prime_v = \sqrt{\bra{\Psi_{vh}(\theta^\prime)}\hat P^{(h)}_+ \ket{\Psi_{vh}(\theta^\prime)}}$. To accurately calculate  matrix $A$ and vector $C$,
it is necessary to determine the normalization factors $\mathcal{N}_v$ and $\mathcal{N}^\prime_v$ according to Eq.~\ref{eq:f1}.  Nevertheless, $\mathcal{N}_v$ is hard to estimate as the cost to estimate scale as  $\mathcal{O}(4^M)$. 
This challenge motivates us to re-formulate the ITDVP for RBM-based NQS ansatz that entirely avoids the normalization factors $\mathcal{N}_v$. Details may be found in the subsection discussing NQS-ITE under Result section and Supplementary IV.


\textbf{Initial state preparation for quantum simulations.} 
The NQS ansatz can be used in conjunction with most HQC simulation algorithms in addition to the time-dependent variational method outlined above. All these methods aim to solve highly non-trivial optimization in which the quality of solutions or the convergence rate depends crucially on the overlap of the initial state with the ground state $\ket{\Psi_{gs}}$.  The two-stage initialization protocols described here gives a systematic approach to guide the preparation of high-quality initial states.  In short, the idea is to selectively optimize a subset of parameters to obtain an approximate solution that could be used as the initial state in a subsequent simulation optimizing over all parameters. 

In the simplest case, one may consider a mean-field approximation, which restricts the considerations to completely factorized product-state wave function $\ket{\Psi_v^{\text{0}}(\vec b)}=\ket{\psi_{v_1}(b_1)}\otimes\cdots\otimes\ket{\psi_{v_N}(b_N)}$ for all visible spins in a quantum simulation.
We then subsequently use $\ket{\Psi^0_v(\vec b)}$ as the starting point of another simulation in which the hidden spins are introduced along with corresponding parameters $\{m_1, \cdots m_M, W_{11}, \cdots W_{NM}\}$ that collectively facilitate the formation of entangled NQS, $\ket{\Psi_v(\theta)}$. We note the mean-field approximation (single-body physics problem) can be easily done on a classical computer.

Nevertheless, for strongly correlated systems, the product states are not guaranteed to support a high overlap with $\ket{\Psi_{gs}}$. Instead of optimizing $\vec b$ (the mean-field approximation) in the first run, it will be beneficial to consider an NQS with specifically designed sparse connectivity.  In the second-stage calculation, the fully-connected architecture will be restored as usual, and the total number of variational parameters scale as $\mathcal{O}(NM)$. An obvious question is how to decide the connectivity of this sparsely connected RBM architecture for the first-stage simulation, which needs to balance the expressive power of the variational ansatz and the complexity of the optimization tasks. For lattice systems, one may consider short-range RBMs that constitute a special class of the well-established entangled-plaquette states \cite{glasser2018neural}. In this case, the total number of variational parameters scale as $\mathcal{O}(N)$.  

\textbf{Simulation details for numerical studies.}
In all our simulations, we use a constant learning rate of 0.01. The variational parameters are initialized as Gaussian random numbers with mean zero and variance 0.01, except in cases where the initial conditions are obtained from mean field solutions (Fig.\ref{fig:mean_field} green dashed lines). 

Hamiltonians of Hydrogen and Lithium Hydride.
We treat the hydrogen and lithium hydride molecules in the minimal STO-3G basis and use PySCF to compute the integrals in the second quantization
The resulting fermionic Hamiltonians are subsequently mapped onto qubit Hamiltonians using Bravyi-Kitaev transformation with OpenFermion~\cite{mcclean2017openfermion}. 
Due to the symmetry in H$_2$, the final Hamiltonian consists of 2 qubits~\cite{o2016scalable}, whereas 
the lithium hydride Hamiltonian contains 4 qubits.

Hamiltonian for TQD.
For TQD simulations, we use the parameters from Ref.~\cite{Delgado2008}, i.e. $t=-0.23$ meV,
$U_i=50|t|$, $V_{ij}=10|t|$. $g^*=-0.44$, $E_i =-|t| $, and $\phi/B =1.25 T^{-1}$. We use Bravyi-Kitaev transformation to map the Hubbard Hamiltonian onto a qubit Hamiltonian using OpenFermion.

\section*{Supplementary Information I: Challenges to prepare complex-valued NQS in quantum devices.}
In prior works, an NQS wave-function amplitude reads
\begin{eqnarray}\label{eq:cl-nnqs}
\bk{\bf v}{\Psi_v(\theta)}
=\left(\sum_{\bf h} \vert Q_{\theta}(\bf v,\bf h)\vert e^{i\arg Q_{\theta}(\bf v,\bf h)}\right)^{1/2},
\end{eqnarray}
where $ Q_{\theta}(\bf v,\bf h)= \left(\sum_{\bf{h}} e^{E_\theta(\bf{v},\bf{h}) }\right)/N_v$ is the joint Boltzmann distribution of visible and hidden spins as governed by the energy function $E_\theta$. When the RBM Hamiltonian is strictly real-valued, then it is obvious that the $\ket{\Psi_v(\theta)}$ can only model positive semi-definite wave functions. While being limited in scope of applications, one can rely on the Marshall Sign Rule to determine when the ground-state wave functions of a spin Hamiltonian should be positive semi-definite. Study on the transverse-field Ising model with positive semi-definite NQS has been reported \cite{gardas2018dwave} for quantum simulations on a D-wave quantum annealer. Nevertheless, there is an advantage of working with this subset of NQS. We note the visible-spin probability density $\vert \bk{\bf v}{\Psi_v(\theta)}\vert^2 =\sum_{\bf h} Q_{\theta}(\bf v,\bf h)$ is simply the marginal of the joint visible-hidden distribution.  Experimentally, one may simply prepare an extended wave function of all visible and hidden spins then simply ``trace over" the hidden spins to estimate $\vert \bk{\bf v}{\Psi_v(\theta)}\vert^2$.  Avoiding the post-selection of $\ket{++\cdots +}_h$ is certainly advantageous. 

However, to model realistic systems of interest, one should at least incorporate the negative signs into wave function. Reference \cite{xia2018rbm} proposed an ingenious approach to generate arbitrary real-valued NQS in quantum circuit by modifying the standard RBM architecture.  The authors added an extra node in the hidden layer with the sole purpose to assign a sign factor to each wave function coefficient. More precisely, the modified NQS read
\begin{eqnarray}\label{eq:cl-nnqs2}
\bk{\bf v}{\Psi_v(\theta)}
=\left(\sum_{\bf h} \vert Q_{\theta}(\bf v,\bf h)\vert e^{i\arg Q_{\theta}(\bf v,\bf h)}\right)^{1/2}s(\mathbf v),
\end{eqnarray}
where $
s(\mathbf v) = s\left( {v _1^z,v _2^z...v _n^z} \right) = \tanh( {\sum_i d_i v _i^z + c})
$ with $d_i$ and $c$ as parameters to be optimized. Given the form of $s(\mathbf v)$, it is clear that it cannot be described simply with the standard RBM Hamiltonian.  Since this extra hidden node does not couple directly to other hidden spins, it does not have to be explicitly incorporated into the state preparation. Rather, when estimating physical properties of a quantum system,the sign functions can be shifted onto the observable operators.  For instance, to measure energy of a quantum system, we prepare a positive semi-definite NQS $\ket{\Psi_v}$ described in the previous paragraph and estimate energy according to 
\begin{eqnarray}
\langle \hat H\rangle = \frac{{\mathop {\sum}\nolimits_{\mathbf z,\mathbf{z}^\prime } {\overline {\Psi_v (\mathbf z)} \overline {s(\mathbf z)} \langle \mathbf z | \hat H|\mathbf{z}^\prime \rangle \Psi_v (\mathbf{z}^\prime )s(\mathbf{z}^\prime )} }}{{\mathop {\sum}\nolimits_{\mathbf z} {|\Psi_v (\mathbf z)s(\mathbf z)|^2} }}.
\end{eqnarray}
While this method has significantly expanded the usefulness of NQS in quantum simulations, the generation of the non-unitary transformation $\exp(W^R_{ij}\hat v^z_i \hat h^z_j)$ cannot be done efficiently in a deterministic fashion.  
The challenge of scalability  and the desire to model complex-valued wave functions with significantly lower amount of resources have led us to propose the state preparation scheme described in the main text.

\section*{Supplementary Information II: Expressive Power of the Unitary-Coupled Boltzmann Machines}
\begin{figure}[h!]
\centering
\includegraphics[width=.8\linewidth]{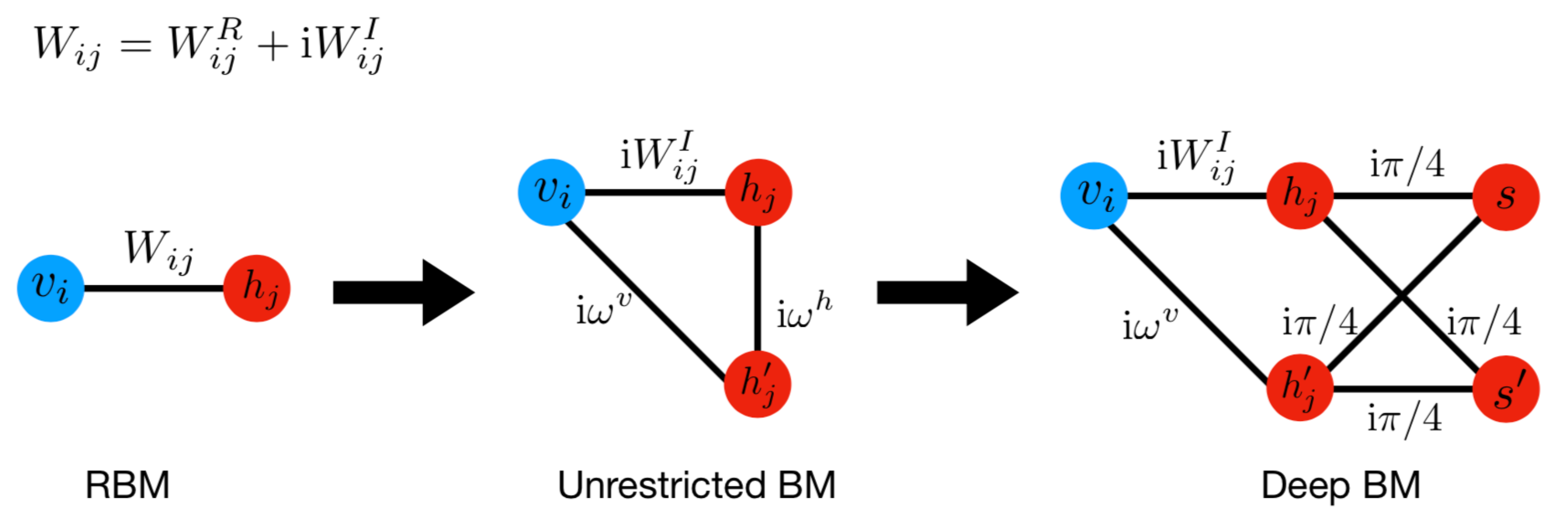}
\caption{Mapping of RBM states onto Unrestricted Boltzmann machines and Deep Boltzmann Machine in order to remove real-valued coupling coefficients in the original RBM network.}
\label{fig:unitary-rbm}
\end{figure}

In order to improve the scalability of the NQS preparation in a quantum circuit, we impose the cross-layer interaction $W_{ij}=iW^I_{ij}$ to be strictly imaginary-valued to avoid probabilistic preparation or a full tomographic characterization of a thermal-like quantum state when $W_{ij}$ have non-vanishing real parts.  In this supplementary, we briefly outline three different schemes to convert an arbitrary complex-valued NQS into variants of Boltzmann Machine inspired wave function ansatz with only imaginary-valued couplings across visible and hidden layers. These conversions clearly show that there is no loss of expressive power with the ``seemingly" restricted form of RBM-NQS we promote in this work. In the following, as illustrated in Fig.~\ref{fig:unitary-rbm}, we present methods to map a standard RBM architecture into two variants of Boltzmann machines as explained below.  

\textbf{Restricted Boltzmann Machine.}
First, we give the most straightforward conversion scheme. 
We note the standard NQS has an unnormalized amplitude,
\begin{eqnarray}\label{eq:nnqs_amp}
      \psi_v(\theta) &=& e^{\sum_{i=1}^N b_i v_i}\prod_{j=1}^M \cosh\left(m_j + \sum_i W_{ij}v_i\right) \nonumber \\
      & = & e^{\sum_{i=1}^N b_i v_i + \sum_{i_1 > i_2} c^2_{i_{1}i_{2}}v_{i_1}v_{i_2} + \sum_{i_1>i_2>i_3} c^3_{i_1 i_2 i_3}v_{i_1}v_{i_2}v_{i_3}
      + \cdots + c^N_{1...N} v_1\dots v_N},
\end{eqnarray}
where the coefficients $c^n_{i_1\dots i_n}$ are obtained from the Taylor expansion of $\log \prod_j \cosh\left(m_j + \sum_i W_{ij}v_i\right)$ with $v_i$ being binary values of $\pm 1$. On the second line of Eq.~\ref{eq:nnqs_amp}, the exponent
is a polynomial of degree N. Each term (with degree greater than 1) of this polynomial can be fully decoupled by the application of this mathematical identity \cite{carleo2018constructing},
\begin{eqnarray}
      e^{\omega \hat v_{1} \hat v_{2}\cdots \hat v_{n}} & = & \frac{C}{4}\prescript{}{s_1}{\bra{+}}e^{(i \tilde m +i\frac{\pi}{4}\sum_{i=1}^n \hat v_i)\hat s_1)}\ket{+}_{s_1}
      \prescript{}{s_2}{\bra{+}}e^{(i\tilde m + i\frac{\pi}{4}\sum_{i=1}^n \hat v_i) \hat s_2)}\ket{+}_{s_2},
\end{eqnarray}
where $\hat s_i$ is a Pauli Z operator acting on state $\ket{+}_{s_i}$, and the other parameters read
\begin{eqnarray}
      \tilde m = \arctan(e^{-\omega})-\frac{\pi}{4} \text{Mod}(2,N) \,\, \text{   and   } \,\, C = \frac{2}{\sin(2\arctan(e^{-\omega}))}.
\end{eqnarray}
In short, every RBM can be exactly turned into another RBM with a fixed imaginary-valued coupling value $i \pi/4$ throughout the network. A potential price to be paid is the exponentially large number of hidden nodes in this alternative (yet fully equivalent) RBM. Certainly, when the coupling parameters $\vert c^n_{i_1 \dots i_n} \vert$ in Eq.~\ref{eq:nnqs_amp} having magnitudes much smaller than that of $b_i$ and $ m_j$ then one can drop many terms in that polynomial expansion. However, even in the lowest order of perturbation with a quadratic polynomial, the alternative RBM still carries $\mathcal{O}(N^2)$ hidden nodes. One may wonder whether the cost to impose strictly unitray couplings across visible and hidden layer is always an astronomical number of extra hidden nodes. 

We remind that the goal of this rather formal analysis is to show that any arbitrary RBM can always be mapped onto a purely imaginary-valued RBM via a simple mapping. By no means, it is the only or the most efficient mapping. We should not be too pessimistic about this. For instance, there are examples \cite{deng2017quantum} of 1D and 2D RBM-NQS (having $M~\mathcal{O}(N)$) exhibiting volume-law entanglements with a constant imaginary-valued coupling coefficient ($i\pi/4$). This implies strongly correlated quantum states can be efficiently created within this subset of RBM-NQS. Furthermore, we hypothesize that if we simply allow the imaginary-valued couplings to vary independently instead of all being held fixed at the value of $i\pi/4$, then there should be viable mapping of arbitrary RBM-NQS onto rather compact architecture of unitary-coupled RBMs.  This intuitive hypothesis is empirically confirmed in our numerical studies reported in the main text. 

\textbf{Unrestricted Boltzmann Machine.}  Given an RBM architecture (composed of N visible nodes and M hidden nodes) with complex-valued couplings $W_{ij}$,  one may exactly transform away the real components, $W^R_{ij}$ via a different approach than the one described above. For instance, the coupling between the $i$-th visible and the $j$-th  hidden spins may be removed with the addition of one more hidden node (labeled by s) according to the relation\cite{carleo2018constructing}, 
\begin{eqnarray}\label{eq:wreal}
\exp(W^R_{ij}\hat v_i \hat h_j)=
 \Delta \left[ \prescript{}{s}{\bra{+}} \exp((i \omega_v \hat v_i + i \omega_h \hat h_j) \hat s) \ket{+}_s \right],
\end{eqnarray}
where $\hat s$ is the Pauli Z operator acting on the additional ancilla qubit in $\ket{+}_s$ state.
The normalization factor is taken to be $\Delta=e^{\vert W^R_{ij} \vert}/2$, and other newly introduced parameters read
\begin{eqnarray}
      \omega_v =\frac{1}{2}\arccos\left(e^{-2\vert W^R_{ij} \vert}\right), \text{  and  }\omega_h =-\frac{1}{2}\arccos\left(e^{-2\vert W^R_{ij} \vert}\right). 
\end{eqnarray}
These parameters $\omega_v$ and $\omega_h$ would be real-valued as expected.  The newly added hidden spin breaks the restriction of no intra-layer couplings.  The total number of hidden spins increase to $\mathcal{O}(MN)$ after removing $W^R_{ij}$. Nevertheless, the modified neural-network architecture is largely compatible with RBM architecture as far as preparing the prescribed quantum states in a quantum circuit is concerned. Although the qubit recycling scheme comes with a caveat that one needs roughly $\mathcal{O}(N)$ extra ancilla qubits (at each sequential stage) to explicitly model the entangled hidden state, which emerge when we transform away $\mathcal{O}(N)$ real-valued couplings between visible nodes and a particular hidden node.

\textbf{Deep Boltzmann Machine.} Finally, we realize that the unrestricted Boltzmann machine can be further mapped onto a deep Boltzmann machine architecture without intra-layer couplings among hidden nodes. This is achieved by invoking the following mathematical identity,
\begin{eqnarray}
      e^{i\omega^h \hat h_i \hat h_j} & = & \frac{C}{4}\prescript{}{s_1}{\bra{+}}e^{(ib+i\frac{\pi}{4}(\hat h_1+ \hat h_2)\hat s_1)}\ket{+}_{s_1}
      \prescript{}{s_2}{\bra{+}}e^{(ib+i\frac{\pi}{4}(\hat h_1+ \hat h_2)\hat s_2)}\ket{+}_{s_2},
\end{eqnarray}
where $\hat s_i$ is a Pauli Z operator acting on state $\ket{+}_{s_i}$ in the deep layer, and the other parameters read
\begin{eqnarray}
      b = \arctan(e^{-i\omega^h})-\frac{\pi}{2} \text{  and  } C = \frac{2}{\sin(2\arctan(e^{-i\omega^h}))}.
\end{eqnarray}
Hence, in order to convert an unrestricted Boltzmann machine into restricted form (i.e. no intra-layer coupling),  every intra-layer couplings between hidden units can be exactly replaced with couplings to extra hidden units in the deep layer. However, using the deep Boltzmann machine for quantum simulation could be difficult as the optimization of the parameters could be hard to converge.  In a follow-up work, we will describe  practical approaches to perform quantum simulation with deep Boltzmann Machines.

\section*{Supplementary Information III: Ensemble State Preparation Protocol}
As shown in Eq.~\ref{eq:qrecycle2}, the post-selected NQS $\ket{\Psi_v(\theta)}$ can be cast as a weighted sum of $2^M$ $\ket{\Psi^{\vec{s}}(\theta)}$.  To facilitate the following discussion, we will work with the un-normalized version of $\ket{\tilde \Psi^{\vec{s}}(\theta)}$ in this section,
\begin{eqnarray}
\ket{\tilde \Psi_v^{\vec{s}}(\theta)} & = &\left[\bra{s_M}
e^{i\mathcal{G}_M}\ket{+} \cdots  \bra{s_1}e^{i\mathcal{G}_1}\ket{+}
e^{\sum_{i=1}^N b_i \hat v^z_i} \right]\ket{++\cdots+}_v, 
\end{eqnarray}
where $\ket{s_i=\pm}=(\ket{0}\pm\ket{1})/\sqrt{2}$ and $\mathcal{G}_j= m_j^I\hat h^z_j + \sum_i W^I_{ij}\hat h^z_j \hat v^z_i$. The normalization factor $N_{\vec s} = \sqrt{\bk{\tilde \Psi_v^{\vec s}(\theta)}{\tilde \Psi_v^{\vec s}(\theta)}}$ is the probability of getting a particular visible-spin wave functions when $M$ hidden spins are observed in the state $\ket{s_1 \cdots s_M}_h$ after being coupled to the visible ones. $N_{\vec s}$ will be brought back to discussion at the
end of this section.

Prior to prove Eq.~\ref{eq:qrecycle2}, we first give a useful identity.  The overlap between $\ket{\tilde \Psi_v^{\vec{s}}}$ 
and $\ket{\tilde \Psi_v^{\vec{s}^\prime}}$ where the two index vectors $\vec s$ and $\vec s^\prime$ differ by $K$ out of $M$ components 
can be cast in the following form, 
\begin{eqnarray}\label{eq:obs-psimk}
\bk{\tilde\Psi_v^{\vec{s}^\prime}}{\mathbf z}\bk{\mathbf z}{\tilde\Psi_v^{\vec{s}}} & = & i^K \bk{\tilde \Psi^{\vec{s}_{M-K}}}{z} V_{i_1}(\mathbf z) \cdots V_{i_K}( \mathbf z) \bk{\mathbf z}{\tilde \Psi^{\vec{s}_{M-K}}} \nonumber \\
& = & i^K \vert \bk{\tilde \Psi_v^{\vec{s}_{M-K}}}{\mathbf z} \vert^2 V_{i_1}(\mathbf z) \cdots V_{i_K}(\mathbf z),
\end{eqnarray}
where $\vec{s}_{M-K}$ is a reduced vector with the components $\{i_1 \cdots i_K\}$ removed from $\vec{s}$ and 
\begin{eqnarray}
V_j(\mathbf z)=s^\prime_j\sin\left(m^I_j+\sum_i W^I_{ij}z_i\right)\cos\left(m^I_j+\sum_i W^I_{ij}z_i\right)
\end{eqnarray}
are real-valued. From Eq.~\ref{eq:obs-psimk}, it is clear that the expression 
\begin{eqnarray}\label{eq:obs-psimk2}
\bk{\tilde \Psi_v^{\vec s^\prime}}{\mathbf z}\bk{\mathbf z}{\tilde\Psi_v^{\vec s}}+\bk{\tilde\Psi_v^{\vec s}}{\mathbf z} \bk{\mathbf z}{\tilde\Psi_v^{\vec s^\prime}}=0,
\end{eqnarray} 
when the index vectors $\vec{s}^\prime$ and $\vec{s}$ differ by an odd number of components, i.e. $K$ is odd. This is because the symmetrized expression above must be real-valued, but Eq.~\ref{eq:obs-psimk} demands pure imaginary numbers when $K$ is odd. Hence, only zero is allowed.
For the cases $\vec{s}^\prime$ and $\vec{s}$ differ by an even number of components, i.e. $K$ is even, we draw attention to a special condition.  Let us first introduce another two states $\ket{\Psi^{\vec t}_v}$ and $\ket{\Psi^{\vec{t}^\prime}_v}$ with the index vectors satisfying (1) $\vec{t}$ and $\vec{s}$ differ by one component, (2) $\vec{t}$ and $\vec{t}^\prime$ differ by $K$ , and (3) $\vec{s}_{M-K}$ and $\vec{t}_{M-K}$ are identical.  Then Eq.~\ref{eq:obs-psimk} implies, 
\begin{eqnarray}\label{eq:obs-psimk3}
\bk{\Psi_v^{\vec s^\prime}}{\mathbf z}\bk{\mathbf z}{\Psi_v^{\vec s}} + \bk{\Psi_v^{\vec t^\prime}}{\mathbf z}\bk{\mathbf z}{\Psi_v^{\vec t}}=0.
\end{eqnarray}
This is because $\prod_{q=1}^K V_{i_q}(\mathbf z)$, as defined in Eq.~\ref{eq:obs-psimk}, carries opposite sign for $\bk{\Psi_v^{\vec s^\prime}}{\mathbf z}\bk{\mathbf z}{\Psi_v^{\vec s}}$ and $\bk{\Psi_v^{\vec t^\prime}}{\mathbf z}\bk{\mathbf z}{\Psi_v^{\vec t}}$, respectively.

Now, consider a Hermitian operator $\hat O$ that is diagonal in the computational basis then the real-valued
expectation value is given by
\begin{eqnarray}\label{eq:obs-app}
\bra{\tilde\Psi_v}\hat O \ket{\tilde\Psi_v} & = & \sum_{\vec{s}^\prime \vec{s}}
\frac{1}{2}\left(\prod_{j=1}^M R_{s^\prime_j}(m^R_j)R_{s_j}(m^R_j)\right) \left(
\bra{\Psi_v^{\vec s^\prime}}\hat O\ket{\Psi_v^{\vec s}}+
\bra{\Psi_v^{\vec s}}\hat O \ket{\Psi_v^{\vec s^\prime}}\right)\nonumber \\
& = & \sum_{\mathbf z} \sum_{\vec{s}^\prime \vec{s}}\frac{O_{\mathbf z \mathbf z}}{2}
\left(\prod_{j=1}^M R_{s^\prime_j}(m^R_j)R_{s_j}(m^R_j)\right)
 \left(\bk{\Psi_v^{\vec s^\prime}}{\mathbf z}\bk{\mathbf z}{\Psi_v^{\vec s}}+
\bk{\Psi_v^{\vec s}}{\mathbf z} \bk{\mathbf z}{\Psi_v^{\vec s^\prime}}\right) \nonumber \\
& = & \sum_{\mathbf z} \sum_{\vec{s}}
O_{\mathbf z \mathbf z}\left(\prod_{j=1}^M R^2_{s_j}(m^R_j)\right)
\bk{\tilde \Psi_v^{\vec s}}{\mathbf z}\bk{\mathbf z}{\tilde \Psi_v^{\vec s}}.
\end{eqnarray}
Substituting Eq.~\ref{eq:obs-psimk} into the epxressions inside the last bracket on the second line of Eq.~\ref{eq:obs-app} and using Eq.~\ref{eq:obs-psimk2}-\ref{eq:obs-psimk3}, we can tremendously simplify the expression to reach the third line above. It is clear how Eq.~\ref{eq:obs-psimk2} helps to suppress terms with $\vec{s}$ and $\vec{s}^\prime$ that differ by an odd number of components.  Next, we note that given a pair of $\vec s$ and $\vec{s}^\prime$, there is always a corresponding pair of $\vec t$ and $\vec{t}^\prime$ under the double summation ($ \sum_{\vec{s}^\prime \vec{s}}$) on the second line of Eq.~\ref{eq:obs-app} that satisfy the assumptions for Eq.~\ref{eq:obs-psimk3}. 

When $\hat O = \mathcal{I}$, then we immediately obtain
\begin{eqnarray}\label{eq:obs-identity}
\bk{\tilde \Psi_v}{\tilde \Psi_v} & = & N_v^2 \nonumber \\
&=& \sum_{\vec{s}}
N^2_{\vec{s}} \left(\prod_{j=1}^M R^2_{s_j}(m^R_j)\right).
\end{eqnarray}
This expression admits a simple interpretation. $N_v^2$ is the success probability of preparing an NQS upon post-selection of $\ket{++\cdots +}_h$.
According to Eq.~\ref{eq:obs-identity}, this probability can be exactly decomposed as a linear combination of conditional probabilities, $N^2_{\vec{s}}\prod_j R^2_{s_j}$, where $N^2_{\vec{s}}$ is the probability of measuring $\ket{s_1 \cdots s_M}_h$ after the hidden spins are coupled to the visible ones, and $\prod_j R^2_{s_j}$ is the conditional probability that the state $\prod_j \exp(m_j^R \hat{h}^z_j)\ket{s_1 \cdots s_M}_h$ can be subsequently projected back to the desired state, $\ket{++\cdots +}_h$.

When $\hat{\mathcal{O}}=\ket{\mathbf{z}}\bra{\mathbf{z}}$ as the projection onto a particular computational state then we obtain
\begin{eqnarray}\label{eq:obs-proj}
\vert \bk{\mathbf{z}}{\tilde{\Psi}_v} \vert^2
 = \sum_{\vec{s}} \left(\prod_{j=1}^M R^2_{s_j}(m^R_j)\right)
\bk{\tilde \Psi_v^{\vec s}}{\mathbf z}\bk{\mathbf z}{\tilde \Psi_v^{\vec s}} 
\end{eqnarray}
By substituting back the normalized factors via $\ket{\tilde \Psi_v} = N_v \ket{\Psi_v}$ and $\ket{\tilde \Psi^{\vec s}_v} = N_{\vec s} \ket{\Psi^{\vec{s}}_v}$ into Eq.~\ref{eq:obs-proj}, we recover Eq.~\ref{eq:mc_obs} in the main text.

\section*{Supplementary Information IV: NQS-ITE Derivations and Implementations}

In Eq.~\ref{eq:chainrule1}, we relate $\partial_{\theta_i}\ket{\Psi_v(\theta)}$ to $\partial_{\theta_i}\ket{\Psi_{vh}(\theta)}$. Alternatively, we may
write 
\begin{eqnarray}
\ket{\Psi_v(\theta)} = \frac{\ket{\tilde\Psi_{vh}(\theta)}}{\sqrt{\bra{\tilde\Psi_{vh}(\theta)}\hat P^{(h)}_+\ket{\tilde\Psi_{vh}(\theta)}}},
\end{eqnarray}
where $\ket{\tilde\Psi_{vh}(\theta)} = \sum_{\mathbf{h}} e^{\hat H_{RBM}(\theta,\mathbf{h})} \ket{++\dots+}_v$ is the un-normalized wave function.
Hence, Eq.~\ref{eq:chainrule1} can be cast in this equivalent form,
\begin{eqnarray}\label{eq:chainrule2}
    \left \vert \frac{\partial \Psi_v}{\partial \theta_n} \right \rangle = \frac{_h\langle ++\dots + |\partial_{\theta_n} \tilde\Psi_{vh}\rangle}{\tilde N_v} 
    - Re \left( \langle \tilde\Psi_{vh} | \hat{P}_+^{(h)}| \partial_{\theta_n} \tilde\Psi_{vh}  \rangle \right)   \frac{_h\langle ++\dots + | {\tilde \Psi_{vh}}  \rangle}{\tilde N_v^3},
\end{eqnarray}
and $\tilde N_v = \sqrt{\bra{\tilde\Psi_{vh}(\theta)}P^{(h)}_+\ket{\tilde\Psi_{vh}(\theta)}}$.  Similarly, we have to provide the derivatives of $\ket{\tilde \Psi_{vh}(\theta)}$,
\begin{eqnarray}\label{eq:unvhwf-deriv}
\begin{array}{ll}
\frac{ \partial |\tilde\Psi_{vh} \rangle }{\partial b^{R}_i} =  \hat{v}^z_i   |\tilde\Psi_{vh} \rangle , &
\frac{ \partial |\tilde\Psi_{vh} \rangle }{\partial m^{R}_j} = \tanh\left(m_j + \sum_i W_{ij} \hat{v}^z_i\right)  |\tilde\Psi_{vh} \rangle, \\
\frac{ \partial |\tilde\Psi_{vh} \rangle }{\partial W^{R}_{ij}} = \hat{v}^z_i \tanh\left(m_j + \sum_i W_{ij} \hat{v}^z_i\right)   |\tilde\Psi_{vh} \rangle, &
\frac{ \partial |\tilde\Psi_{vh} \rangle }{\partial b^{I}_i} =  i\hat{v}^z_i | \tilde \Psi_{vh} \rangle, \\
\frac{ \partial |\tilde\Psi_{vh} \rangle }{\partial m^{I}_j} =  i \tanh\left(m_j + \sum_i W_{ij} \hat{v}^z_i\right)   |\tilde\Psi_{vh} \rangle, &
\frac{ \partial |\tilde\Psi_{vh} \rangle }{\partial W^{I}_{ij}} =  i  \hat{v}^z_i \tanh\left(m_j + \sum_i W_{ij} \hat{v}^z_i\right) |\tilde\Psi_{vh} \rangle.
\end{array}
\end{eqnarray}
Substituting Eqs.~\ref{eq:chainrule2}-\ref{eq:unvhwf-deriv} into A matrix and C vector in Eq.~\ref{eq:IT-EOM}, we derive Eq.~\ref{eq:mtxAvecC} in the main text. To facilitate the following discussions, we denote $O_j = \tanh\left(m_j + \sum_i W_{ij} \hat{v}^z_i\right)$.

Since $O_j$ are not Hermitian operators and not directly measurable, an experimental scheme for measuring $\langle O_j \rangle_v$ (implied in Eqs.~\ref{eq:mtxAvecC}-\ref{eq:stochreconfigO}) should be clearly given. We propose the following strategy using the fact that $[O_j,\hat{v}^z_i]=0$,
\begin{eqnarray}
\langle O^\dag_j O_{j^{\prime} }\rangle_v & = & \bra{\Psi_v(\theta)}  O^\dag_j O_{j^{\prime}}  \ket{\Psi_v(\theta)} \nonumber \\
& = &  \sum_{\mathbf{z}_v} \vert \bk{\Psi_v(\theta)}{\mathbf{z}_v} \vert^2 O^\dag_j(\mathbf{z}_v) O_{j^{\prime}}(\mathbf{z}_v)  \nonumber \\
& \xrightarrow[\text{according to } P_v(\mathbf{z}_v)]{\text{Monte Carlo sampling}} & \sum_{k=1}^{N_{\text{exp}}} \frac{O^\dag_j(\mathbf{z}_v^k) O_{j^{\prime}}(\mathbf{z}_v^k) }{N_{\text{exp}}},
\end{eqnarray}
where $P_v(\mathbf{z}_v) = \vert \bk{\Psi_v(\theta)}{\mathbf{z}_v} \vert^2$ is a probability density, $\mathbf{z}_v = [z_{v,1}, \cdots, z_{v,N}]$ is a 
length-N binary string, and $O_{j} = \tanh(m_j + \sum_i W_{ij} z_{v,i})$. The second line in the equation above implies that simple projective measurements (in the computational basis) in a quantum circuit that prepares $\ket{\Psi_v(\theta)}$ state is an efficient approach to sample $\mathbf{z}^k_v$ binary strings from $P_v(\mathbf{z}_v)$.  $N_{\text{exp}}$ samples of [$\mathbf{z}_v^{(k=1)} \cdots  \mathbf{z}_v^{(k=N_\text{exp})}$] are obtained from the quantum circuit to estimate $O^\dag_j(\mathbf{z}_v^k) O_{j^{\prime}}(\mathbf{z}_v^k)$, according to the third line (the Monte Carlo method) in equation above.  The expression $O^\dag_j(\mathbf{z}_v^k) O_{j^{\prime}}(\mathbf{z}_v^k)$ can be evaluated efficiently once a computational state $\mathbf{z}^k_v$ is specified.

$\langle O_j \rangle_v$ can be similarly estimated. On the hard, the evaluation of $\langle O^\dag_j \hat H \rangle_v$ requires further explications,
\begin{eqnarray}\label{eq:mc2}
\langle O^\dag_j \hat H \rangle_v & = &  \bra{\Psi_v(\theta)}  O^\dag_j O_{j^{\prime}}  \ket{\Psi_v(\theta)} \nonumber \\
& = &   \sum_{\mathbf{z}_v,\mathbf{y}_v} \bk{\Psi_v(\theta)}{\mathbf{z}_v}  O^\dag_j(\mathbf{z}_v) \hat H(\mathbf{z}_v,\mathbf{y}_v)
\bk{\mathbf{y}_v}{\Psi_v(\theta)} \nonumber \\
& = & \sum_{\mathbf{z}_v} \vert \bk{\Psi_v(\theta)}{\mathbf{z}_v} \vert^2 \left( \sum_{\mathbf{y}_v}   O^\dag_j(\mathbf{z}_v) \hat H(\mathbf{z}_v,\mathbf{y}_v)
\frac{\bk{\mathbf{y}_v}{\Psi_v(\theta)}}{\bk{\mathbf{z}_v}{\Psi_v(\theta)}}\right), \nonumber \\
& \xrightarrow[\text{according to } P_v(\mathbf{z}_v)]{\text{Monte Carlo sampling}} & \sum_{k=1}^{N_{\text{exp}}} \frac{1}{N_{\text{exp}}}
\left( \sum_{j}   O^\dag_j(\mathbf{z}^k_v) \hat H(\mathbf{z}^k_v,\mathbf{y}^{k,j}_v)\frac{\bk{\mathbf{y}^{k,j}_v}{\Psi_v(\theta)}}{\bk{\mathbf{z}^k_v}{\Psi_v(\theta)}}\right),
\end{eqnarray}
where $\hat H(\mathbf{z}_v^k, \mathbf{y}^{k,j}_v) = \langle \mathbf{z}^k_v \vert \hat H \vert \mathbf{y}^{k,j}_v \rangle$. When the Hamiltonian,$ \hat H=\sum_l w_l 
\hat P_l$, is  a linear combination of Pauli strings, $\hat H$ is a sparse matrix such that each computational state $\ket{\mathbf{z}^k_v}$ is only connected to 
a few other states $\ket{\mathbf{y}_v^{k,j}}$. On the third line of Eq.~\ref{eq:mc2}, the expression in the bracket should be evaluated classically. $O^
\dag_j(\mathbf{z}^k_v) \hat H(\mathbf{z}^k_v,\mathbf{y}^{k,j}_v)$ can be done efficiently, but not so much with the individual wave function amplitude $
\ket{\Psi_v(\theta)}$ which contains the normalization constant $\tilde N_v$ that is \#P-hard to compute in general. However, we only need the un-normalized wave function amplitudes as the quantity appearing in the bracket is the ratio 
of the wave function amplitudes projected onto two computational basis states, $\ket{\mathbf{z}_v^k}$ and $\ket{\mathbf{y}_v^{k,j}}$, respectively.


\section*{Supplementary Information V: Preparing a complex-valued NQS in Quantum Circuits} 
In the main text, we present a scalable method to prepare unitary-coupled RBM-NQS in quantum circuits. Now we discuss
an extension that generates complex-valued NQS with arbitrary complex-valued $W_{ij}=W^R_{ij}+iW^I_{ij}$. First, 
we remark that the qubit-recycled approach to prepare an NQS is still applicable when coupling coefficients $W_{ij}$ are promoted to 
unrestricted complex numbers,
\begin{eqnarray}\label{eq:qrecycle-app}
\ket{\Psi_v(\theta)} & = & \frac{1}{N_v}
       \left[\prescript{}{}{\bra{+}}\left[e^{\hat h^z_M \left(m_M + \sum_i W_{iM}\hat v^z_i\right)}\right]\ket{+}\right]_{M}
       \left[\prescript{}{}{\bra{+}}\left[e^{\hat h^z_{M-1}  \left( m_{M-1} + \sum_i W_{iM-1} \hat v^z_i\right)}\right]\ket{+}\right]_{M-1} \cdots 
       \nonumber \\
       & & \times \left[\prescript{}{}{\bra{+}}\left[e^{\hat h^z_1  \left(m_1 + \sum_i W_{i1}\hat v^z_i\right)}\right]\ket{+}\right]_{1}\,\,
       e^{\sum_i b_i \hat v^z_i} \ket{++\cdots+}_v.
\end{eqnarray}
Similarly, the ensemble state preparation scheme does not have to be modified when we deal with unrestricted couplings, $W_{ij}$.
This conclusion is obvious as the crucial decomposition of the $j$-th block, $\left[\prescript{}{}{\bra{+}}[ \cdots ]\ket{+}\right]_{j}$, in Eq.~\ref{eq:qrecycle-app} still holds,
\begin{eqnarray}\label{eq:block-decomp2}
 \prescript{}{}{\bra{+}}\left[e^{\hat h^z_j \left(m_j + \sum_i W_{ij}\hat v^z_i\right)}\right]\ket{+}
& = & \sum_{s=\pm} \bra{+} e^{m^R_j \hat h^z_j} \ket{s} \bra{s} e^{\left(im^I_j+\sum_i W_{ij}\hat v^z_i\right) \hat h^z_j} \ket{+} \nonumber \\
& = & \sum_{s=\pm} R_{s}(m_j^R) \bra{s} e^{\sum_i W^R_{ij}\hat v^z_i\hat h^z_j}e^{\left(im^I_j+\sum_i W^I_{ij}\hat v^z_i\right) \hat h^z_j} \ket{+}.
\end{eqnarray}

The new task is to efficiently map the non-unitary operators $\exp\left(\sum_i W^R_{ij}\hat v^z_i \hat h^z_j \right)$ into a set of appropriate unitary operations.
We adapt the method introduced in Ref.~\cite{motta2019quantum} to get this mapping done without using post-selections. We briefly the summarize the key ideas and refer readers to Ref.~\cite{motta2019quantum} for full details.  Starting with the decomposition $\exp\left(\sum_i W^R_{ij}\hat v^z_i \hat h^z_j \right)=\exp\left((1/K) \sum_i W^R_{ij} \hat v^z_i \hat h^z_j \right)^K$ , the single non-unitary operation may be turned into K successive rotations via the following relation,
\begin{eqnarray}
\ket{\tilde{\psi}} = e^{(1/K)W_{ij}^R\hat{v}^z_{i}\hat{h}^z_j}\ket{\psi} = c e^{i (1/K) A}\ket{\psi},
\end{eqnarray}
where $c=\sqrt{\bk{\tilde\psi}{\tilde\psi}}$ and $A=\sum_{s_1,\cdots,s_D,t} a_{s_1\cdots s_D t}\hat v^{s_1}_1\cdots\hat v^{s_D}_D \hat h^t_1$.
The coefficients $a_{s_1\cdots s_D t}$ can be determined by minimizing the differences between two states, $\vert \vert W^R_{ij}
\hat{v}^z_{i}\hat{h}^z_j\ket{\psi}-icA\ket{\psi} \vert\vert$.  

In order to solve the minimization and determine $A$, one has to perform tomography on the quantum state $\ket{\psi}$. This is certainly a limiting step as the number of measurements required for a precise tomography scales exponentially with the system size. However, the situation could be 
eaiser to deal with if we assume that the correlation length of a typical many-body quantum system has a finite cut-off range such that the operator $A$ acts at most on $D$ visible spins and a single hidden spin. With this assumption\cite{motta2019quantum}, the restricted tomography mitigates the experimental costs.


\begin{thebibliography}{95}%
\makeatletter
\providecommand \@ifxundefined [1]{%
 \@ifx{#1\undefined}
}%
\providecommand \@ifnum [1]{%
 \ifnum #1\expandafter \@firstoftwo
 \else \expandafter \@secondoftwo
 \fi
}%
\providecommand \@ifx [1]{%
 \ifx #1\expandafter \@firstoftwo
 \else \expandafter \@secondoftwo
 \fi
}%
\providecommand \natexlab [1]{#1}%
\providecommand \enquote  [1]{``#1''}%
\providecommand \bibnamefont  [1]{#1}%
\providecommand \bibfnamefont [1]{#1}%
\providecommand \citenamefont [1]{#1}%
\providecommand \href@noop [0]{\@secondoftwo}%
\providecommand \href [0]{\begingroup \@sanitize@url \@href}%
\providecommand \@href[1]{\@@startlink{#1}\@@href}%
\providecommand \@@href[1]{\endgroup#1\@@endlink}%
\providecommand \@sanitize@url [0]{\catcode `\\12\catcode `\$12\catcode
  `\&12\catcode `\#12\catcode `\^12\catcode `\_12\catcode `\%12\relax}%
\providecommand \@@startlink[1]{}%
\providecommand \@@endlink[0]{}%
\providecommand \url  [0]{\begingroup\@sanitize@url \@url }%
\providecommand \@url [1]{\endgroup\@href {#1}{\urlprefix }}%
\providecommand \urlprefix  [0]{URL }%
\providecommand \Eprint [0]{\href }%
\providecommand \doibase [0]{http://dx.doi.org/}%
\providecommand \selectlanguage [0]{\@gobble}%
\providecommand \bibinfo  [0]{\@secondoftwo}%
\providecommand \bibfield  [0]{\@secondoftwo}%
\providecommand \translation [1]{[#1]}%
\providecommand \BibitemOpen [0]{}%
\providecommand \bibitemStop [0]{}%
\providecommand \bibitemNoStop [0]{.\EOS\space}%
\providecommand \EOS [0]{\spacefactor3000\relax}%
\providecommand \BibitemShut  [1]{\csname bibitem#1\endcsname}%
\let\auto@bib@innerbib\@empty
\bibitem [{\citenamefont {McClean}\ \emph {et~al.}(2016)\citenamefont
  {McClean}, \citenamefont {Romero}, \citenamefont {Babbush},\ and\
  \citenamefont {Aspuru-Guzik}}]{mcclean2016theory}%
  \BibitemOpen
  \bibfield  {author} {\bibinfo {author} {\bibfnamefont {J.~R.}\ \bibnamefont
  {McClean}}, \bibinfo {author} {\bibfnamefont {J.}~\bibnamefont {Romero}},
  \bibinfo {author} {\bibfnamefont {R.}~\bibnamefont {Babbush}}, \ and\
  \bibinfo {author} {\bibfnamefont {A.}~\bibnamefont {Aspuru-Guzik}},\
  }\href@noop {} {\bibfield  {journal} {\bibinfo  {journal} {New J. Phys.}\
  }\textbf {\bibinfo {volume} {18}},\ \bibinfo {pages} {023023} (\bibinfo
  {year} {2016})}\BibitemShut {NoStop}%
\bibitem [{\citenamefont {Preskill}(2018)}]{preskill2018quantum}%
  \BibitemOpen
  \bibfield  {author} {\bibinfo {author} {\bibfnamefont {J.}~\bibnamefont
  {Preskill}},\ }\href@noop {} {\bibfield  {journal} {\bibinfo  {journal}
  {Quantum}\ }\textbf {\bibinfo {volume} {2}},\ \bibinfo {pages} {79} (\bibinfo
  {year} {2018})}\BibitemShut {NoStop}%
\bibitem [{\citenamefont {Peruzzo}\ \emph {et~al.}(2014)\citenamefont
  {Peruzzo}, \citenamefont {McClean}, \citenamefont {Shadbolt}, \citenamefont
  {Yung}, \citenamefont {Zhou}, \citenamefont {Love}, \citenamefont
  {Aspuru-Guzik},\ and\ \citenamefont {O’Brien}}]{peruzzo_natcomm_14}%
  \BibitemOpen
  \bibfield  {author} {\bibinfo {author} {\bibfnamefont {A.}~\bibnamefont
  {Peruzzo}}, \bibinfo {author} {\bibfnamefont {J.}~\bibnamefont {McClean}},
  \bibinfo {author} {\bibfnamefont {P.}~\bibnamefont {Shadbolt}}, \bibinfo
  {author} {\bibfnamefont {M.-H.}\ \bibnamefont {Yung}}, \bibinfo {author}
  {\bibfnamefont {X.-Q.}\ \bibnamefont {Zhou}}, \bibinfo {author}
  {\bibfnamefont {P.~J.}\ \bibnamefont {Love}}, \bibinfo {author}
  {\bibfnamefont {A.}~\bibnamefont {Aspuru-Guzik}}, \ and\ \bibinfo {author}
  {\bibfnamefont {J.~L.}\ \bibnamefont {O’Brien}},\ }\href@noop {} {\bibfield
   {journal} {\bibinfo  {journal} {Nat. Commun.}\ }\textbf {\bibinfo {volume}
  {5}},\ \bibinfo {pages} {4213} (\bibinfo {year} {2014})}\BibitemShut
  {NoStop}%
\bibitem [{\citenamefont {Farhi}\ \emph {et~al.}(2014)\citenamefont {Farhi},
  \citenamefont {Goldstone},\ and\ \citenamefont {Gutmann}}]{Farhi2014}%
  \BibitemOpen
  \bibfield  {author} {\bibinfo {author} {\bibfnamefont {E.}~\bibnamefont
  {Farhi}}, \bibinfo {author} {\bibfnamefont {J.}~\bibnamefont {Goldstone}}, \
  and\ \bibinfo {author} {\bibfnamefont {S.}~\bibnamefont {Gutmann}},\
  }\href@noop {} {\bibfield  {journal} {\bibinfo  {journal} {arXiv preprint
  arXiv:1411.4028}\ } (\bibinfo {year} {2014})}\BibitemShut {NoStop}%
\bibitem [{\citenamefont {Kandala}\ \emph {et~al.}(2017)\citenamefont
  {Kandala}, \citenamefont {Mezzacapo}, \citenamefont {Temme}, \citenamefont
  {Takita}, \citenamefont {Brink}, \citenamefont {Chow},\ and\ \citenamefont
  {Gambetta}}]{kandala2017hardware}%
  \BibitemOpen
  \bibfield  {author} {\bibinfo {author} {\bibfnamefont {A.}~\bibnamefont
  {Kandala}}, \bibinfo {author} {\bibfnamefont {A.}~\bibnamefont {Mezzacapo}},
  \bibinfo {author} {\bibfnamefont {K.}~\bibnamefont {Temme}}, \bibinfo
  {author} {\bibfnamefont {M.}~\bibnamefont {Takita}}, \bibinfo {author}
  {\bibfnamefont {M.}~\bibnamefont {Brink}}, \bibinfo {author} {\bibfnamefont
  {J.~M.}\ \bibnamefont {Chow}}, \ and\ \bibinfo {author} {\bibfnamefont
  {J.~M.}\ \bibnamefont {Gambetta}},\ }\href@noop {} {\bibfield  {journal}
  {\bibinfo  {journal} {Nature}\ }\textbf {\bibinfo {volume} {549}},\ \bibinfo
  {pages} {242} (\bibinfo {year} {2017})}\BibitemShut {NoStop}%
\bibitem [{\citenamefont {Hempel}\ \emph {et~al.}(2018)\citenamefont {Hempel},
  \citenamefont {Maier}, \citenamefont {Romero}, \citenamefont {McClean},
  \citenamefont {Monz}, \citenamefont {Shen}, \citenamefont {Jurcevic},
  \citenamefont {Lanyon}, \citenamefont {Love}, \citenamefont {Babbush} \emph
  {et~al.}}]{hempel2018quantum}%
  \BibitemOpen
  \bibfield  {author} {\bibinfo {author} {\bibfnamefont {C.}~\bibnamefont
  {Hempel}}, \bibinfo {author} {\bibfnamefont {C.}~\bibnamefont {Maier}},
  \bibinfo {author} {\bibfnamefont {J.}~\bibnamefont {Romero}}, \bibinfo
  {author} {\bibfnamefont {J.}~\bibnamefont {McClean}}, \bibinfo {author}
  {\bibfnamefont {T.}~\bibnamefont {Monz}}, \bibinfo {author} {\bibfnamefont
  {H.}~\bibnamefont {Shen}}, \bibinfo {author} {\bibfnamefont {P.}~\bibnamefont
  {Jurcevic}}, \bibinfo {author} {\bibfnamefont {B.~P.}\ \bibnamefont
  {Lanyon}}, \bibinfo {author} {\bibfnamefont {P.}~\bibnamefont {Love}},
  \bibinfo {author} {\bibfnamefont {R.}~\bibnamefont {Babbush}},  \emph
  {et~al.},\ }\href@noop {} {\bibfield  {journal} {\bibinfo  {journal} {Phys.
  Rev. X}\ }\textbf {\bibinfo {volume} {8}},\ \bibinfo {pages} {031022}
  (\bibinfo {year} {2018})}\BibitemShut {NoStop}%
\bibitem [{\citenamefont {Colless}\ \emph {et~al.}(2018)\citenamefont
  {Colless}, \citenamefont {Ramasesh}, \citenamefont {Dahlen}, \citenamefont
  {Blok}, \citenamefont {Kimchi-Schwartz}, \citenamefont {McClean},
  \citenamefont {Carter}, \citenamefont {De~Jong},\ and\ \citenamefont
  {Siddiqi}}]{colless2018computation}%
  \BibitemOpen
  \bibfield  {author} {\bibinfo {author} {\bibfnamefont {J.~I.}\ \bibnamefont
  {Colless}}, \bibinfo {author} {\bibfnamefont {V.~V.}\ \bibnamefont
  {Ramasesh}}, \bibinfo {author} {\bibfnamefont {D.}~\bibnamefont {Dahlen}},
  \bibinfo {author} {\bibfnamefont {M.~S.}\ \bibnamefont {Blok}}, \bibinfo
  {author} {\bibfnamefont {M.}~\bibnamefont {Kimchi-Schwartz}}, \bibinfo
  {author} {\bibfnamefont {J.}~\bibnamefont {McClean}}, \bibinfo {author}
  {\bibfnamefont {J.}~\bibnamefont {Carter}}, \bibinfo {author} {\bibfnamefont
  {W.}~\bibnamefont {De~Jong}}, \ and\ \bibinfo {author} {\bibfnamefont
  {I.}~\bibnamefont {Siddiqi}},\ }\href@noop {} {\bibfield  {journal} {\bibinfo
   {journal} {Phys. Rev. X}\ }\textbf {\bibinfo {volume} {8}},\ \bibinfo
  {pages} {011021} (\bibinfo {year} {2018})}\BibitemShut {NoStop}%
\bibitem [{\citenamefont {Sagastizabal}\ \emph {et~al.}(2019)\citenamefont
  {Sagastizabal}, \citenamefont {Bonet-Monroig}, \citenamefont {Singh},
  \citenamefont {Rol}, \citenamefont {Bultink}, \citenamefont {Fu},
  \citenamefont {Price}, \citenamefont {Ostroukh}, \citenamefont
  {Muthusubramanian}, \citenamefont {Bruno} \emph
  {et~al.}}]{sagastizabal2019experimental}%
  \BibitemOpen
  \bibfield  {author} {\bibinfo {author} {\bibfnamefont {R.}~\bibnamefont
  {Sagastizabal}}, \bibinfo {author} {\bibfnamefont {X.}~\bibnamefont
  {Bonet-Monroig}}, \bibinfo {author} {\bibfnamefont {M.}~\bibnamefont
  {Singh}}, \bibinfo {author} {\bibfnamefont {M.}~\bibnamefont {Rol}}, \bibinfo
  {author} {\bibfnamefont {C.}~\bibnamefont {Bultink}}, \bibinfo {author}
  {\bibfnamefont {X.}~\bibnamefont {Fu}}, \bibinfo {author} {\bibfnamefont
  {C.}~\bibnamefont {Price}}, \bibinfo {author} {\bibfnamefont
  {V.}~\bibnamefont {Ostroukh}}, \bibinfo {author} {\bibfnamefont
  {N.}~\bibnamefont {Muthusubramanian}}, \bibinfo {author} {\bibfnamefont
  {A.}~\bibnamefont {Bruno}},  \emph {et~al.},\ }\href@noop {} {\bibfield
  {journal} {\bibinfo  {journal} {Phys. Rev. A}\ }\textbf {\bibinfo {volume}
  {100}},\ \bibinfo {pages} {010302} (\bibinfo {year} {2019})}\BibitemShut
  {NoStop}%
\bibitem [{\citenamefont {Aspuru-Guzik}\ \emph {et~al.}(2005)\citenamefont
  {Aspuru-Guzik}, \citenamefont {Dutoi}, \citenamefont {Love},\ and\
  \citenamefont {Head-Gordon}}]{aspuru2005simulated}%
  \BibitemOpen
  \bibfield  {author} {\bibinfo {author} {\bibfnamefont {A.}~\bibnamefont
  {Aspuru-Guzik}}, \bibinfo {author} {\bibfnamefont {A.~D.}\ \bibnamefont
  {Dutoi}}, \bibinfo {author} {\bibfnamefont {P.~J.}\ \bibnamefont {Love}}, \
  and\ \bibinfo {author} {\bibfnamefont {M.}~\bibnamefont {Head-Gordon}},\
  }\href@noop {} {\bibfield  {journal} {\bibinfo  {journal} {Science}\ }\textbf
  {\bibinfo {volume} {309}},\ \bibinfo {pages} {1704} (\bibinfo {year}
  {2005})}\BibitemShut {NoStop}%
\bibitem [{\citenamefont {Li}\ \emph {et~al.}(2019{\natexlab{a}})\citenamefont
  {Li}, \citenamefont {Hu}, \citenamefont {Zhang}, \citenamefont {Song},\ and\
  \citenamefont {Yung}}]{li2019variational}%
  \BibitemOpen
  \bibfield  {author} {\bibinfo {author} {\bibfnamefont {Y.}~\bibnamefont
  {Li}}, \bibinfo {author} {\bibfnamefont {J.}~\bibnamefont {Hu}}, \bibinfo
  {author} {\bibfnamefont {X.-M.}\ \bibnamefont {Zhang}}, \bibinfo {author}
  {\bibfnamefont {Z.}~\bibnamefont {Song}}, \ and\ \bibinfo {author}
  {\bibfnamefont {M.-H.}\ \bibnamefont {Yung}},\ }\href@noop {} {\bibfield
  {journal} {\bibinfo  {journal} {Adv. Theor. Simul.}\ }\textbf {\bibinfo
  {volume} {2}},\ \bibinfo {pages} {1800182} (\bibinfo {year}
  {2019}{\natexlab{a}})}\BibitemShut {NoStop}%
\bibitem [{\citenamefont {Cao}\ \emph {et~al.}(2018)\citenamefont {Cao},
  \citenamefont {Romero}, \citenamefont {Olson}, \citenamefont {Degroote},
  \citenamefont {Johnson}, \citenamefont {Kieferov{\'a}}, \citenamefont
  {Kivlichan}, \citenamefont {Menke}, \citenamefont {Peropadre}, \citenamefont
  {Sawaya} \emph {et~al.}}]{cao2018quantum}%
  \BibitemOpen
  \bibfield  {author} {\bibinfo {author} {\bibfnamefont {Y.}~\bibnamefont
  {Cao}}, \bibinfo {author} {\bibfnamefont {J.}~\bibnamefont {Romero}},
  \bibinfo {author} {\bibfnamefont {J.~P.}\ \bibnamefont {Olson}}, \bibinfo
  {author} {\bibfnamefont {M.}~\bibnamefont {Degroote}}, \bibinfo {author}
  {\bibfnamefont {P.~D.}\ \bibnamefont {Johnson}}, \bibinfo {author}
  {\bibfnamefont {M.}~\bibnamefont {Kieferov{\'a}}}, \bibinfo {author}
  {\bibfnamefont {I.~D.}\ \bibnamefont {Kivlichan}}, \bibinfo {author}
  {\bibfnamefont {T.}~\bibnamefont {Menke}}, \bibinfo {author} {\bibfnamefont
  {B.}~\bibnamefont {Peropadre}}, \bibinfo {author} {\bibfnamefont {N.~P.}\
  \bibnamefont {Sawaya}},  \emph {et~al.},\ }\href@noop {} {\bibfield
  {journal} {\bibinfo  {journal} {arXiv preprint arXiv:1812.09976}\ } (\bibinfo
  {year} {2018})}\BibitemShut {NoStop}%
\bibitem [{\citenamefont {McArdle}\ \emph {et~al.}(2018)\citenamefont
  {McArdle}, \citenamefont {Endo}, \citenamefont {Aspuru-Guzik}, \citenamefont
  {Benjamin},\ and\ \citenamefont {Yuan}}]{mcardle2018quantum}%
  \BibitemOpen
  \bibfield  {author} {\bibinfo {author} {\bibfnamefont {S.}~\bibnamefont
  {McArdle}}, \bibinfo {author} {\bibfnamefont {S.}~\bibnamefont {Endo}},
  \bibinfo {author} {\bibfnamefont {A.}~\bibnamefont {Aspuru-Guzik}}, \bibinfo
  {author} {\bibfnamefont {S.}~\bibnamefont {Benjamin}}, \ and\ \bibinfo
  {author} {\bibfnamefont {X.}~\bibnamefont {Yuan}},\ }\href@noop {} {\bibfield
   {journal} {\bibinfo  {journal} {arXiv preprint arXiv:1808.10402}\ }
  (\bibinfo {year} {2018})}\BibitemShut {NoStop}%
\bibitem [{\citenamefont {Childs}\ \emph {et~al.}(2018)\citenamefont {Childs},
  \citenamefont {Maslov}, \citenamefont {Nam}, \citenamefont {Ross},\ and\
  \citenamefont {Su}}]{childs2018toward}%
  \BibitemOpen
  \bibfield  {author} {\bibinfo {author} {\bibfnamefont {A.~M.}\ \bibnamefont
  {Childs}}, \bibinfo {author} {\bibfnamefont {D.}~\bibnamefont {Maslov}},
  \bibinfo {author} {\bibfnamefont {Y.}~\bibnamefont {Nam}}, \bibinfo {author}
  {\bibfnamefont {N.~J.}\ \bibnamefont {Ross}}, \ and\ \bibinfo {author}
  {\bibfnamefont {Y.}~\bibnamefont {Su}},\ }\href@noop {} {\bibfield  {journal}
  {\bibinfo  {journal} {Proc. Natl. Acad. Sci.}\ }\textbf {\bibinfo {volume}
  {115}},\ \bibinfo {pages} {9456} (\bibinfo {year} {2018})}\BibitemShut
  {NoStop}%
\bibitem [{\citenamefont {Wecker}\ \emph {et~al.}(2015)\citenamefont {Wecker},
  \citenamefont {Hastings},\ and\ \citenamefont {Troyer}}]{wecker_pra2015}%
  \BibitemOpen
  \bibfield  {author} {\bibinfo {author} {\bibfnamefont {D.}~\bibnamefont
  {Wecker}}, \bibinfo {author} {\bibfnamefont {M.~B.}\ \bibnamefont
  {Hastings}}, \ and\ \bibinfo {author} {\bibfnamefont {M.}~\bibnamefont
  {Troyer}},\ }\href {\doibase 10.1103/PhysRevA.92.042303} {\bibfield
  {journal} {\bibinfo  {journal} {Phys. Rev. A}\ }\textbf {\bibinfo {volume}
  {92}},\ \bibinfo {pages} {042303} (\bibinfo {year} {2015})}\BibitemShut
  {NoStop}%
\bibitem [{\citenamefont {K{\"u}hn}\ \emph {et~al.}(2018)\citenamefont
  {K{\"u}hn}, \citenamefont {Zanker}, \citenamefont {Deglmann}, \citenamefont
  {Marthaler},\ and\ \citenamefont {Wei{\ss}}}]{kuhn2018accuracy}%
  \BibitemOpen
  \bibfield  {author} {\bibinfo {author} {\bibfnamefont {M.}~\bibnamefont
  {K{\"u}hn}}, \bibinfo {author} {\bibfnamefont {S.}~\bibnamefont {Zanker}},
  \bibinfo {author} {\bibfnamefont {P.}~\bibnamefont {Deglmann}}, \bibinfo
  {author} {\bibfnamefont {M.}~\bibnamefont {Marthaler}}, \ and\ \bibinfo
  {author} {\bibfnamefont {H.}~\bibnamefont {Wei{\ss}}},\ }\href@noop {}
  {\bibfield  {journal} {\bibinfo  {journal} {arXiv preprint arXiv:1812.06814}\
  } (\bibinfo {year} {2018})}\BibitemShut {NoStop}%
\bibitem [{\citenamefont {Li}\ \emph {et~al.}(2019{\natexlab{b}})\citenamefont
  {Li}, \citenamefont {Li}, \citenamefont {Dattani}, \citenamefont {Umrigar},\
  and\ \citenamefont {Chan}}]{li2019electronic}%
  \BibitemOpen
  \bibfield  {author} {\bibinfo {author} {\bibfnamefont {Z.}~\bibnamefont
  {Li}}, \bibinfo {author} {\bibfnamefont {J.}~\bibnamefont {Li}}, \bibinfo
  {author} {\bibfnamefont {N.~S.}\ \bibnamefont {Dattani}}, \bibinfo {author}
  {\bibfnamefont {C.}~\bibnamefont {Umrigar}}, \ and\ \bibinfo {author}
  {\bibfnamefont {G.~K.-L.}\ \bibnamefont {Chan}},\ }\href@noop {} {\bibfield
  {journal} {\bibinfo  {journal} {J. Chem. Phys.}\ }\textbf {\bibinfo {volume}
  {150}},\ \bibinfo {pages} {024302} (\bibinfo {year}
  {2019}{\natexlab{b}})}\BibitemShut {NoStop}%
\bibitem [{\citenamefont {Reiher}\ \emph {et~al.}(2017)\citenamefont {Reiher},
  \citenamefont {Wiebe}, \citenamefont {Svore}, \citenamefont {Wecker},\ and\
  \citenamefont {Troyer}}]{reiher2017elucidating}%
  \BibitemOpen
  \bibfield  {author} {\bibinfo {author} {\bibfnamefont {M.}~\bibnamefont
  {Reiher}}, \bibinfo {author} {\bibfnamefont {N.}~\bibnamefont {Wiebe}},
  \bibinfo {author} {\bibfnamefont {K.~M.}\ \bibnamefont {Svore}}, \bibinfo
  {author} {\bibfnamefont {D.}~\bibnamefont {Wecker}}, \ and\ \bibinfo {author}
  {\bibfnamefont {M.}~\bibnamefont {Troyer}},\ }\href@noop {} {\bibfield
  {journal} {\bibinfo  {journal} {Proceedings of the National Academy of
  Sciences}\ }\textbf {\bibinfo {volume} {114}},\ \bibinfo {pages} {7555}
  (\bibinfo {year} {2017})}\BibitemShut {NoStop}%
\bibitem [{\citenamefont {Kivlichan}\ \emph {et~al.}(2018)\citenamefont
  {Kivlichan}, \citenamefont {McClean}, \citenamefont {Wiebe}, \citenamefont
  {Gidney}, \citenamefont {Aspuru-Guzik}, \citenamefont {Chan},\ and\
  \citenamefont {Babbush}}]{kivlichan2018quantum}%
  \BibitemOpen
  \bibfield  {author} {\bibinfo {author} {\bibfnamefont {I.~D.}\ \bibnamefont
  {Kivlichan}}, \bibinfo {author} {\bibfnamefont {J.}~\bibnamefont {McClean}},
  \bibinfo {author} {\bibfnamefont {N.}~\bibnamefont {Wiebe}}, \bibinfo
  {author} {\bibfnamefont {C.}~\bibnamefont {Gidney}}, \bibinfo {author}
  {\bibfnamefont {A.}~\bibnamefont {Aspuru-Guzik}}, \bibinfo {author}
  {\bibfnamefont {G.~K.-L.}\ \bibnamefont {Chan}}, \ and\ \bibinfo {author}
  {\bibfnamefont {R.}~\bibnamefont {Babbush}},\ }\href@noop {} {\bibfield
  {journal} {\bibinfo  {journal} {Physical review letters}\ }\textbf {\bibinfo
  {volume} {120}},\ \bibinfo {pages} {110501} (\bibinfo {year}
  {2018})}\BibitemShut {NoStop}%
\bibitem [{\citenamefont {Babbush}\ \emph {et~al.}(2018)\citenamefont
  {Babbush}, \citenamefont {Wiebe}, \citenamefont {McClean}, \citenamefont
  {McClain}, \citenamefont {Neven},\ and\ \citenamefont
  {Chan}}]{babbush2018low}%
  \BibitemOpen
  \bibfield  {author} {\bibinfo {author} {\bibfnamefont {R.}~\bibnamefont
  {Babbush}}, \bibinfo {author} {\bibfnamefont {N.}~\bibnamefont {Wiebe}},
  \bibinfo {author} {\bibfnamefont {J.}~\bibnamefont {McClean}}, \bibinfo
  {author} {\bibfnamefont {J.}~\bibnamefont {McClain}}, \bibinfo {author}
  {\bibfnamefont {H.}~\bibnamefont {Neven}}, \ and\ \bibinfo {author}
  {\bibfnamefont {G.~K.-L.}\ \bibnamefont {Chan}},\ }\href@noop {} {\bibfield
  {journal} {\bibinfo  {journal} {Physical Review X}\ }\textbf {\bibinfo
  {volume} {8}},\ \bibinfo {pages} {011044} (\bibinfo {year}
  {2018})}\BibitemShut {NoStop}%
\bibitem [{\citenamefont {Ryabinkin}\ \emph {et~al.}(2018)\citenamefont
  {Ryabinkin}, \citenamefont {Yen}, \citenamefont {Genin},\ and\ \citenamefont
  {Izmaylov}}]{ryabinkin2018qubit}%
  \BibitemOpen
  \bibfield  {author} {\bibinfo {author} {\bibfnamefont {I.~G.}\ \bibnamefont
  {Ryabinkin}}, \bibinfo {author} {\bibfnamefont {T.-C.}\ \bibnamefont {Yen}},
  \bibinfo {author} {\bibfnamefont {S.~N.}\ \bibnamefont {Genin}}, \ and\
  \bibinfo {author} {\bibfnamefont {A.~F.}\ \bibnamefont {Izmaylov}},\
  }\href@noop {} {\bibfield  {journal} {\bibinfo  {journal} {J. Chem. Theory
  Comput.}\ }\textbf {\bibinfo {volume} {14}},\ \bibinfo {pages} {6317}
  (\bibinfo {year} {2018})}\BibitemShut {NoStop}%
\bibitem [{\citenamefont {Ryabinkin}\ and\ \citenamefont
  {Genin}(2019)}]{ryabinkin2019iterative}%
  \BibitemOpen
  \bibfield  {author} {\bibinfo {author} {\bibfnamefont {I.~G.}\ \bibnamefont
  {Ryabinkin}}\ and\ \bibinfo {author} {\bibfnamefont {S.~N.}\ \bibnamefont
  {Genin}},\ }\href@noop {} {\bibfield  {journal} {\bibinfo  {journal} {arXiv
  preprint arXiv:1906.11192}\ } (\bibinfo {year} {2019})}\BibitemShut {NoStop}%
\bibitem [{\citenamefont {Dallaire-Demers}\ \emph {et~al.}(2019)\citenamefont
  {Dallaire-Demers}, \citenamefont {Fontalvo}, \citenamefont {Veis},
  \citenamefont {Sim},\ and\ \citenamefont {Aspuru-Guzik}}]{dallaire2019low}%
  \BibitemOpen
  \bibfield  {author} {\bibinfo {author} {\bibfnamefont {P.-L.}\ \bibnamefont
  {Dallaire-Demers}}, \bibinfo {author} {\bibfnamefont {J.~R.}\ \bibnamefont
  {Fontalvo}}, \bibinfo {author} {\bibfnamefont {L.}~\bibnamefont {Veis}},
  \bibinfo {author} {\bibfnamefont {S.}~\bibnamefont {Sim}}, \ and\ \bibinfo
  {author} {\bibfnamefont {A.}~\bibnamefont {Aspuru-Guzik}},\ }\href@noop {}
  {\bibfield  {journal} {\bibinfo  {journal} {Quantum Science and Technology}\
  } (\bibinfo {year} {2019})}\BibitemShut {NoStop}%
\bibitem [{\citenamefont {Romero}\ \emph {et~al.}(2018)\citenamefont {Romero},
  \citenamefont {Babbush}, \citenamefont {McClean}, \citenamefont {Hempel},
  \citenamefont {Love},\ and\ \citenamefont
  {Aspuru-Guzik}}]{romero2018strategies}%
  \BibitemOpen
  \bibfield  {author} {\bibinfo {author} {\bibfnamefont {J.}~\bibnamefont
  {Romero}}, \bibinfo {author} {\bibfnamefont {R.}~\bibnamefont {Babbush}},
  \bibinfo {author} {\bibfnamefont {J.~R.}\ \bibnamefont {McClean}}, \bibinfo
  {author} {\bibfnamefont {C.}~\bibnamefont {Hempel}}, \bibinfo {author}
  {\bibfnamefont {P.~J.}\ \bibnamefont {Love}}, \ and\ \bibinfo {author}
  {\bibfnamefont {A.}~\bibnamefont {Aspuru-Guzik}},\ }\href@noop {} {\bibfield
  {journal} {\bibinfo  {journal} {Quantum Sci. Technol.}\ }\textbf {\bibinfo
  {volume} {4}},\ \bibinfo {pages} {014008} (\bibinfo {year}
  {2018})}\BibitemShut {NoStop}%
\bibitem [{\citenamefont {O'Brien}\ \emph {et~al.}(2019)\citenamefont
  {O'Brien}, \citenamefont {Senjean}, \citenamefont {Sagastizabal},
  \citenamefont {Bonet-Monroig}, \citenamefont {Dutkiewicz}, \citenamefont
  {Buda}, \citenamefont {DiCarlo},\ and\ \citenamefont
  {Visscher}}]{o2019calculating}%
  \BibitemOpen
  \bibfield  {author} {\bibinfo {author} {\bibfnamefont {T.}~\bibnamefont
  {O'Brien}}, \bibinfo {author} {\bibfnamefont {B.}~\bibnamefont {Senjean}},
  \bibinfo {author} {\bibfnamefont {R.}~\bibnamefont {Sagastizabal}}, \bibinfo
  {author} {\bibfnamefont {X.}~\bibnamefont {Bonet-Monroig}}, \bibinfo {author}
  {\bibfnamefont {A.}~\bibnamefont {Dutkiewicz}}, \bibinfo {author}
  {\bibfnamefont {F.}~\bibnamefont {Buda}}, \bibinfo {author} {\bibfnamefont
  {L.}~\bibnamefont {DiCarlo}}, \ and\ \bibinfo {author} {\bibfnamefont
  {L.}~\bibnamefont {Visscher}},\ }\href@noop {} {\bibfield  {journal}
  {\bibinfo  {journal} {arXiv preprint arXiv:1905.03742}\ } (\bibinfo {year}
  {2019})}\BibitemShut {NoStop}%
\bibitem [{\citenamefont {Benfenati}\ \emph {et~al.}(2019)\citenamefont
  {Benfenati}, \citenamefont {Guidoni}, \citenamefont {Mazzola}, \citenamefont
  {Barkoutsos}, \citenamefont {Ollitrault},\ and\ \citenamefont
  {Tavernelli}}]{benfenati2019extended}%
  \BibitemOpen
  \bibfield  {author} {\bibinfo {author} {\bibfnamefont {F.}~\bibnamefont
  {Benfenati}}, \bibinfo {author} {\bibfnamefont {L.}~\bibnamefont {Guidoni}},
  \bibinfo {author} {\bibfnamefont {G.}~\bibnamefont {Mazzola}}, \bibinfo
  {author} {\bibfnamefont {P.}~\bibnamefont {Barkoutsos}}, \bibinfo {author}
  {\bibfnamefont {P.}~\bibnamefont {Ollitrault}}, \ and\ \bibinfo {author}
  {\bibfnamefont {I.}~\bibnamefont {Tavernelli}},\ }in\ \href@noop {} {\emph
  {\bibinfo {booktitle} {Quantum Information and Measurement}}}\ (\bibinfo
  {organization} {Optical Society of America},\ \bibinfo {year} {2019})\ pp.\
  \bibinfo {pages} {F5A--36}\BibitemShut {NoStop}%
\bibitem [{\citenamefont {Zhao}\ \emph {et~al.}(2019)\citenamefont {Zhao},
  \citenamefont {Tranter}, \citenamefont {Kirby}, \citenamefont {Ung},
  \citenamefont {Miyake},\ and\ \citenamefont {Love}}]{zhao2019measurement}%
  \BibitemOpen
  \bibfield  {author} {\bibinfo {author} {\bibfnamefont {A.}~\bibnamefont
  {Zhao}}, \bibinfo {author} {\bibfnamefont {A.}~\bibnamefont {Tranter}},
  \bibinfo {author} {\bibfnamefont {W.~M.}\ \bibnamefont {Kirby}}, \bibinfo
  {author} {\bibfnamefont {S.~F.}\ \bibnamefont {Ung}}, \bibinfo {author}
  {\bibfnamefont {A.}~\bibnamefont {Miyake}}, \ and\ \bibinfo {author}
  {\bibfnamefont {P.}~\bibnamefont {Love}},\ }\href@noop {} {\bibfield
  {journal} {\bibinfo  {journal} {arXiv preprint arXiv:1908.08067}\ } (\bibinfo
  {year} {2019})}\BibitemShut {NoStop}%
\bibitem [{\citenamefont {Rubin}\ \emph {et~al.}(2018)\citenamefont {Rubin},
  \citenamefont {Babbush},\ and\ \citenamefont
  {McClean}}]{rubin2018application}%
  \BibitemOpen
  \bibfield  {author} {\bibinfo {author} {\bibfnamefont {N.~C.}\ \bibnamefont
  {Rubin}}, \bibinfo {author} {\bibfnamefont {R.}~\bibnamefont {Babbush}}, \
  and\ \bibinfo {author} {\bibfnamefont {J.}~\bibnamefont {McClean}},\
  }\href@noop {} {\bibfield  {journal} {\bibinfo  {journal} {New J. Phys.}\
  }\textbf {\bibinfo {volume} {20}},\ \bibinfo {pages} {053020} (\bibinfo
  {year} {2018})}\BibitemShut {NoStop}%
\bibitem [{\citenamefont {Crawford}\ \emph {et~al.}(2019)\citenamefont
  {Crawford}, \citenamefont {van Straaten}, \citenamefont {Wang}, \citenamefont
  {Parks}, \citenamefont {Campbell},\ and\ \citenamefont
  {Brierley}}]{crawford2019efficient}%
  \BibitemOpen
  \bibfield  {author} {\bibinfo {author} {\bibfnamefont {O.}~\bibnamefont
  {Crawford}}, \bibinfo {author} {\bibfnamefont {B.}~\bibnamefont {van
  Straaten}}, \bibinfo {author} {\bibfnamefont {D.}~\bibnamefont {Wang}},
  \bibinfo {author} {\bibfnamefont {T.}~\bibnamefont {Parks}}, \bibinfo
  {author} {\bibfnamefont {E.}~\bibnamefont {Campbell}}, \ and\ \bibinfo
  {author} {\bibfnamefont {S.}~\bibnamefont {Brierley}},\ }\href@noop {}
  {\bibfield  {journal} {\bibinfo  {journal} {arXiv preprint arXiv:1908.06942}\
  } (\bibinfo {year} {2019})}\BibitemShut {NoStop}%
\bibitem [{\citenamefont {Izmaylov}\ \emph
  {et~al.}(2019{\natexlab{a}})\citenamefont {Izmaylov}, \citenamefont {Yen},\
  and\ \citenamefont {Ryabinkin}}]{izmaylov2019revising}%
  \BibitemOpen
  \bibfield  {author} {\bibinfo {author} {\bibfnamefont {A.~F.}\ \bibnamefont
  {Izmaylov}}, \bibinfo {author} {\bibfnamefont {T.-C.}\ \bibnamefont {Yen}}, \
  and\ \bibinfo {author} {\bibfnamefont {I.~G.}\ \bibnamefont {Ryabinkin}},\
  }\href@noop {} {\bibfield  {journal} {\bibinfo  {journal} {Chem. Sci.}\
  }\textbf {\bibinfo {volume} {10}},\ \bibinfo {pages} {3746} (\bibinfo {year}
  {2019}{\natexlab{a}})}\BibitemShut {NoStop}%
\bibitem [{\citenamefont {Izmaylov}\ \emph
  {et~al.}(2019{\natexlab{b}})\citenamefont {Izmaylov}, \citenamefont {Yen},
  \citenamefont {Lang},\ and\ \citenamefont
  {Verteletskyi}}]{izmaylov2019unitary}%
  \BibitemOpen
  \bibfield  {author} {\bibinfo {author} {\bibfnamefont {A.~F.}\ \bibnamefont
  {Izmaylov}}, \bibinfo {author} {\bibfnamefont {T.-C.}\ \bibnamefont {Yen}},
  \bibinfo {author} {\bibfnamefont {R.~A.}\ \bibnamefont {Lang}}, \ and\
  \bibinfo {author} {\bibfnamefont {V.}~\bibnamefont {Verteletskyi}},\
  }\href@noop {} {\bibfield  {journal} {\bibinfo  {journal} {arXiv preprint
  arXiv:1907.09040}\ } (\bibinfo {year} {2019}{\natexlab{b}})}\BibitemShut
  {NoStop}%
\bibitem [{\citenamefont {Verteletskyi}\ \emph {et~al.}(2019)\citenamefont
  {Verteletskyi}, \citenamefont {Yen},\ and\ \citenamefont
  {Izmaylov}}]{verteletskyi2019measurement}%
  \BibitemOpen
  \bibfield  {author} {\bibinfo {author} {\bibfnamefont {V.}~\bibnamefont
  {Verteletskyi}}, \bibinfo {author} {\bibfnamefont {T.-C.}\ \bibnamefont
  {Yen}}, \ and\ \bibinfo {author} {\bibfnamefont {A.~F.}\ \bibnamefont
  {Izmaylov}},\ }\href@noop {} {\bibfield  {journal} {\bibinfo  {journal}
  {arXiv preprint arXiv:1907.03358}\ } (\bibinfo {year} {2019})}\BibitemShut
  {NoStop}%
\bibitem [{\citenamefont {Huggins}\ \emph
  {et~al.}(2019{\natexlab{a}})\citenamefont {Huggins}, \citenamefont {McClean},
  \citenamefont {Rubin}, \citenamefont {Jiang}, \citenamefont {Wiebe},
  \citenamefont {Whaley},\ and\ \citenamefont
  {Babbush}}]{huggins2019efficient}%
  \BibitemOpen
  \bibfield  {author} {\bibinfo {author} {\bibfnamefont {W.~J.}\ \bibnamefont
  {Huggins}}, \bibinfo {author} {\bibfnamefont {J.}~\bibnamefont {McClean}},
  \bibinfo {author} {\bibfnamefont {N.}~\bibnamefont {Rubin}}, \bibinfo
  {author} {\bibfnamefont {Z.}~\bibnamefont {Jiang}}, \bibinfo {author}
  {\bibfnamefont {N.}~\bibnamefont {Wiebe}}, \bibinfo {author} {\bibfnamefont
  {K.~B.}\ \bibnamefont {Whaley}}, \ and\ \bibinfo {author} {\bibfnamefont
  {R.}~\bibnamefont {Babbush}},\ }\href@noop {} {\bibfield  {journal} {\bibinfo
   {journal} {arXiv preprint arXiv:1907.13117}\ } (\bibinfo {year}
  {2019}{\natexlab{a}})}\BibitemShut {NoStop}%
\bibitem [{\citenamefont {Mitarai}\ and\ \citenamefont
  {Fujii}(2019)}]{mitarai2019methodology}%
  \BibitemOpen
  \bibfield  {author} {\bibinfo {author} {\bibfnamefont {K.}~\bibnamefont
  {Mitarai}}\ and\ \bibinfo {author} {\bibfnamefont {K.}~\bibnamefont
  {Fujii}},\ }\href@noop {} {\bibfield  {journal} {\bibinfo  {journal}
  {Physical Review Research}\ }\textbf {\bibinfo {volume} {1}},\ \bibinfo
  {pages} {013006} (\bibinfo {year} {2019})}\BibitemShut {NoStop}%
\bibitem [{\citenamefont {McArdle}\ \emph {et~al.}(2019)\citenamefont
  {McArdle}, \citenamefont {Jones}, \citenamefont {Endo}, \citenamefont {Li},
  \citenamefont {Benjamin},\ and\ \citenamefont {Yuan}}]{mcardle2018img}%
  \BibitemOpen
  \bibfield  {author} {\bibinfo {author} {\bibfnamefont {S.}~\bibnamefont
  {McArdle}}, \bibinfo {author} {\bibfnamefont {T.}~\bibnamefont {Jones}},
  \bibinfo {author} {\bibfnamefont {S.}~\bibnamefont {Endo}}, \bibinfo {author}
  {\bibfnamefont {Y.}~\bibnamefont {Li}}, \bibinfo {author} {\bibfnamefont
  {S.~C.}\ \bibnamefont {Benjamin}}, \ and\ \bibinfo {author} {\bibfnamefont
  {X.}~\bibnamefont {Yuan}},\ }\href@noop {} {\bibfield  {journal} {\bibinfo
  {journal} {npj Quantum Inf.}\ }\textbf {\bibinfo {volume} {5}},\ \bibinfo
  {pages} {1} (\bibinfo {year} {2019})}\BibitemShut {NoStop}%
\bibitem [{\citenamefont {Yang}\ \emph {et~al.}(2017)\citenamefont {Yang},
  \citenamefont {Rahmani}, \citenamefont {Shabani}, \citenamefont {Neven},\
  and\ \citenamefont {Chamon}}]{yang2017optimizing}%
  \BibitemOpen
  \bibfield  {author} {\bibinfo {author} {\bibfnamefont {Z.-C.}\ \bibnamefont
  {Yang}}, \bibinfo {author} {\bibfnamefont {A.}~\bibnamefont {Rahmani}},
  \bibinfo {author} {\bibfnamefont {A.}~\bibnamefont {Shabani}}, \bibinfo
  {author} {\bibfnamefont {H.}~\bibnamefont {Neven}}, \ and\ \bibinfo {author}
  {\bibfnamefont {C.}~\bibnamefont {Chamon}},\ }\href@noop {} {\bibfield
  {journal} {\bibinfo  {journal} {Phys. Rev. X}\ }\textbf {\bibinfo {volume}
  {7}},\ \bibinfo {pages} {021027} (\bibinfo {year} {2017})}\BibitemShut
  {NoStop}%
\bibitem [{\citenamefont {Zhu}\ \emph {et~al.}(2018)\citenamefont {Zhu},
  \citenamefont {Linke}, \citenamefont {Benedetti}, \citenamefont {Landsman},
  \citenamefont {Nguyen}, \citenamefont {Alderete}, \citenamefont
  {Perdomo-Ortiz}, \citenamefont {Korda}, \citenamefont {Garfoot},
  \citenamefont {Brecque} \emph {et~al.}}]{zhu2018training}%
  \BibitemOpen
  \bibfield  {author} {\bibinfo {author} {\bibfnamefont {D.}~\bibnamefont
  {Zhu}}, \bibinfo {author} {\bibfnamefont {N.}~\bibnamefont {Linke}}, \bibinfo
  {author} {\bibfnamefont {M.}~\bibnamefont {Benedetti}}, \bibinfo {author}
  {\bibfnamefont {K.}~\bibnamefont {Landsman}}, \bibinfo {author}
  {\bibfnamefont {N.}~\bibnamefont {Nguyen}}, \bibinfo {author} {\bibfnamefont
  {C.}~\bibnamefont {Alderete}}, \bibinfo {author} {\bibfnamefont
  {A.}~\bibnamefont {Perdomo-Ortiz}}, \bibinfo {author} {\bibfnamefont
  {N.}~\bibnamefont {Korda}}, \bibinfo {author} {\bibfnamefont
  {A.}~\bibnamefont {Garfoot}}, \bibinfo {author} {\bibfnamefont
  {C.}~\bibnamefont {Brecque}},  \emph {et~al.},\ }\href@noop {} {\bibfield
  {journal} {\bibinfo  {journal} {arXiv preprint arXiv:1812.08862}\ } (\bibinfo
  {year} {2018})}\BibitemShut {NoStop}%
\bibitem [{\citenamefont {Shaydulin}\ \emph {et~al.}(2019)\citenamefont
  {Shaydulin}, \citenamefont {Safro},\ and\ \citenamefont
  {Larson}}]{shaydulin2019multistart}%
  \BibitemOpen
  \bibfield  {author} {\bibinfo {author} {\bibfnamefont {R.}~\bibnamefont
  {Shaydulin}}, \bibinfo {author} {\bibfnamefont {I.}~\bibnamefont {Safro}}, \
  and\ \bibinfo {author} {\bibfnamefont {J.}~\bibnamefont {Larson}},\
  }\href@noop {} {\bibfield  {journal} {\bibinfo  {journal} {arXiv preprint
  arXiv:1905.08768}\ } (\bibinfo {year} {2019})}\BibitemShut {NoStop}%
\bibitem [{\citenamefont {Guerreschi}\ and\ \citenamefont
  {Smelyanskiy}(2017)}]{guerreschi2017practical}%
  \BibitemOpen
  \bibfield  {author} {\bibinfo {author} {\bibfnamefont {G.~G.}\ \bibnamefont
  {Guerreschi}}\ and\ \bibinfo {author} {\bibfnamefont {M.}~\bibnamefont
  {Smelyanskiy}},\ }\href@noop {} {\bibfield  {journal} {\bibinfo  {journal}
  {arXiv preprint arXiv:1701.01450}\ } (\bibinfo {year} {2017})}\BibitemShut
  {NoStop}%
\bibitem [{\citenamefont {Nakanishi}\ \emph {et~al.}(2019)\citenamefont
  {Nakanishi}, \citenamefont {Fujii},\ and\ \citenamefont
  {Todo}}]{nakanishi2019sequential}%
  \BibitemOpen
  \bibfield  {author} {\bibinfo {author} {\bibfnamefont {K.~M.}\ \bibnamefont
  {Nakanishi}}, \bibinfo {author} {\bibfnamefont {K.}~\bibnamefont {Fujii}}, \
  and\ \bibinfo {author} {\bibfnamefont {S.}~\bibnamefont {Todo}},\ }\href@noop
  {} {\bibfield  {journal} {\bibinfo  {journal} {arXiv preprint
  arXiv:1903.12166}\ } (\bibinfo {year} {2019})}\BibitemShut {NoStop}%
\bibitem [{\citenamefont {Parrish}\ \emph
  {et~al.}(2019{\natexlab{a}})\citenamefont {Parrish}, \citenamefont {Iosue},
  \citenamefont {Ozaeta},\ and\ \citenamefont {McMahon}}]{parrish2019jacobi}%
  \BibitemOpen
  \bibfield  {author} {\bibinfo {author} {\bibfnamefont {R.~M.}\ \bibnamefont
  {Parrish}}, \bibinfo {author} {\bibfnamefont {J.~T.}\ \bibnamefont {Iosue}},
  \bibinfo {author} {\bibfnamefont {A.}~\bibnamefont {Ozaeta}}, \ and\ \bibinfo
  {author} {\bibfnamefont {P.~L.}\ \bibnamefont {McMahon}},\ }\href@noop {}
  {\bibfield  {journal} {\bibinfo  {journal} {arXiv preprint arXiv:1904.03206}\
  } (\bibinfo {year} {2019}{\natexlab{a}})}\BibitemShut {NoStop}%
\bibitem [{\citenamefont {Parrish}\ \emph
  {et~al.}(2019{\natexlab{b}})\citenamefont {Parrish}, \citenamefont
  {Hohenstein}, \citenamefont {McMahon},\ and\ \citenamefont
  {Martinez}}]{parrish2019hybrid}%
  \BibitemOpen
  \bibfield  {author} {\bibinfo {author} {\bibfnamefont {R.~M.}\ \bibnamefont
  {Parrish}}, \bibinfo {author} {\bibfnamefont {E.~G.}\ \bibnamefont
  {Hohenstein}}, \bibinfo {author} {\bibfnamefont {P.~L.}\ \bibnamefont
  {McMahon}}, \ and\ \bibinfo {author} {\bibfnamefont {T.~J.}\ \bibnamefont
  {Martinez}},\ }\href@noop {} {\bibfield  {journal} {\bibinfo  {journal}
  {arXiv preprint arXiv:1906.08728}\ } (\bibinfo {year}
  {2019}{\natexlab{b}})}\BibitemShut {NoStop}%
\bibitem [{\citenamefont {Schuld}\ \emph {et~al.}(2019)\citenamefont {Schuld},
  \citenamefont {Bergholm}, \citenamefont {Gogolin}, \citenamefont {Izaac},\
  and\ \citenamefont {Killoran}}]{schuld2019evaluating}%
  \BibitemOpen
  \bibfield  {author} {\bibinfo {author} {\bibfnamefont {M.}~\bibnamefont
  {Schuld}}, \bibinfo {author} {\bibfnamefont {V.}~\bibnamefont {Bergholm}},
  \bibinfo {author} {\bibfnamefont {C.}~\bibnamefont {Gogolin}}, \bibinfo
  {author} {\bibfnamefont {J.}~\bibnamefont {Izaac}}, \ and\ \bibinfo {author}
  {\bibfnamefont {N.}~\bibnamefont {Killoran}},\ }\href@noop {} {\bibfield
  {journal} {\bibinfo  {journal} {Phys. Rev. A}\ }\textbf {\bibinfo {volume}
  {99}},\ \bibinfo {pages} {032331} (\bibinfo {year} {2019})}\BibitemShut
  {NoStop}%
\bibitem [{\citenamefont {Moseley}\ \emph {et~al.}()\citenamefont {Moseley},
  \citenamefont {Osborne},\ and\ \citenamefont {Benjamin}}]{moseley3bayesian}%
  \BibitemOpen
  \bibfield  {author} {\bibinfo {author} {\bibfnamefont {B.}~\bibnamefont
  {Moseley}}, \bibinfo {author} {\bibfnamefont {M.}~\bibnamefont {Osborne}}, \
  and\ \bibinfo {author} {\bibfnamefont {S.}~\bibnamefont {Benjamin}},\
  }\href@noop {} {\bibfield  {journal} {\bibinfo  {journal} {quantum}\ }\textbf
  {\bibinfo {volume} {3}},\ \bibinfo {pages} {4}}\BibitemShut {NoStop}%
\bibitem [{\citenamefont {Sarma}\ \emph {et~al.}(2019)\citenamefont {Sarma},
  \citenamefont {Deng},\ and\ \citenamefont {Duan}}]{sarma2019machine}%
  \BibitemOpen
  \bibfield  {author} {\bibinfo {author} {\bibfnamefont {S.~D.}\ \bibnamefont
  {Sarma}}, \bibinfo {author} {\bibfnamefont {D.-L.}\ \bibnamefont {Deng}}, \
  and\ \bibinfo {author} {\bibfnamefont {L.-M.}\ \bibnamefont {Duan}},\
  }\href@noop {} {\bibfield  {journal} {\bibinfo  {journal} {Physics Today}\
  }\textbf {\bibinfo {volume} {72}},\ \bibinfo {pages} {48} (\bibinfo {year}
  {2019})}\BibitemShut {NoStop}%
\bibitem [{\citenamefont {Melko}\ \emph {et~al.}(2019)\citenamefont {Melko},
  \citenamefont {Carleo}, \citenamefont {Carrasquilla},\ and\ \citenamefont
  {Cirac}}]{melko2019restricted}%
  \BibitemOpen
  \bibfield  {author} {\bibinfo {author} {\bibfnamefont {R.~G.}\ \bibnamefont
  {Melko}}, \bibinfo {author} {\bibfnamefont {G.}~\bibnamefont {Carleo}},
  \bibinfo {author} {\bibfnamefont {J.}~\bibnamefont {Carrasquilla}}, \ and\
  \bibinfo {author} {\bibfnamefont {J.~I.}\ \bibnamefont {Cirac}},\ }\href@noop
  {} {\bibfield  {journal} {\bibinfo  {journal} {Nat. Phys.}\ ,\ \bibinfo
  {pages} {1}} (\bibinfo {year} {2019})}\BibitemShut {NoStop}%
\bibitem [{\citenamefont {Jia}\ \emph {et~al.}(2019{\natexlab{a}})\citenamefont
  {Jia}, \citenamefont {Yi}, \citenamefont {Zhai}, \citenamefont {Wu},
  \citenamefont {Guo},\ and\ \citenamefont {Guo}}]{jia2019quantum}%
  \BibitemOpen
  \bibfield  {author} {\bibinfo {author} {\bibfnamefont {Z.-A.}\ \bibnamefont
  {Jia}}, \bibinfo {author} {\bibfnamefont {B.}~\bibnamefont {Yi}}, \bibinfo
  {author} {\bibfnamefont {R.}~\bibnamefont {Zhai}}, \bibinfo {author}
  {\bibfnamefont {Y.-C.}\ \bibnamefont {Wu}}, \bibinfo {author} {\bibfnamefont
  {G.-C.}\ \bibnamefont {Guo}}, \ and\ \bibinfo {author} {\bibfnamefont
  {G.-P.}\ \bibnamefont {Guo}},\ }\href@noop {} {\bibfield  {journal} {\bibinfo
   {journal} {Adv. Quantum Technol.}\ } (\bibinfo {year}
  {2019}{\natexlab{a}})}\BibitemShut {NoStop}%
\bibitem [{\citenamefont {Torlai}\ and\ \citenamefont
  {Melko}(2016)}]{torlai2016learning}%
  \BibitemOpen
  \bibfield  {author} {\bibinfo {author} {\bibfnamefont {G.}~\bibnamefont
  {Torlai}}\ and\ \bibinfo {author} {\bibfnamefont {R.~G.}\ \bibnamefont
  {Melko}},\ }\href@noop {} {\bibfield  {journal} {\bibinfo  {journal} {Phys.
  Rev. B}\ }\textbf {\bibinfo {volume} {94}},\ \bibinfo {pages} {165134}
  (\bibinfo {year} {2016})}\BibitemShut {NoStop}%
\bibitem [{\citenamefont {Carrasquilla}\ and\ \citenamefont
  {Melko}(2017)}]{carrasquilla2017machine}%
  \BibitemOpen
  \bibfield  {author} {\bibinfo {author} {\bibfnamefont {J.}~\bibnamefont
  {Carrasquilla}}\ and\ \bibinfo {author} {\bibfnamefont {R.~G.}\ \bibnamefont
  {Melko}},\ }\href@noop {} {\bibfield  {journal} {\bibinfo  {journal} {Nat.
  Phys}\ }\textbf {\bibinfo {volume} {13}},\ \bibinfo {pages} {431} (\bibinfo
  {year} {2017})}\BibitemShut {NoStop}%
\bibitem [{\citenamefont {Kaubruegger}\ \emph {et~al.}(2018)\citenamefont
  {Kaubruegger}, \citenamefont {Pastori},\ and\ \citenamefont
  {Budich}}]{kaubruegger2018chiral}%
  \BibitemOpen
  \bibfield  {author} {\bibinfo {author} {\bibfnamefont {R.}~\bibnamefont
  {Kaubruegger}}, \bibinfo {author} {\bibfnamefont {L.}~\bibnamefont
  {Pastori}}, \ and\ \bibinfo {author} {\bibfnamefont {J.~C.}\ \bibnamefont
  {Budich}},\ }\href@noop {} {\bibfield  {journal} {\bibinfo  {journal} {Phys.
  Rev. B}\ }\textbf {\bibinfo {volume} {97}},\ \bibinfo {pages} {195136}
  (\bibinfo {year} {2018})}\BibitemShut {NoStop}%
\bibitem [{\citenamefont {Koch-Janusz}\ and\ \citenamefont
  {Ringel}(2018)}]{koch2018mutual}%
  \BibitemOpen
  \bibfield  {author} {\bibinfo {author} {\bibfnamefont {M.}~\bibnamefont
  {Koch-Janusz}}\ and\ \bibinfo {author} {\bibfnamefont {Z.}~\bibnamefont
  {Ringel}},\ }\href@noop {} {\bibfield  {journal} {\bibinfo  {journal} {Nat.
  Phys.}\ }\textbf {\bibinfo {volume} {14}},\ \bibinfo {pages} {578} (\bibinfo
  {year} {2018})}\BibitemShut {NoStop}%
\bibitem [{\citenamefont {Czischek}\ \emph {et~al.}(2018)\citenamefont
  {Czischek}, \citenamefont {G{\"a}rttner},\ and\ \citenamefont
  {Gasenzer}}]{czischek2018quenches}%
  \BibitemOpen
  \bibfield  {author} {\bibinfo {author} {\bibfnamefont {S.}~\bibnamefont
  {Czischek}}, \bibinfo {author} {\bibfnamefont {M.}~\bibnamefont
  {G{\"a}rttner}}, \ and\ \bibinfo {author} {\bibfnamefont {T.}~\bibnamefont
  {Gasenzer}},\ }\href@noop {} {\bibfield  {journal} {\bibinfo  {journal}
  {Phys. Rev. B}\ }\textbf {\bibinfo {volume} {98}},\ \bibinfo {pages} {024311}
  (\bibinfo {year} {2018})}\BibitemShut {NoStop}%
\bibitem [{\citenamefont {Lu}\ \emph {et~al.}(2019)\citenamefont {Lu},
  \citenamefont {Gao},\ and\ \citenamefont {Duan}}]{lu2019efficient}%
  \BibitemOpen
  \bibfield  {author} {\bibinfo {author} {\bibfnamefont {S.}~\bibnamefont
  {Lu}}, \bibinfo {author} {\bibfnamefont {X.}~\bibnamefont {Gao}}, \ and\
  \bibinfo {author} {\bibfnamefont {L.-M.}\ \bibnamefont {Duan}},\ }\href@noop
  {} {\bibfield  {journal} {\bibinfo  {journal} {Phys. Rev. B}\ }\textbf
  {\bibinfo {volume} {99}},\ \bibinfo {pages} {155136} (\bibinfo {year}
  {2019})}\BibitemShut {NoStop}%
\bibitem [{\citenamefont {Xu}\ and\ \citenamefont {Xu}(2018)}]{xu2018neural}%
  \BibitemOpen
  \bibfield  {author} {\bibinfo {author} {\bibfnamefont {Q.}~\bibnamefont
  {Xu}}\ and\ \bibinfo {author} {\bibfnamefont {S.}~\bibnamefont {Xu}},\
  }\href@noop {} {\bibfield  {journal} {\bibinfo  {journal} {arXiv preprint
  arXiv:1811.06654}\ } (\bibinfo {year} {2018})}\BibitemShut {NoStop}%
\bibitem [{\citenamefont {Torlai}\ \emph {et~al.}(2019)\citenamefont {Torlai},
  \citenamefont {Timar}, \citenamefont {van Nieuwenburg}, \citenamefont
  {Levine}, \citenamefont {Omran}, \citenamefont {Keesling}, \citenamefont
  {Bernien}, \citenamefont {Greiner}, \citenamefont {Vuleti{\'c}},
  \citenamefont {Lukin} \emph {et~al.}}]{torlai2019integrating}%
  \BibitemOpen
  \bibfield  {author} {\bibinfo {author} {\bibfnamefont {G.}~\bibnamefont
  {Torlai}}, \bibinfo {author} {\bibfnamefont {B.}~\bibnamefont {Timar}},
  \bibinfo {author} {\bibfnamefont {E.~P.}\ \bibnamefont {van Nieuwenburg}},
  \bibinfo {author} {\bibfnamefont {H.}~\bibnamefont {Levine}}, \bibinfo
  {author} {\bibfnamefont {A.}~\bibnamefont {Omran}}, \bibinfo {author}
  {\bibfnamefont {A.}~\bibnamefont {Keesling}}, \bibinfo {author}
  {\bibfnamefont {H.}~\bibnamefont {Bernien}}, \bibinfo {author} {\bibfnamefont
  {M.}~\bibnamefont {Greiner}}, \bibinfo {author} {\bibfnamefont
  {V.}~\bibnamefont {Vuleti{\'c}}}, \bibinfo {author} {\bibfnamefont {M.~D.}\
  \bibnamefont {Lukin}},  \emph {et~al.},\ }\href@noop {} {\bibfield  {journal}
  {\bibinfo  {journal} {arXiv preprint arXiv:1904.08441}\ } (\bibinfo {year}
  {2019})}\BibitemShut {NoStop}%
\bibitem [{\citenamefont {Huang}\ \emph {et~al.}(2017)\citenamefont {Huang},
  \citenamefont {Yang},\ and\ \citenamefont {Wang}}]{huang2017recommender}%
  \BibitemOpen
  \bibfield  {author} {\bibinfo {author} {\bibfnamefont {L.}~\bibnamefont
  {Huang}}, \bibinfo {author} {\bibfnamefont {Y.-f.}\ \bibnamefont {Yang}}, \
  and\ \bibinfo {author} {\bibfnamefont {L.}~\bibnamefont {Wang}},\ }\href@noop
  {} {\bibfield  {journal} {\bibinfo  {journal} {Phys. Rev. E}\ }\textbf
  {\bibinfo {volume} {95}},\ \bibinfo {pages} {031301} (\bibinfo {year}
  {2017})}\BibitemShut {NoStop}%
\bibitem [{\citenamefont {Wang}(2017)}]{wang2017exploring}%
  \BibitemOpen
  \bibfield  {author} {\bibinfo {author} {\bibfnamefont {L.}~\bibnamefont
  {Wang}},\ }\href@noop {} {\bibfield  {journal} {\bibinfo  {journal} {Phys.
  Rev. E}\ }\textbf {\bibinfo {volume} {96}},\ \bibinfo {pages} {051301}
  (\bibinfo {year} {2017})}\BibitemShut {NoStop}%
\bibitem [{\citenamefont {Inack}\ \emph {et~al.}(2018)\citenamefont {Inack},
  \citenamefont {Santoro}, \citenamefont {Dell'Anna},\ and\ \citenamefont
  {Pilati}}]{inack2018projective}%
  \BibitemOpen
  \bibfield  {author} {\bibinfo {author} {\bibfnamefont {E.}~\bibnamefont
  {Inack}}, \bibinfo {author} {\bibfnamefont {G.}~\bibnamefont {Santoro}},
  \bibinfo {author} {\bibfnamefont {L.}~\bibnamefont {Dell'Anna}}, \ and\
  \bibinfo {author} {\bibfnamefont {S.}~\bibnamefont {Pilati}},\ }\href@noop {}
  {\bibfield  {journal} {\bibinfo  {journal} {Phys. Rev. B}\ }\textbf {\bibinfo
  {volume} {98}},\ \bibinfo {pages} {235145} (\bibinfo {year}
  {2018})}\BibitemShut {NoStop}%
\bibitem [{\citenamefont {Torlai}\ and\ \citenamefont
  {Melko}(2017)}]{torlai2017neural}%
  \BibitemOpen
  \bibfield  {author} {\bibinfo {author} {\bibfnamefont {G.}~\bibnamefont
  {Torlai}}\ and\ \bibinfo {author} {\bibfnamefont {R.~G.}\ \bibnamefont
  {Melko}},\ }\href@noop {} {\bibfield  {journal} {\bibinfo  {journal} {Phys.
  Rev. Lett.}\ }\textbf {\bibinfo {volume} {119}},\ \bibinfo {pages} {030501}
  (\bibinfo {year} {2017})}\BibitemShut {NoStop}%
\bibitem [{\citenamefont {Bausch}\ and\ \citenamefont
  {Leditzky}(2018)}]{bausch2018quantum}%
  \BibitemOpen
  \bibfield  {author} {\bibinfo {author} {\bibfnamefont {J.}~\bibnamefont
  {Bausch}}\ and\ \bibinfo {author} {\bibfnamefont {F.}~\bibnamefont
  {Leditzky}},\ }\href@noop {} {\bibfield  {journal} {\bibinfo  {journal}
  {arXiv preprint arXiv:1806.08781}\ } (\bibinfo {year} {2018})}\BibitemShut
  {NoStop}%
\bibitem [{\citenamefont {Zhang}\ \emph {et~al.}(2018)\citenamefont {Zhang},
  \citenamefont {Jia}, \citenamefont {Wu},\ and\ \citenamefont
  {Guo}}]{zhang2018efficient}%
  \BibitemOpen
  \bibfield  {author} {\bibinfo {author} {\bibfnamefont {Y.-H.}\ \bibnamefont
  {Zhang}}, \bibinfo {author} {\bibfnamefont {Z.-A.}\ \bibnamefont {Jia}},
  \bibinfo {author} {\bibfnamefont {Y.-C.}\ \bibnamefont {Wu}}, \ and\ \bibinfo
  {author} {\bibfnamefont {G.-C.}\ \bibnamefont {Guo}},\ }\href@noop {}
  {\bibfield  {journal} {\bibinfo  {journal} {arXiv preprint arXiv:1809.08631}\
  } (\bibinfo {year} {2018})}\BibitemShut {NoStop}%
\bibitem [{\citenamefont {Jia}\ \emph {et~al.}(2019{\natexlab{b}})\citenamefont
  {Jia}, \citenamefont {Zhang}, \citenamefont {Wu}, \citenamefont {Kong},
  \citenamefont {Guo},\ and\ \citenamefont {Guo}}]{jia2019efficient}%
  \BibitemOpen
  \bibfield  {author} {\bibinfo {author} {\bibfnamefont {Z.-A.}\ \bibnamefont
  {Jia}}, \bibinfo {author} {\bibfnamefont {Y.-H.}\ \bibnamefont {Zhang}},
  \bibinfo {author} {\bibfnamefont {Y.-C.}\ \bibnamefont {Wu}}, \bibinfo
  {author} {\bibfnamefont {L.}~\bibnamefont {Kong}}, \bibinfo {author}
  {\bibfnamefont {G.-C.}\ \bibnamefont {Guo}}, \ and\ \bibinfo {author}
  {\bibfnamefont {G.-P.}\ \bibnamefont {Guo}},\ }\href@noop {} {\bibfield
  {journal} {\bibinfo  {journal} {Phys. Rev. A}\ }\textbf {\bibinfo {volume}
  {99}},\ \bibinfo {pages} {012307} (\bibinfo {year}
  {2019}{\natexlab{b}})}\BibitemShut {NoStop}%
\bibitem [{\citenamefont {Carleo}\ and\ \citenamefont
  {Troyer}(2017)}]{carleo2017solving}%
  \BibitemOpen
  \bibfield  {author} {\bibinfo {author} {\bibfnamefont {G.}~\bibnamefont
  {Carleo}}\ and\ \bibinfo {author} {\bibfnamefont {M.}~\bibnamefont
  {Troyer}},\ }\href@noop {} {\bibfield  {journal} {\bibinfo  {journal}
  {Science}\ }\textbf {\bibinfo {volume} {355}},\ \bibinfo {pages} {602}
  (\bibinfo {year} {2017})}\BibitemShut {NoStop}%
\bibitem [{\citenamefont {Deng}\ \emph
  {et~al.}(2017{\natexlab{a}})\citenamefont {Deng}, \citenamefont {Li},\ and\
  \citenamefont {Sarma}}]{deng2017machine}%
  \BibitemOpen
  \bibfield  {author} {\bibinfo {author} {\bibfnamefont {D.-L.}\ \bibnamefont
  {Deng}}, \bibinfo {author} {\bibfnamefont {X.}~\bibnamefont {Li}}, \ and\
  \bibinfo {author} {\bibfnamefont {S.~D.}\ \bibnamefont {Sarma}},\ }\href@noop
  {} {\bibfield  {journal} {\bibinfo  {journal} {Phys. Rev. B}\ }\textbf
  {\bibinfo {volume} {96}},\ \bibinfo {pages} {195145} (\bibinfo {year}
  {2017}{\natexlab{a}})}\BibitemShut {NoStop}%
\bibitem [{\citenamefont {Saito}(2018)}]{saito2018method}%
  \BibitemOpen
  \bibfield  {author} {\bibinfo {author} {\bibfnamefont {H.}~\bibnamefont
  {Saito}},\ }\href@noop {} {\bibfield  {journal} {\bibinfo  {journal} {J.
  Phys. Soc. Jpn.}\ }\textbf {\bibinfo {volume} {87}},\ \bibinfo {pages}
  {074002} (\bibinfo {year} {2018})}\BibitemShut {NoStop}%
\bibitem [{\citenamefont {Hartmann}\ and\ \citenamefont
  {Carleo}(2019)}]{hartmann2019neural}%
  \BibitemOpen
  \bibfield  {author} {\bibinfo {author} {\bibfnamefont {M.~J.}\ \bibnamefont
  {Hartmann}}\ and\ \bibinfo {author} {\bibfnamefont {G.}~\bibnamefont
  {Carleo}},\ }\href@noop {} {\bibfield  {journal} {\bibinfo  {journal} {Phys.
  Rev. Lett.}\ }\textbf {\bibinfo {volume} {122}},\ \bibinfo {pages} {250502}
  (\bibinfo {year} {2019})}\BibitemShut {NoStop}%
\bibitem [{\citenamefont {Nagy}\ and\ \citenamefont
  {Savona}(2019)}]{nagy2019variational}%
  \BibitemOpen
  \bibfield  {author} {\bibinfo {author} {\bibfnamefont {A.}~\bibnamefont
  {Nagy}}\ and\ \bibinfo {author} {\bibfnamefont {V.}~\bibnamefont {Savona}},\
  }\href@noop {} {\bibfield  {journal} {\bibinfo  {journal} {Phys. Rev. Lett.}\
  }\textbf {\bibinfo {volume} {122}},\ \bibinfo {pages} {250501} (\bibinfo
  {year} {2019})}\BibitemShut {NoStop}%
\bibitem [{\citenamefont {Yoshioka}\ and\ \citenamefont
  {Hamazaki}(2019)}]{yoshioka2019constructing}%
  \BibitemOpen
  \bibfield  {author} {\bibinfo {author} {\bibfnamefont {N.}~\bibnamefont
  {Yoshioka}}\ and\ \bibinfo {author} {\bibfnamefont {R.}~\bibnamefont
  {Hamazaki}},\ }\href@noop {} {\bibfield  {journal} {\bibinfo  {journal}
  {Phys. Rev. B}\ }\textbf {\bibinfo {volume} {99}},\ \bibinfo {pages} {214306}
  (\bibinfo {year} {2019})}\BibitemShut {NoStop}%
\bibitem [{\citenamefont {Luo}\ and\ \citenamefont
  {Clark}(2019)}]{luo2019backflow}%
  \BibitemOpen
  \bibfield  {author} {\bibinfo {author} {\bibfnamefont {D.}~\bibnamefont
  {Luo}}\ and\ \bibinfo {author} {\bibfnamefont {B.~K.}\ \bibnamefont
  {Clark}},\ }\href@noop {} {\bibfield  {journal} {\bibinfo  {journal} {Phys.
  Rev. Lett.}\ }\textbf {\bibinfo {volume} {122}},\ \bibinfo {pages} {226401}
  (\bibinfo {year} {2019})}\BibitemShut {NoStop}%
\bibitem [{\citenamefont {Torlai}\ and\ \citenamefont
  {Melko}(2018)}]{torlai2018latent}%
  \BibitemOpen
  \bibfield  {author} {\bibinfo {author} {\bibfnamefont {G.}~\bibnamefont
  {Torlai}}\ and\ \bibinfo {author} {\bibfnamefont {R.~G.}\ \bibnamefont
  {Melko}},\ }\href@noop {} {\bibfield  {journal} {\bibinfo  {journal} {Phys.
  Rev. Lett.}\ }\textbf {\bibinfo {volume} {120}},\ \bibinfo {pages} {240503}
  (\bibinfo {year} {2018})}\BibitemShut {NoStop}%
\bibitem [{\citenamefont {Nomura}\ \emph {et~al.}(2017)\citenamefont {Nomura},
  \citenamefont {Darmawan}, \citenamefont {Yamaji},\ and\ \citenamefont
  {Imada}}]{nomura2017restricted}%
  \BibitemOpen
  \bibfield  {author} {\bibinfo {author} {\bibfnamefont {Y.}~\bibnamefont
  {Nomura}}, \bibinfo {author} {\bibfnamefont {A.~S.}\ \bibnamefont
  {Darmawan}}, \bibinfo {author} {\bibfnamefont {Y.}~\bibnamefont {Yamaji}}, \
  and\ \bibinfo {author} {\bibfnamefont {M.}~\bibnamefont {Imada}},\
  }\href@noop {} {\bibfield  {journal} {\bibinfo  {journal} {Phys. Rev. B}\
  }\textbf {\bibinfo {volume} {96}},\ \bibinfo {pages} {205152} (\bibinfo
  {year} {2017})}\BibitemShut {NoStop}%
\bibitem [{\citenamefont {J{\'o}nsson}\ \emph {et~al.}(2018)\citenamefont
  {J{\'o}nsson}, \citenamefont {Bauer},\ and\ \citenamefont
  {Carleo}}]{jonsson2018neural}%
  \BibitemOpen
  \bibfield  {author} {\bibinfo {author} {\bibfnamefont {B.}~\bibnamefont
  {J{\'o}nsson}}, \bibinfo {author} {\bibfnamefont {B.}~\bibnamefont {Bauer}},
  \ and\ \bibinfo {author} {\bibfnamefont {G.}~\bibnamefont {Carleo}},\
  }\href@noop {} {\bibfield  {journal} {\bibinfo  {journal} {arXiv preprint
  arXiv:1808.05232}\ } (\bibinfo {year} {2018})}\BibitemShut {NoStop}%
\bibitem [{\citenamefont {Glasser}\ \emph {et~al.}(2018)\citenamefont
  {Glasser}, \citenamefont {Pancotti}, \citenamefont {August}, \citenamefont
  {Rodriguez},\ and\ \citenamefont {Cirac}}]{glasser2018neural}%
  \BibitemOpen
  \bibfield  {author} {\bibinfo {author} {\bibfnamefont {I.}~\bibnamefont
  {Glasser}}, \bibinfo {author} {\bibfnamefont {N.}~\bibnamefont {Pancotti}},
  \bibinfo {author} {\bibfnamefont {M.}~\bibnamefont {August}}, \bibinfo
  {author} {\bibfnamefont {I.~D.}\ \bibnamefont {Rodriguez}}, \ and\ \bibinfo
  {author} {\bibfnamefont {J.~I.}\ \bibnamefont {Cirac}},\ }\href@noop {}
  {\bibfield  {journal} {\bibinfo  {journal} {Phys. Rev. X}\ }\textbf {\bibinfo
  {volume} {8}},\ \bibinfo {pages} {011006} (\bibinfo {year}
  {2018})}\BibitemShut {NoStop}%
\bibitem [{\citenamefont {Deng}\ \emph
  {et~al.}(2017{\natexlab{b}})\citenamefont {Deng}, \citenamefont {Li},\ and\
  \citenamefont {Sarma}}]{deng2017quantum}%
  \BibitemOpen
  \bibfield  {author} {\bibinfo {author} {\bibfnamefont {D.-L.}\ \bibnamefont
  {Deng}}, \bibinfo {author} {\bibfnamefont {X.}~\bibnamefont {Li}}, \ and\
  \bibinfo {author} {\bibfnamefont {S.~D.}\ \bibnamefont {Sarma}},\ }\href@noop
  {} {\bibfield  {journal} {\bibinfo  {journal} {Phys. Rev. X}\ }\textbf
  {\bibinfo {volume} {7}},\ \bibinfo {pages} {021021} (\bibinfo {year}
  {2017}{\natexlab{b}})}\BibitemShut {NoStop}%
\bibitem [{\citenamefont {Huang}\ and\ \citenamefont
  {Moore}(2017)}]{huang2017neural}%
  \BibitemOpen
  \bibfield  {author} {\bibinfo {author} {\bibfnamefont {Y.}~\bibnamefont
  {Huang}}\ and\ \bibinfo {author} {\bibfnamefont {J.~E.}\ \bibnamefont
  {Moore}},\ }\href@noop {} {\bibfield  {journal} {\bibinfo  {journal} {arXiv
  preprint arXiv:1701.06246}\ } (\bibinfo {year} {2017})}\BibitemShut {NoStop}%
\bibitem [{\citenamefont {Gao}\ and\ \citenamefont
  {Duan}(2017{\natexlab{a}})}]{Gao2017}%
  \BibitemOpen
  \bibfield  {author} {\bibinfo {author} {\bibfnamefont {X.}~\bibnamefont
  {Gao}}\ and\ \bibinfo {author} {\bibfnamefont {L.-M.}\ \bibnamefont {Duan}},\
  }\href {https://doi.org/10.1038/s41467-017-00705-2} {\bibfield  {journal}
  {\bibinfo  {journal} {Nat. Commun.}\ }\textbf {\bibinfo {volume} {8}},\
  \bibinfo {pages} {662} (\bibinfo {year} {2017}{\natexlab{a}})}\BibitemShut
  {NoStop}%
\bibitem [{\citenamefont {Clark}(2018)}]{clark2018unifying}%
  \BibitemOpen
  \bibfield  {author} {\bibinfo {author} {\bibfnamefont {S.~R.}\ \bibnamefont
  {Clark}},\ }\href@noop {} {\bibfield  {journal} {\bibinfo  {journal} {J.
  Phys. A Math. Theor.}\ }\textbf {\bibinfo {volume} {51}},\ \bibinfo {pages}
  {135301} (\bibinfo {year} {2018})}\BibitemShut {NoStop}%
\bibitem [{\citenamefont {Chen}\ \emph {et~al.}(2018)\citenamefont {Chen},
  \citenamefont {Cheng}, \citenamefont {Xie}, \citenamefont {Wang},\ and\
  \citenamefont {Xiang}}]{chen2018equivalence}%
  \BibitemOpen
  \bibfield  {author} {\bibinfo {author} {\bibfnamefont {J.}~\bibnamefont
  {Chen}}, \bibinfo {author} {\bibfnamefont {S.}~\bibnamefont {Cheng}},
  \bibinfo {author} {\bibfnamefont {H.}~\bibnamefont {Xie}}, \bibinfo {author}
  {\bibfnamefont {L.}~\bibnamefont {Wang}}, \ and\ \bibinfo {author}
  {\bibfnamefont {T.}~\bibnamefont {Xiang}},\ }\href@noop {} {\bibfield
  {journal} {\bibinfo  {journal} {Phys. Rev. B}\ }\textbf {\bibinfo {volume}
  {97}},\ \bibinfo {pages} {085104} (\bibinfo {year} {2018})}\BibitemShut
  {NoStop}%
\bibitem [{\citenamefont {Xia}\ and\ \citenamefont {Kais}(2018)}]{xia2018rbm}%
  \BibitemOpen
  \bibfield  {author} {\bibinfo {author} {\bibfnamefont {R.}~\bibnamefont
  {Xia}}\ and\ \bibinfo {author} {\bibfnamefont {S.}~\bibnamefont {Kais}},\
  }\href@noop {} {\bibfield  {journal} {\bibinfo  {journal} {Nat. Commun.}\
  }\textbf {\bibinfo {volume} {9}},\ \bibinfo {pages} {4195} (\bibinfo {year}
  {2018})}\BibitemShut {NoStop}%
\bibitem [{\citenamefont {Gardas}\ \emph {et~al.}(2018)\citenamefont {Gardas},
  \citenamefont {Rams},\ and\ \citenamefont {Dziarmaga}}]{gardas2018dwave}%
  \BibitemOpen
  \bibfield  {author} {\bibinfo {author} {\bibfnamefont {B.}~\bibnamefont
  {Gardas}}, \bibinfo {author} {\bibfnamefont {M.~M.}\ \bibnamefont {Rams}}, \
  and\ \bibinfo {author} {\bibfnamefont {J.}~\bibnamefont {Dziarmaga}},\
  }\href@noop {} {\bibfield  {journal} {\bibinfo  {journal} {Phys. Rev. B}\
  }\textbf {\bibinfo {volume} {98}},\ \bibinfo {pages} {184304} (\bibinfo
  {year} {2018})}\BibitemShut {NoStop}%
\bibitem [{\citenamefont {Liu}\ \emph {et~al.}(2019)\citenamefont {Liu},
  \citenamefont {Zhang}, \citenamefont {Wan},\ and\ \citenamefont
  {Wang}}]{liu2019recycle}%
  \BibitemOpen
  \bibfield  {author} {\bibinfo {author} {\bibfnamefont {J.-G.}\ \bibnamefont
  {Liu}}, \bibinfo {author} {\bibfnamefont {Y.-H.}\ \bibnamefont {Zhang}},
  \bibinfo {author} {\bibfnamefont {Y.}~\bibnamefont {Wan}}, \ and\ \bibinfo
  {author} {\bibfnamefont {L.}~\bibnamefont {Wang}},\ }\href@noop {} {\bibfield
   {journal} {\bibinfo  {journal} {Phys. Rev. Research}\ }\textbf {\bibinfo
  {volume} {1}},\ \bibinfo {pages} {023025} (\bibinfo {year}
  {2019})}\BibitemShut {NoStop}%
\bibitem [{\citenamefont {Huggins}\ \emph
  {et~al.}(2019{\natexlab{b}})\citenamefont {Huggins}, \citenamefont {Patil},
  \citenamefont {Mitchell}, \citenamefont {Whaley},\ and\ \citenamefont
  {Stoudenmire}}]{Huggins_2019}%
  \BibitemOpen
  \bibfield  {author} {\bibinfo {author} {\bibfnamefont {W.}~\bibnamefont
  {Huggins}}, \bibinfo {author} {\bibfnamefont {P.}~\bibnamefont {Patil}},
  \bibinfo {author} {\bibfnamefont {B.}~\bibnamefont {Mitchell}}, \bibinfo
  {author} {\bibfnamefont {K.~B.}\ \bibnamefont {Whaley}}, \ and\ \bibinfo
  {author} {\bibfnamefont {E.~M.}\ \bibnamefont {Stoudenmire}},\ }\href@noop {}
  {\bibfield  {journal} {\bibinfo  {journal} {Quantum Sci. Technol.}\ }\textbf
  {\bibinfo {volume} {4}},\ \bibinfo {pages} {024001} (\bibinfo {year}
  {2019}{\natexlab{b}})}\BibitemShut {NoStop}%
\bibitem [{\citenamefont {Brassard}\ \emph {et~al.}(2002)\citenamefont
  {Brassard}, \citenamefont {Hoyer}, \citenamefont {Mosca},\ and\ \citenamefont
  {Tapp}}]{brassard2002quantum}%
  \BibitemOpen
  \bibfield  {author} {\bibinfo {author} {\bibfnamefont {G.}~\bibnamefont
  {Brassard}}, \bibinfo {author} {\bibfnamefont {P.}~\bibnamefont {Hoyer}},
  \bibinfo {author} {\bibfnamefont {M.}~\bibnamefont {Mosca}}, \ and\ \bibinfo
  {author} {\bibfnamefont {A.}~\bibnamefont {Tapp}},\ }\href@noop {} {\bibfield
   {journal} {\bibinfo  {journal} {Contemporary Mathematics}\ }\textbf
  {\bibinfo {volume} {305}},\ \bibinfo {pages} {53} (\bibinfo {year}
  {2002})}\BibitemShut {NoStop}%
\bibitem [{\citenamefont {Berry}\ \emph {et~al.}(2014)\citenamefont {Berry},
  \citenamefont {Childs}, \citenamefont {Cleve}, \citenamefont {Kothari},\ and\
  \citenamefont {Somma}}]{Berry_oaa_2014}%
  \BibitemOpen
  \bibfield  {author} {\bibinfo {author} {\bibfnamefont {D.~W.}\ \bibnamefont
  {Berry}}, \bibinfo {author} {\bibfnamefont {A.~M.}\ \bibnamefont {Childs}},
  \bibinfo {author} {\bibfnamefont {R.}~\bibnamefont {Cleve}}, \bibinfo
  {author} {\bibfnamefont {R.}~\bibnamefont {Kothari}}, \ and\ \bibinfo
  {author} {\bibfnamefont {R.~D.}\ \bibnamefont {Somma}},\ }in\ \href {\doibase
  10.1145/2591796.2591854} {\emph {\bibinfo {booktitle} {Proceedings of the
  Forty-sixth Annual ACM Symposium on Theory of Computing}}},\ \bibinfo {series
  and number} {STOC '14}\ (\bibinfo  {publisher} {ACM},\ \bibinfo {address}
  {New York, NY, USA},\ \bibinfo {year} {2014})\ pp.\ \bibinfo {pages}
  {283--292}\BibitemShut {NoStop}%
\bibitem [{\citenamefont {Seeley}\ \emph {et~al.}(2012)\citenamefont {Seeley},
  \citenamefont {Richard},\ and\ \citenamefont {Love}}]{seeley2012bravyi}%
  \BibitemOpen
  \bibfield  {author} {\bibinfo {author} {\bibfnamefont {J.~T.}\ \bibnamefont
  {Seeley}}, \bibinfo {author} {\bibfnamefont {M.~J.}\ \bibnamefont {Richard}},
  \ and\ \bibinfo {author} {\bibfnamefont {P.~J.}\ \bibnamefont {Love}},\
  }\href@noop {} {\bibfield  {journal} {\bibinfo  {journal} {J. Chem. Phys.}\
  }\textbf {\bibinfo {volume} {137}},\ \bibinfo {pages} {224109} (\bibinfo
  {year} {2012})}\BibitemShut {NoStop}%
\bibitem [{\citenamefont {Sun}\ \emph {et~al.}(2018)\citenamefont {Sun},
  \citenamefont {Berkelbach}, \citenamefont {Blunt}, \citenamefont {Booth},
  \citenamefont {Guo}, \citenamefont {Li}, \citenamefont {Liu}, \citenamefont
  {McClain}, \citenamefont {Sayfutyarova}, \citenamefont {Sharma} \emph
  {et~al.}}]{sun2018pyscf}%
  \BibitemOpen
  \bibfield  {author} {\bibinfo {author} {\bibfnamefont {Q.}~\bibnamefont
  {Sun}}, \bibinfo {author} {\bibfnamefont {T.~C.}\ \bibnamefont {Berkelbach}},
  \bibinfo {author} {\bibfnamefont {N.~S.}\ \bibnamefont {Blunt}}, \bibinfo
  {author} {\bibfnamefont {G.~H.}\ \bibnamefont {Booth}}, \bibinfo {author}
  {\bibfnamefont {S.}~\bibnamefont {Guo}}, \bibinfo {author} {\bibfnamefont
  {Z.}~\bibnamefont {Li}}, \bibinfo {author} {\bibfnamefont {J.}~\bibnamefont
  {Liu}}, \bibinfo {author} {\bibfnamefont {J.~D.}\ \bibnamefont {McClain}},
  \bibinfo {author} {\bibfnamefont {E.~R.}\ \bibnamefont {Sayfutyarova}},
  \bibinfo {author} {\bibfnamefont {S.}~\bibnamefont {Sharma}},  \emph
  {et~al.},\ }\href@noop {} {\bibfield  {journal} {\bibinfo  {journal} {Wiley
  Interdiscip. Rev. Comput. Mol. Sci.}\ }\textbf {\bibinfo {volume} {8}},\
  \bibinfo {pages} {e1340} (\bibinfo {year} {2018})}\BibitemShut {NoStop}%
\bibitem [{\citenamefont {McClean}\ \emph {et~al.}(2017)\citenamefont
  {McClean}, \citenamefont {Sung}, \citenamefont {Kivlichan}, \citenamefont
  {Cao}, \citenamefont {Dai}, \citenamefont {Fried}, \citenamefont {Gidney},
  \citenamefont {Gimby}, \citenamefont {Gokhale}, \citenamefont {H{\"a}ner}
  \emph {et~al.}}]{mcclean2017openfermion}%
  \BibitemOpen
  \bibfield  {author} {\bibinfo {author} {\bibfnamefont {J.~R.}\ \bibnamefont
  {McClean}}, \bibinfo {author} {\bibfnamefont {K.~J.}\ \bibnamefont {Sung}},
  \bibinfo {author} {\bibfnamefont {I.~D.}\ \bibnamefont {Kivlichan}}, \bibinfo
  {author} {\bibfnamefont {Y.}~\bibnamefont {Cao}}, \bibinfo {author}
  {\bibfnamefont {C.}~\bibnamefont {Dai}}, \bibinfo {author} {\bibfnamefont
  {E.~S.}\ \bibnamefont {Fried}}, \bibinfo {author} {\bibfnamefont
  {C.}~\bibnamefont {Gidney}}, \bibinfo {author} {\bibfnamefont
  {B.}~\bibnamefont {Gimby}}, \bibinfo {author} {\bibfnamefont
  {P.}~\bibnamefont {Gokhale}}, \bibinfo {author} {\bibfnamefont
  {T.}~\bibnamefont {H{\"a}ner}},  \emph {et~al.},\ }\href@noop {} {\bibfield
  {journal} {\bibinfo  {journal} {arXiv preprint arXiv:1710.07629}\ } (\bibinfo
  {year} {2017})}\BibitemShut {NoStop}%
\bibitem [{\citenamefont {Hsieh}\ \emph {et~al.}(2012)\citenamefont {Hsieh},
  \citenamefont {Shim}, \citenamefont {Korkusinski},\ and\ \citenamefont
  {Hawrylak}}]{Hsieh2012}%
  \BibitemOpen
  \bibfield  {author} {\bibinfo {author} {\bibfnamefont {C.-Y.}\ \bibnamefont
  {Hsieh}}, \bibinfo {author} {\bibfnamefont {Y.-P.}\ \bibnamefont {Shim}},
  \bibinfo {author} {\bibfnamefont {M.}~\bibnamefont {Korkusinski}}, \ and\
  \bibinfo {author} {\bibfnamefont {P.}~\bibnamefont {Hawrylak}},\ }\href@noop
  {} {\bibfield  {journal} {\bibinfo  {journal} {Rep. Prog. Phys.}\ }\textbf
  {\bibinfo {volume} {75}},\ \bibinfo {pages} {114501} (\bibinfo {year}
  {2012})}\BibitemShut {NoStop}%
\bibitem [{\citenamefont {Delgado}\ \emph {et~al.}(2008)\citenamefont
  {Delgado}, \citenamefont {Shim}, \citenamefont {Korkusinski}, \citenamefont
  {Gaudreau}, \citenamefont {Studenikin}, \citenamefont {Sachrajda},\ and\
  \citenamefont {Hawrylak}}]{Delgado2008}%
  \BibitemOpen
  \bibfield  {author} {\bibinfo {author} {\bibfnamefont {F.}~\bibnamefont
  {Delgado}}, \bibinfo {author} {\bibfnamefont {Y.-P.}\ \bibnamefont {Shim}},
  \bibinfo {author} {\bibfnamefont {M.}~\bibnamefont {Korkusinski}}, \bibinfo
  {author} {\bibfnamefont {L.}~\bibnamefont {Gaudreau}}, \bibinfo {author}
  {\bibfnamefont {S.~A.}\ \bibnamefont {Studenikin}}, \bibinfo {author}
  {\bibfnamefont {A.~S.}\ \bibnamefont {Sachrajda}}, \ and\ \bibinfo {author}
  {\bibfnamefont {P.}~\bibnamefont {Hawrylak}},\ }\href@noop {} {\bibfield
  {journal} {\bibinfo  {journal} {Phys. Rev. Lett.}\ }\textbf {\bibinfo
  {volume} {101}},\ \bibinfo {pages} {226810} (\bibinfo {year}
  {2008})}\BibitemShut {NoStop}%
\bibitem [{\citenamefont {Brown}\ \emph {et~al.}(2016)\citenamefont {Brown},
  \citenamefont {Kim},\ and\ \citenamefont {Monroe}}]{brown2016iontrap}%
  \BibitemOpen
  \bibfield  {author} {\bibinfo {author} {\bibfnamefont {K.~R.}\ \bibnamefont
  {Brown}}, \bibinfo {author} {\bibfnamefont {J.}~\bibnamefont {Kim}}, \ and\
  \bibinfo {author} {\bibfnamefont {C.}~\bibnamefont {Monroe}},\ }\href@noop {}
  {\bibfield  {journal} {\bibinfo  {journal} {npj Quantum Inf.}\ }\textbf
  {\bibinfo {volume} {2}},\ \bibinfo {pages} {16034} (\bibinfo {year}
  {2016})}\BibitemShut {NoStop}%
\bibitem [{\citenamefont {Bruzewicz}\ \emph {et~al.}(2019)\citenamefont
  {Bruzewicz}, \citenamefont {Chiaverini}, \citenamefont {McConnell},\ and\
  \citenamefont {Sage}}]{bruzewicz2019iontrapreview}%
  \BibitemOpen
  \bibfield  {author} {\bibinfo {author} {\bibfnamefont {C.~D.}\ \bibnamefont
  {Bruzewicz}}, \bibinfo {author} {\bibfnamefont {J.}~\bibnamefont
  {Chiaverini}}, \bibinfo {author} {\bibfnamefont {R.}~\bibnamefont
  {McConnell}}, \ and\ \bibinfo {author} {\bibfnamefont {J.~M.}\ \bibnamefont
  {Sage}},\ }\href@noop {} {\bibfield  {journal} {\bibinfo  {journal} {Appl.
  Phys. Rev.}\ }\textbf {\bibinfo {volume} {6}},\ \bibinfo {pages} {021314}
  (\bibinfo {year} {2019})}\BibitemShut {NoStop}%
\bibitem [{\citenamefont {Gao}\ and\ \citenamefont
  {Duan}(2017{\natexlab{b}})}]{gao2017efficient}%
  \BibitemOpen
  \bibfield  {author} {\bibinfo {author} {\bibfnamefont {X.}~\bibnamefont
  {Gao}}\ and\ \bibinfo {author} {\bibfnamefont {L.-M.}\ \bibnamefont {Duan}},\
  }\href@noop {} {\bibfield  {journal} {\bibinfo  {journal} {Nat. Commun.}\
  }\textbf {\bibinfo {volume} {8}},\ \bibinfo {pages} {662} (\bibinfo {year}
  {2017}{\natexlab{b}})}\BibitemShut {NoStop}%
\bibitem [{\citenamefont {O’Malley}\ \emph {et~al.}(2016)\citenamefont
  {O’Malley}, \citenamefont {Babbush}, \citenamefont {Kivlichan},
  \citenamefont {Romero}, \citenamefont {McClean}, \citenamefont {Barends},
  \citenamefont {Kelly}, \citenamefont {Roushan}, \citenamefont {Tranter},
  \citenamefont {Ding} \emph {et~al.}}]{o2016scalable}%
  \BibitemOpen
  \bibfield  {author} {\bibinfo {author} {\bibfnamefont {P.~J.}\ \bibnamefont
  {O’Malley}}, \bibinfo {author} {\bibfnamefont {R.}~\bibnamefont {Babbush}},
  \bibinfo {author} {\bibfnamefont {I.~D.}\ \bibnamefont {Kivlichan}}, \bibinfo
  {author} {\bibfnamefont {J.}~\bibnamefont {Romero}}, \bibinfo {author}
  {\bibfnamefont {J.~R.}\ \bibnamefont {McClean}}, \bibinfo {author}
  {\bibfnamefont {R.}~\bibnamefont {Barends}}, \bibinfo {author} {\bibfnamefont
  {J.}~\bibnamefont {Kelly}}, \bibinfo {author} {\bibfnamefont
  {P.}~\bibnamefont {Roushan}}, \bibinfo {author} {\bibfnamefont
  {A.}~\bibnamefont {Tranter}}, \bibinfo {author} {\bibfnamefont
  {N.}~\bibnamefont {Ding}},  \emph {et~al.},\ }\href@noop {} {\bibfield
  {journal} {\bibinfo  {journal} {Phys. Rev. X}\ }\textbf {\bibinfo {volume}
  {6}},\ \bibinfo {pages} {031007} (\bibinfo {year} {2016})}\BibitemShut
  {NoStop}%
\bibitem [{\citenamefont {Carleo}\ \emph {et~al.}(2018)\citenamefont {Carleo},
  \citenamefont {Nomura},\ and\ \citenamefont
  {Imada}}]{carleo2018constructing}%
  \BibitemOpen
  \bibfield  {author} {\bibinfo {author} {\bibfnamefont {G.}~\bibnamefont
  {Carleo}}, \bibinfo {author} {\bibfnamefont {Y.}~\bibnamefont {Nomura}}, \
  and\ \bibinfo {author} {\bibfnamefont {M.}~\bibnamefont {Imada}},\
  }\href@noop {} {\bibfield  {journal} {\bibinfo  {journal} {Nat. Commun.}\
  }\textbf {\bibinfo {volume} {9}},\ \bibinfo {pages} {5322} (\bibinfo {year}
  {2018})}\BibitemShut {NoStop}%
\bibitem [{\citenamefont {Motta}\ \emph {et~al.}(2019)\citenamefont {Motta},
  \citenamefont {Sun}, \citenamefont {Tan}, \citenamefont {O’Rourke},
  \citenamefont {Ye}, \citenamefont {Minnich}, \citenamefont {Brand{\~a}o},\
  and\ \citenamefont {Chan}}]{motta2019quantum}%
  \BibitemOpen
  \bibfield  {author} {\bibinfo {author} {\bibfnamefont {M.}~\bibnamefont
  {Motta}}, \bibinfo {author} {\bibfnamefont {C.}~\bibnamefont {Sun}}, \bibinfo
  {author} {\bibfnamefont {A.~T.}\ \bibnamefont {Tan}}, \bibinfo {author}
  {\bibfnamefont {M.~J.}\ \bibnamefont {O’Rourke}}, \bibinfo {author}
  {\bibfnamefont {E.}~\bibnamefont {Ye}}, \bibinfo {author} {\bibfnamefont
  {A.~J.}\ \bibnamefont {Minnich}}, \bibinfo {author} {\bibfnamefont {F.~G.}\
  \bibnamefont {Brand{\~a}o}}, \ and\ \bibinfo {author} {\bibfnamefont
  {G.~K.-L.}\ \bibnamefont {Chan}},\ }\href@noop {} {\bibfield  {journal}
  {\bibinfo  {journal} {Nat. Phys.}\ ,\ \bibinfo {pages} {1}} (\bibinfo {year}
  {2019})}\BibitemShut {NoStop}%
\end{thebibliography}
\end{document}